\documentclass[11pt,a4paper]{article}
\pdfoutput=1
\usepackage{amssymb,amsmath,amsfonts, mathtools, mathrsfs}
\usepackage[utf8]{inputenc} 
\usepackage[dvipsnames]{xcolor}
\usepackage{float}
\usepackage{soul}
\usepackage{cancel}

\newif\ifnatbibsort\natbibsorttrue

\relax

\ifnatbibsort\RequirePackage[numbers,sort&compress]{natbib}\else\RequirePackage[numbers,compress]{natbib}\fi
\RequirePackage[colorlinks=true
,urlcolor=blue
,anchorcolor=blue
,citecolor=blue
,filecolor=blue
,linkcolor=blue
,menucolor=blue
,pagecolor=blue
,linktocpage=true
,pdfproducer=medialab
,pdfa=true
]{hyperref}

\usepackage{graphicx}
\usepackage{caption}
\usepackage{tensor}
\usepackage{subfigure}
\usepackage{enumerate}
\usepackage{dsfont}
\usepackage{verbatim}
\usepackage{amsfonts,euscript,amssymb,stmaryrd,braket}
\usepackage{tikz}
\usetikzlibrary{arrows,decorations.markings,patterns}
\usepackage{comment}
\usepackage{slashed}
\usepackage[normalem]{ulem}

\def\clock{{\count0=\time
		\divide\count0 60
		\ifnum\count0<10 0\fi\the\count0
		\multiply\count0 -60 \advance\count0 \time
		:\ifnum\count0<10 0\fi \the\count0
}}
\newcommand{\timestamp}{{\small\vbox{\hbox{\tt\jobname.tex}
			\hbox{\the\day/\the\month/\the\year, \clock}}}}

\newcommand{\be}{\begin{equation}}
\newcommand{\ee}{\end{equation}}
\newcommand{\bea}{\begin{eqnarray}}
\newcommand{\eea}{\end{eqnarray}}
\newcommand{\nn}{\nonumber}

\makeatletter
\let\old@startsection=\@startsection
\let\oldl@section=\l@section
\renewcommand{\@startsection}[6]{\old@startsection{#1}{#2}{#3}{#4}{#5}{#6\mathversion{bold}}}
\renewcommand{\l@section}[2]{\oldl@section{\mathversion{bold}#1}{#2}}
\makeatother

\numberwithin{equation}{section}

\def \RR {{\mathbb R}}
\def \CC {{\mathbb C}}

\def \ZZ {{\mathbb Z}}
\def \NN {{\mathbb N}}
\def\ri {{\rm i}}
\def\rd {{\rm d}}
\def\e {{\rm e}}
\def\prt {{\partial}}

\def \EG \mathrm{\tiny E}

\begin{document}
	\renewcommand{\thefootnote}{\arabic{footnote}}

	\overfullrule=0pt
	\parskip=2pt
	\parindent=12pt
	\headheight=0in \headsep=0in \topmargin=0in \oddsidemargin=0in

	\vspace{ -3cm} \thispagestyle{empty} \vspace{-1cm}
	\begin{flushright} 
		\footnotesize
		\textcolor{red}{\phantom{print-report}}
	\end{flushright}

	\begin{center}
	\vspace{.0cm}

	{\Large\bf \mathversion{bold}
	Entanglement entropies of an interval
	}
	\\
	\vspace{.25cm}
	\noindent
	{\Large\bf \mathversion{bold}
	in the free Schr\"odinger field theory on the half line}

	\vspace{0.8cm} {
		Mihail Mintchev$^{\,a}$,
		Diego Pontello$^{\,b}$
		and Erik Tonni$^{\,b}$
	}
	\vskip  0.7cm
	
	\small
	{\em
		$^{a}\,$Dipartimento di Fisica, Universit\'a di Pisa and INFN Sezione di Pisa, \\
		largo Bruno Pontecorvo 3, 56127 Pisa, Italy
		\vskip 0.05cm
		$^{b}\,$SISSA and INFN Sezione di Trieste, via Bonomea 265, 34136, Trieste, Italy 
	}
	\normalsize

\end{center}

\vspace{0.3cm}
\begin{abstract} 

We study the entanglement entropies of an interval adjacent to the boundary
of the half line for the free fermionic spinless Schr\"odinger field theory at finite density and zero temperature,
with either Neumann or Dirichlet boundary conditions.
They are finite functions of the dimensionless parameter
given by the product of the Fermi momentum and the length of the interval. 
The entanglement entropy displays an oscillatory behaviour, 
differently from the case of the interval on the whole line.
This behaviour is related to the Friedel oscillations 
of the mean particle density on the half line at the entangling point. 
We find analytic expressions for the expansions of the entanglement entropies
in the regimes of small and large values of the dimensionless parameter.
They display a remarkable agreement with the curves obtained numerically.
The analysis is extended to a family of free fermionic Lifshitz models
labelled by their integer Lifshitz exponent,
whose parity determines the properties of the entanglement entropies. 
The cumulants of the local charge operator 
and the Schatten norms of the underlying kernels are also explored. 
\end{abstract}

\newpage

\tableofcontents

\section{Introduction}

The bipartite entanglement corresponding to a spatial bipartition has been 
intensively investigated in the past three decades 
by employing methods of quantum field theories, 
quantum may-body systems and quantum gravity
(see e.g. the reviews \cite{Calabrese-Doyon, EislerPeschel:2009review, Casini:2009sr, Eisert:2008ur,Rangamani:2016dms, Headrick:2019eth, Tonni:2020bjq}).

Consider a quantum system in a state $\rho$ and a bipartition 
of the space $A \cup B$ which provides a corresponding factorisation of the Hilbert space $\mathcal{H} = \mathcal{H}_A \otimes \mathcal{H}_B$.
When $\rho$ is pure, the bipartite entanglement is measured by the entanglement entropy $S_A$, which is defined as the von Neumann entropy of the reduced density matrix $\rho_A \equiv \textrm{Tr}_{\mathcal{H}_B}(\rho)$, namely
\be
\label{ee-def-intro}
S_A
\,\equiv\,
  -\,\textrm{Tr}\big(\rho_A \log \rho_A\big) 
  \,=\,
  \lim_{\alpha \to 1} S_A^{(\alpha)} 
\ee
(hereafter the notation
$\textrm{Tr} (\dots)\equiv \textrm{Tr}_{\mathcal{H}_A} (\dots)$ is adopted). 
The entanglement entropy can be obtained also 
through the replica limit, i.e. 
the analytic continuation $\alpha \to 1$ 
of the R\'enyi entropies 
\be
\label{renyi-def-intro}
S_A^{(\alpha)} 
\equiv \frac{1}{1-\alpha}\,\log\!\big[\textrm{Tr}(\rho_A^\alpha)\big]
\ee
where $\alpha \neq 1$ is a real and positive parameter;
hence we  identify $S_A^{(1)} \equiv S_A$.
The single copy entanglement \cite{peschel-05,eisert-05,orus-06}
is obtained as the limit $\alpha \to +\infty$ of the 
R\'enyi entropies (\ref{renyi-def-intro}),
where $\textrm{Tr}(\rho_A^\alpha) = \sum_j \lambda_j^\alpha$
in terms of the eigenvalues $\lambda_j \in [0,1]$ of $\rho_A$;
hence $S_A^{(\infty)}= - \log (\lambda_{\textrm{\tiny max}})$,
with $\lambda_{\textrm{\tiny max}} $ being the largest eigenvalue of $\rho_A$.
The entanglement entropies include the entanglement entropy $S_A$,
the R\'enyi entropies $S_A^{(\alpha)} $
and the single copy entanglement $S_A^{(\infty)}$.

For relativistic quantum field theories in $d+1$ spacetime dimensions and  in their ground state,
the entanglement entropies of a region $A$ are divergent quantities 
as the ultraviolet (UV) cutoff $\epsilon$ vanishes
and the leading divergence 
$S_A^{(\alpha)} \propto \textrm{Area}(\partial A)/\epsilon^{d-1} +\cdots$ 
provides the area law 
\cite{Bombelli:1986rw, Srednicki:1993im,Ryu:2006bv,Ryu:2006ef},
where the dots denote subleading terms as $\epsilon\to 0$.
An important exception to this behavior is observed for conformal field theories in $d=1$,
where for the entanglement entropies of an interval $A = [-R,R]$ on the line
we have $S_A^{(\alpha)} = \tfrac{c}{6} (1+\frac{1}{\alpha}) \log(2R/\epsilon) + \cdots$ as $\epsilon\to 0$, being $c$ the central charge of the model
\cite{Callan:1994py, Holzhey:1994we, Calabrese:2004eu}.
In the presence of spatial boundaries,
the entanglement entropies depend also on the boundary conditions (b.c.).
For instance, in a $d=1$ boundary conformal field theory 
on the half line $x\geqslant 0$ and in its ground state, 
for the entanglement entropies of an interval $A = [0,R]$ adjacent to the boundary
it has been found \cite{Calabrese:2004eu} that
$S_A^{(\alpha)} = \tfrac{c}{12} (1+\frac{1}{\alpha}) \log(2R/\epsilon) + \cdots$
and that the subleading constant term contains the 
Affleck-Ludwig boundary entropy \cite{Affleck:1991tk},
which encodes the boundary conditions and
provides a monotonic function along a boundary renormalization group flow 
\cite{Friedan:2003yc, Casini:2016fgb, Casini:2018nym}.

The properties of the bipartite entanglement in a quantum field theory 
depend on the nature of the spacetime symmetry.
In order to gain some new insights on this relation, it is worth investigating
the bipartite entanglement in non-relativistic quantum field theories. 
Insightful non-relativistic models exhibit the Lifshitz invariance
\cite{Hertz-76, Niederer-72, Hagen-72, Henkel:1993sg, 
Nishida:2007pj, Hartong:2014pma, Hartong-15},
where the time and space coordinates scale in a different way,
characterised by the Lifshitz exponent $z>0$
(relativistic field theories have $z=1$).
Various quantities in these models have been studied,
also in higher dimensions,
including the entanglement entropies
\cite{Ardonne:2003wa, Gioev:2006zz, Wolf:2006zzb,
Fradkin:2006mb, Hsu:2008af, Fradkin:2009dus, Solodukhin:2009sk, 
Spitzer-14, Keranen:2016ija}.
A remarkable property of the entanglement entropies for the free
fermions at finite density in generic dimension 
is the violation of the area law \cite{Gioev:2006zz, Wolf:2006zzb,Sobolev-10}.

We focus on the $d=1$ free fermionic spinless Schr\"odinger field theory 
at zero temperature and finite density $\mu$.
This is a free non-relativistic quantum field theory with $z=2$
which describes the dilute spinless Fermi gas in $d=1$ 
\cite{Benfatto-book,sachdev_book}.
When this model is defined on the line, 
the entanglement entropies of an interval $[-R,R] \subset \mathbb{R}$ 
have been studied in \cite{Mintchev:2022xqh},
finding that they are finite functions of the dimensionless parameter 
$\eta \equiv R\, k_{\textrm{\tiny F}} \geqslant 0$, where $k_{\textrm{\tiny F}}$ is the Fermi momentum,
and that the entanglement entropy $S_A$ is a monotonically increasing function
of $\eta$\,.
The $\mu = 0$ case has been considered earlier in 
\cite{Pal:2017ntk, Hason-17, Hartmann:2021vrt}.

In this manuscript we investigate the above mentioned Schr\"odinger field theory on the half line $x\geqslant 0$ with scale invariant boundary conditions (that are of either Neumann or Dirichlet type) imposed at the origin $x=0$.
In these models the mean value of the particle density exhibits Friedel oscillations,
which depend on the boundary conditions 
and decay with the distance from the boundary
\cite{Friedel-52}.
We study the entanglement entropies of the interval $A=[0,R]$
adjacent to the boundary of the half line. 
We find that also these entanglement entropies are finite functions of 
the dimensionless parameter $\eta \equiv R\, k_{\textrm{\tiny F}} \geqslant 0$.
In these models the entanglement entropy displays an oscillatory behaviour,
differently from the entanglement entropy of the interval on the line
considered in \cite{Mintchev:2022xqh}.
We remark that in our analyses
the dispersion relation $\omega(k)\propto k^2$ is not approximated 
through a linear dispersion relation at the Fermi points (Tomonaga’s approximation) \cite{Giamarchi_book, Glazman_12}.

The finiteness of the entanglement entropies
in these models on the half line is a consequence of 
the analogous property which holds  
for the entanglement entropies of an interval on the line \cite{Mintchev:2022xqh}.
The latter follows from the properties of the 
solution of the sine kernel spectral problem in the interval on the line,
which has been found in a series of seminal papers by Slepian, Pollak and Landau
\cite{Slepian-part-1, Slepian-part-2, Slepian-part-3, Slepian-part-4}
and it is written in terms of the prolate spheroidal wave functions (PSWF) of order zero
(see also the overview \cite{Slepian-83} and the recent book \cite{Rokhlin-book}).
Also the numerical evaluation of these functions has been carefully investigated 
(see \cite{Rokhlin-book} and references therein).
The relevance of this spectral problem
for the entanglement in free fermionic systems
has been highlighted in \cite{EislerPeschelProlate}.

The procedure described in \cite{Mintchev:2022xqh}
for the entanglement entropies of the interval on the line,
which follows the one discussed in 
\cite{Jin_2004,Keating_04, Calabrese:2009us, Calabrese-Essler-10}
for some lattice models,
can be adapted to the entanglement entropies of an interval 
adjacent to the boundary of the half line in a straightforward way
and this leads us to write analytic expressions 
for the expansions of the entanglement entropies 
in the regimes of small and large values of $\eta$.
These results are based on the 
expansions of the Bessel kernel tau function
reported in
\cite{Tracy:1993xj, Gamayun:2013auu, Bonelli:2016qwg, Bothner_2019},
specialised to two specific values of the parameter 
in the Bessel kernel. 
We remark that the complete expansions found in \cite{Gamayun:2013auu, Bonelli:2016qwg}
have been obtained by applying to the Painlev\'e $\textrm{III}_1$ equation 
the method (Kyiv formula) introduced in \cite{Gamayun:2012ma} for the Painlev\'e VI equation.
Also some results \cite{Basor-01,Basor2002,Basor2008,Deift-11,Fagotti:2010cc} obtained in lattice models are relevant for our analyses.

The outline of this manuscript is as follows. 
In Sec.\,\ref{sec_model} we briefly describe
the free fermionic spinless Schr\"odinger field theory on the half line
at finite density and finite temperature, focussing on the zero temperature limit 
and on the scale invariant boundary conditions. 
The entanglement entropies of the interval $A=[0,R]$ for this model,
which are the main results of this manuscript,
are discussed in Sec.\,\ref{sec_entropies}.
In Sec.\,\ref{sec_lifshitz} we extend the analysis 
to a hierarchy of Lifshitz fermion fields 
with integer Lifshitz exponents $z\geqslant 1$.
The expansions of the entanglement entropies 
as $\eta \to 0$ and $\eta \to \infty$ are investigated 
in Sec.\,\ref{sec_small_eta} and Sec.\,\ref{sec_large_eta} 
respectively. 
In Sec.\,\ref{sec_cumulants_ee} we explore the Schatten norms
and the relation between $S_A$ and the charge cumulants 
\cite{Klich-Levitov-09, Klich-Song-11, Klich-Laflorencie-12}.
Some conclusions are drawn in Sec.\,\ref{sec_conclusions}.
The Appendices\;\ref{app_bessel}, \ref{app_small_eta} and \ref{app_large_eta}
contain the derivations of some results reported in the main text 
and also further technical details.

\section{Free Schr\"odinger field theory on the half line at finite density}
\label{sec_model}

The dynamics of the free fermionic Schr\"odinger field theory on the half line $x\geqslant0$ is defined by 
the equation of motion 
\begin{equation}
\left (\ri\, \prt_t +\frac{1}{2m} \,\prt_x^2\right )\psi (t,x) = 0 
\label{e1}
\end{equation} 
where $m>0$ is the mass and $\psi$ is a complex quantum field.
This field satisfies the equal-time canonical anticommutation relations 
\be
\label{e2a-b}
\big\{\psi (t,x_1)\, ,\, \psi^*(t,x_2) \big\} = \delta(x_1-x_2) 
\;\;\qquad\;\;
\big\{\psi (t,x_1)\, ,\, \psi (t,x_2) \big\} = \big\{\psi^* (t,x_1)\, ,\, \psi^* (t,x_2) \big\} = 0
\ee
and the boundary condition 
\begin{equation} 
\lim_{x\to 0^+} \!\big(\,\prt_x -\vartheta \,\big)\psi (t,x) = 0
\label{h1} 
\end{equation} 
where the parameter $\vartheta$ has dimension of mass and parametrizes all self-adjoint 
extensions of the Hamiltonian $- \tfrac{1}{2m}\, \prt_x^2$ on the half line \cite{ReedSimon-book}.

The solution of the
boundary value problem defined by \eqref{e1}-\eqref{h1} for $\vartheta \geqslant 0$ reads
\begin{equation} 
\label{h2} 
\psi (t,x)  = \int_{0}^{\infty} \!
\e^{-\ri \omega (k) t} \left (\e^{\ri k x}+\frac{k+\ri \vartheta}{k-\ri \vartheta}\; \e^{-\ri k x}\right ) a (k) \; \frac{\rd k}{2\pi }
\;\;\; \qquad \;\;\;
\omega(k) = \frac{k^2}{2m} 
\end{equation} 
where the oscillators $\{a (k)\,:\,k \geqslant0\}$ and their Hermitian conjugates 
$\{a^*(k)\, :\, k \geqslant0\}$ generate a standard canonical anticommutation relation algebra $\mathcal{A}$ 
\begin{equation}
\big\{a(k)\, ,\, a^*(p)\big\} = 2\pi\, \delta(k-p)
\;\;\;\qquad\;\;\;
\big\{a(k)\, ,\, a(p) \big\} = \big\{a^*(k)\, ,\, a^*(p) \big\} = 0 \,.
\label{e5}
\end{equation} 
The phase factor $\frac{k+\ri \vartheta}{k-\ri \vartheta}$ in the integrand of \eqref{h2} describes the reflection from the boundary at $x=0$. 
For $\vartheta < 0$, 
in addition to the scattering states there exists a bound state $\e^{\vartheta x}$ with energy 
$\omega_{\textrm{\tiny b}}(\vartheta ) = \vartheta^2/(2m)$. 
Since in this paper we focus on the scale invariant points 
$\vartheta =0$ and $\vartheta = \infty$, 
for details in treating this bound state we refer to \cite{Mintchev:2016dmf}.

In order to implement the finite density condition, we adopt the Gibbs representation 
of the algebra $\mathcal{A}$. 
In this representation, the basic two-point correlators are \cite{Brattelli2}
\bea
\label{g1} 
\langle a^*(p)\, a(k) \rangle_{\beta,\mu}  
&=&
\frac{1}{1 + \e^{\beta [\omega(k) - \mu]}} \;
2\pi \, \delta (k-p)   
\\
\rule{0pt}{.8cm}
\label{g2} 
\langle a(p)\, a^*(k) \rangle_{\beta,\mu}  
&=&
\frac{\e^{\beta [\omega(k) - \mu]}}{1 + \e^{\beta [\omega(k) - \mu]}} \;
2\pi \, \delta (k-p)   
\eea
where $\beta >0$ is the inverse temperature and $\mu$ is the chemical potential in the Fermi distribution 
in \eqref{g1}. 
Combining (\ref{h2}) with
 \eqref{g1} and \eqref{g2}, one obtains the following two-point functions 
\bea
\label{cfnd1-theta}
\langle \psi^* (t_1,x_1) \,\psi (t_2,x_2)\rangle_{\beta, \mu} 
\!&=&\!
\int_{-\infty}^\infty \frac{\e^{\ri \omega(k) (t_{12}-\ri \varepsilon)} }{
1 + \e^{\beta [\omega(k) - \mu]}}
\bigg(
\textrm{e}^{-\ri k x_{12}} + \textrm{e}^{-\ri k \widetilde{x}_{12}} 
+
\frac{2\ri\,\vartheta }{k - \ri \vartheta} \; \textrm{e}^{-\ri k \widetilde{x}_{12}} 
\bigg)
\frac{\rd k}{2\pi }
\hspace{1cm}
\\
\rule{0pt}{.9cm}
\label{cfnd2-theta}
\langle \psi (t_1,x_1) \,\psi^* (t_2,x_2)\rangle_{\beta, \mu} 
\!&=&\!
\int_{-\infty}^\infty \frac{\e^{-\ri \omega(k) (t_{12}-\ri \varepsilon)} \, \e^{\beta [\omega(k) - \mu]}}{
1 + \e^{\beta [\omega(k) - \mu]}}
\bigg(
\textrm{e}^{\ri k x_{12}} + \textrm{e}^{\ri k \widetilde{x}_{12}} 
-
\frac{2\ri\,\vartheta }{k + \ri \vartheta} \; \textrm{e}^{\ri k \widetilde{x}_{12}} 
\bigg)
\frac{\rd k}{2\pi }
\nn
\\
& &
\eea
where 
\be 
t_{12} \equiv t_1-t_2 \;\; \qquad \;\; x_{12} \equiv x_1-x_2 \;\; \qquad \;\;  \widetilde{x}_{12}\equiv x_1 +x_2 \,.
\label{notation}
\ee

As mentioned above, in this manuscript we study 
the limiting regimes where $\vartheta =0$ and $\vartheta \to  \infty$, 
which are scale invariant and
define respectively the Neumann $(+)$ and Dirichlet $(-)$ boundary conditions, i.e.
\begin{equation} 
\label{h12} 
\lim_{x\to 0^+} \prt_x \psi_+ (t,x) = 0 
\;\;\;\qquad\;\;\; 
\lim_{x\to 0^+} \psi_- (t,x) = 0\,.
\end{equation} 
From \eqref{h2} one gets 
\begin{equation} 
\psi_\pm (t,x)  = \int_{0}^{\infty}  \!
\e^{-\ri \omega (k) t} \Big(\e^{\ri k x}\pm \e^{-\ri k x}\Big)\, a (k) \;\frac{\rd k}{2\pi }\,.
\label{h3} 
\end{equation} 
Taking the limits $\vartheta \to 0$ and $\vartheta \to \infty$ 
in (\ref{cfnd1-theta}) and (\ref{cfnd2-theta}), one finds
\bea
\label{cfnd1}
\langle \psi^*_\pm (t_1,x_1) \,\psi_\pm (t_2,x_2)\rangle_{\beta, \mu} 
&=&
\int_{-\infty}^{\infty}  \! \e^{\ri \omega(k) (t_{12}-\ri \varepsilon)} \;
\frac{\left (\e^{-\ri k x_{12}} \pm \e^{-\ri k \widetilde{x}_{12}} \right )}{1 + \e^{\beta [\omega(k) - \mu]}} 
\;\frac{\rd k}{2\pi }
\\
\rule{0pt}{.8cm}
\label{cfnd2}
\langle \psi_\pm (t_1,x_1) \, \psi^*_\pm(t_2,x_2)\rangle_{\beta, \mu} 
&=&
\int_{-\infty}^{\infty} \! \e^{-\ri \omega(k) (t_{12}-\ri \varepsilon)} \;
\frac{\left (\e^{\ri k x_{12}} \pm \e^{\ri k \widetilde{x}_{12}} \right )
\e^{\beta [\omega(k) - \mu]} }{1 + \e^{\beta [\omega(k) - \mu]}} 
\;\frac{\rd k}{2\pi }\,.
\hspace{1cm}
\eea

At equal times $t_1=t_2 \equiv t$ and in the zero temperature limit $\beta \to \infty$, 
the integration over $k$ in \eqref{cfnd1} and \eqref{cfnd2} can be easily performed and gives 
\bea
\label{cfnd12}
\langle \psi^*_\pm(t,x_1)\,\psi_\pm(t,x_2)\rangle_{\infty, \mu} 
&=&
 \frac{\sin (k_{\textrm{\tiny F}} x_{12})}{\pi \,x_{12}}  \pm 
\frac{\sin (k_{\textrm{\tiny F}} \widetilde{x}_{12})}{\pi \,\widetilde{x}_{12}} 
\\
\label{cfnd22}
\rule{0pt}{.8cm}
\langle \psi_\pm(t,x_1)\,\psi^*_\pm(t,x_2)\rangle_{\infty, \mu} 
&=&
 \delta (x_{12}) - 
\left[\,
 \frac{\sin (k_{\textrm{\tiny F}} x_{12})}{\pi \,x_{12}}  \pm 
\frac{\sin (k_{\textrm{\tiny F}} \widetilde{x}_{12})}{\pi \,\widetilde{x}_{12}} 
\,\right]
\eea
where $k_{\textrm{\tiny F}}$ is the Fermi momentum 
\be 
k_{\textrm{\tiny F}} \equiv \sqrt {2m\mu} \,.
\label{Fm}
\ee 
In this regime, the correlators \eqref{cfnd1-theta} and \eqref{cfnd2-theta} can be expressed as
\bea
\langle \psi^*(t,x_1)\,\psi(t,x_2)\rangle_{\infty, \mu} 
&=&
\langle \psi^*_+(t,x_1)\,\psi_+(t,x_2)\rangle_{\infty, \mu} 
+
2\ri\,\vartheta
\int_{-k_{\textrm{\tiny F}}}^{k_{\textrm{\tiny F}}}
\frac{ \textrm{e}^{-\ri k \widetilde{x}_{12}} }{k - \ri \vartheta} \; 
\frac{\rd k}{2\pi }
\\
\rule{0pt}{.8cm}
\langle \psi(t,x_1)\,\psi^*(t,x_2)\rangle_{\infty, \mu} 
&=&
\langle \psi_+(t,x_1)\,\psi^*_+(t,x_2)\rangle_{\infty, \mu} 
-
2\ri\,\vartheta
\int_{-k_{\textrm{\tiny F}}}^{k_{\textrm{\tiny F}}}
\frac{ \textrm{e}^{\ri k \widetilde{x}_{12}} }{k + \ri \vartheta} \; 
\frac{\rd k}{2\pi }\,.
\eea

The expression \eqref{cfnd12} allows us to evaluate 
the mean value of the particle densities for the two b.c.'s that we are considering. 
The result is
\begin{equation}
\langle \varrho_\pm (t,x)\rangle_{\infty,\mu} 
\equiv \langle \psi_\pm^*(t,x) \, \psi_\pm (t,x)\rangle_{\infty,\mu} 
=  \frac{k_{\textrm{\tiny F}}}{\pi}   \pm \frac{\sin(2k_{\textrm{\tiny F}} x)}{2\pi \,x} 
=  \langle \varrho(t,x)\rangle_{\infty,\mu} 
 \pm \frac{\sin(2k_{\textrm{\tiny F}} x)}{2\pi \,x} 
\label{dnd}
\end{equation} 
where $\langle \varrho(t,x)\rangle_{\infty,\mu} = k_{\textrm{\tiny F}} / \pi  $ is the mean value of the particle density on the whole line,
which is independent of the position. 
Thus, \eqref{dnd} shows the Friedel-type oscillations \cite{Friedel-52}   
around the particle density on the line,
whose amplitude decays with the distance from the boundary. 
The densities vanish for $\mu=0$, as expected. 
We find it worth considering the following normalised densities
\be
\label{normalised-densities}
\frac{\langle \varrho_\pm (t,x)\rangle_{\infty,\mu} }{k_{\textrm{\tiny F}} } 
= 
\frac{\langle \varrho (t,x)\rangle_{\infty,\mu} }{k_{\textrm{\tiny F}} } 
\pm \frac{\sin(2 \chi)}{2\pi \, \chi}
\;\;\;\qquad\;\;\;
\chi \equiv k_{\textrm{\tiny F}} x
\ee
which are functions of the dimensionless parameter $\chi$.

In this paper we study the entanglement entropies 
of the bipartition $[0,R]\cup [R,\infty)$ of the half line
for the system described above. 
This bipartition of the half line naturally leads us to consider
the normalised densities \eqref{normalised-densities} evaluated at the entangling point $x=R$, which read
\be
\label{normalised-densities-at-R}
\frac{\langle \varrho_\pm (t,R)\rangle_{\infty,\mu} }{k_{\textrm{\tiny F}} } 
= 
\frac{\langle \varrho (t,R)\rangle_{\infty,\mu} }{k_{\textrm{\tiny F}} } 
\pm \frac{\sin(2 \eta)}{2\pi \, \eta}
\;\;\;\qquad\;\;\;
\eta \equiv k_{\textrm{\tiny F}} R\,.
\ee

Another natural quantity to introduce is 
the mean particle number $N_{A, \pm}$ in the interval $A=[0,R]$, namely
\be
\label{mpn1}
N_{A, \pm}
\equiv 
\int_0^R\!  \langle \varrho_\pm (t,x)\rangle_{\infty,\mu}\, \rd x
\,=\,
\frac{\eta}{\pi} 
\pm
\frac{\textrm{Si}(2\eta)}{2\pi} 
\ee
where $\textrm{Si}(z) \equiv \int_0^z \tfrac{\sin(t)}{t}\, \rd t$ is the sine integral function. 
The dimensionless parameter $\eta$ plays a fundamental role throughout our analysis. 
In the regime of large $\eta$, for \eqref{mpn1} we have\footnote{The expansion of $\mathrm{Si}(z)$ as $z \to \infty$ reads \cite{NIST:DLMF}
\be
\mathrm{Si}(z) = \frac{\pi}{2}
- \frac{\cos(z)}{z} \sum_{n=0}^{\infty}\frac{(-1)^{n}(2n)!}{z^{2n}}
-\frac{\sin(z)}{z^2} \sum_{n=0}^{\infty}\frac{(-1)^{n}(2n+1)!}{z^{2n}} 
\ee
where the two series are asymptotic.}
\be
\label{mpn3}
N_{A, \pm}  = \frac{\eta}{\pi} \pm \left(\frac{1}{4}-\frac{\cos(2\eta)}{4\pi\eta}\right)+O(1/\eta^{2}) \hspace{1cm} \eta \rightarrow \infty \,.
\ee

Since for the Schr\"odinger problem on the whole line  \cite{Mintchev:2022xqh}
the mean particle number in the interval $[-R,R]$ is $N_{2A \,\subset \,\mathbb{R}} = 2\eta /\pi$, one can rewrite \eqref{mpn1} as
\be
\label{mpn2}
N_{A, \pm} - \frac{1}{2} \,N_{2A \,\subset \,\mathbb{R}} 
\,=\,
\pm \, \frac{\textrm{Si}(2\eta)}{2\pi}\,.
\ee 
The expressions in \eqref{normalised-densities}, \eqref{mpn2} and \eqref{mpn3} provide the red and blue curves in Fig.\,\ref{fig:friedel-ee}.

\section{Entanglement entropies}
\label{sec_entropies}

The main quantities investigated  in this manuscript are
the entanglement entropies 
(see (\ref{ee-def-intro}) and (\ref{renyi-def-intro}))
for the free Schr\"odinger field 
at finite density and zero temperature
on the half line $x \geqslant 0$ 
when the spatial bipartition 
is given by the interval $A = [0,R]$ and its complement. 
The Gaussian nature of the state in this free fermionic model
allows to compute these entanglement entropies 
through the spectra associated to the spectral problems 
described in Sec.\,\ref{sec_spectral}. 
The entanglement entropies 
are then evaluated in Sec.\,\ref{sec_sub_ee}.

\subsection{Spectral problems}
\label{sec_spectral}

Since we are dealing with a free fermionic model, 
the entanglement entropies can be evaluated from the two-point functions 
on the half line for either Neumann ($+$) or Dirichlet ($-$) b.c., namely
(see \eqref{cfnd12})
\be
\label{kernel-pm-def}
K_\pm(k_{\textrm{\tiny F}};x,y) \equiv \frac{\sin[k_{\textrm{\tiny F}} (x-y)]}{\pi(x-y)} \pm \frac{\sin[k_{\textrm{\tiny F}} (x+y)]}{\pi(x+y)}\,.
\ee
These kernels satisfy  
\be 
\int_0^\infty \! K_\pm(k_{\textrm{\tiny F}};x,z)\, K_\pm(k_{\textrm{\tiny F}};z,y)\, \rd z
 = K_\pm(k_{\textrm{\tiny F}};x,y) 
 \;\;\qquad\;\;
 x, y \geqslant  0
\label{projection}
\ee
and therefore define projection operators on the half line. This property implies  that the finite 
density states, which generate  the correlation functions \eqref{cfnd12} and \eqref{cfnd22}, 
are pure states \cite{powers-70}.

It is straightforward to observe that the sine kernel, 
which provides the two-point function of the same model on the line, 
is related to the kernels \eqref{kernel-pm-def} as follows
\be
K_{\textrm{\tiny sine}}(k_{\textrm{\tiny F}};x,y)  
\equiv 
\frac{\sin[k_{\textrm{\tiny F}}(x-y)]}{\pi(x-y)} 
\,=\,\frac{K_+(k_{\textrm{\tiny F}};x,y)   + K_-(k_{\textrm{\tiny F}};x,y)  }{2}\,.
\ee

Considering the kernels \eqref{kernel-pm-def} reduced to $A = [0,R] \subset \RR^+$,
after rescaling $R$, the corresponding spectral problems read
\be 
\label{k1}
\int_0^1 K_\pm (\eta; x,y) \, f^\pm_n(\eta; y) \, \rd y
\,=\,
 \gamma_n^\pm\, f^\pm_n(\eta ; x) 
 \;\;\qquad\;\;
 x \in \big[0,1\big]
\ee
where $\gamma_n^\pm = \gamma_n^\pm(\eta)$ are functions of $\eta$.
In order to solve \eqref{k1}, first we consider the auxiliary spectral problem 
associated to the sine kernel, i.e. 
\be 
\label{spectral-problem-v2}
\int_{-1}^1   K_{\textrm{\tiny sine}}(\eta ;x,y)  \, f_n(\eta; y) \, \rd y \,=\,  \gamma_n\, f_n(\eta; x) 
 \;\;\qquad\;\;
 x \in \big[\!-\!1,1\big]
\ee 
whose eigenvalues and eigenfuctions can be expressed in terms of the prolate spheroidal wave functions (PSWF) \cite{Morse-Feshbach-book, Flammer-book, abramowitz}. The eigenvalues in \eqref{spectral-problem-v2} can be written in terms of the radial PSWF of zero order $\mathcal{R}_{0n}$ \cite{Slepian-part-1,Rokhlin-book, CadaMoore04}
\be
\label{eigenvalues} 
\gamma_n
\,=\, 
\frac{2\eta}{\pi}\; \mathcal{R}_{0n}(\eta,1)^2 
\;\;\;\qquad\;\;\;
n \in \mathbb{N}_0 
\ee
while the corresponding eigenfunctions are expressed through 
the  angular PSWF of zero order $\mathcal{S}_{0n}$ as follows 
\be
 \label{PSWF-def}
f_n(\eta;x) \,=\, \sqrt{n+\frac{1}{2}} \; \mathcal{S}_{0n}(\eta, x) 
\ee
which also satisfy 
\be 
f_n(\eta;-x) = (-1)^n f_n(\eta ; x) \,.
\label{parity}
\ee

The spectral problems  \eqref{k1} can be related to the sine kernel spectral problem \eqref{spectral-problem-v2} by first rewriting the latter one in the form 
\be 
\label{a}
\int_{-1}^0 
\frac{\sin [\eta (x-y)]}{\pi (x-y)}\; f_n(\eta; y)\, \rd y 
\,+
 \int_0^1 \frac{\sin [\eta (x-y)]}{\pi (x-y)} \; f_n(\eta; y) \, \rd y
\,=\, \gamma_n\, f_n(\eta; x)
 \qquad
 x \in \big[\!-\!1,1\big]\,.
\ee
Changing the variable $y \mapsto -y$  in the first integral of \eqref{a} and using the parity condition \eqref{parity}, one obtains  
\be 
(-1)^n \int_0^1 \frac{\sin [\eta (x+y)]}{\pi (x+y)} \; f_n(\eta; y) \, \rd y
\,+
 \int_0^1  \frac{\sin [\eta (x-y)]}{\pi (x-y)} \;\ f_n(\eta; y) \, \rd y
\,=\, \gamma_n\,  f_n(\eta; x)\,.
\label{k2}
\ee
Finally, by comparing \eqref{k1} and \eqref{k2}, 
for Neumann b.c. we have 
\be
\label{soln-spectral-problem-neumann}
\gamma_n^+ = \gamma_{2n}
\;\;\;\;\qquad\;\;\;\;
f_n^+(\eta;x) = f_{2n}(\eta;x)
\;\;\;\;\;\;\qquad\;\;\;\;
n \in \mathbb{N}_0
\qquad
 x \in \big[0,1\big]
\ee
while for Dirichlet b.c. one gets
\be
\label{soln-spectral-problem-dirichlet}
\gamma_n^- = \gamma_{2n+1}
\;\;\;\qquad\;\;\;
f_n^-(\eta;x) = f_{2n+1}(\eta;x)
\;\;\;\qquad\;\;\;
n \in \mathbb{N}_0
\qquad
 x \in \big[0,1\big]\,.
\ee

The spectrum $\{ \gamma_n \}$ in \eqref{spectral-problem-v2} 
has been extensively discussed in the literature \cite{Rokhlin-book}.
In particular, $\gamma_n \in (0,1)$ for any $n \in \mathbb{N}$ and any $\eta >0$.
For a fixed value of $\eta$, 
these eigenvalues are non-degenerate and decrease with $n$.
Furthermore, $\gamma_n \to 0$ as $n \to \infty$ in a super-exponential way. 
The critical index 
\be
\label{n0-critical}
n_0 \equiv \left\lfloor \frac{2\eta}{\pi} \right\rfloor \in \NN_0
\ee
can be identified where $\gamma_{n_0} \simeq 1/2$.
This critical index allows to partition the spectrum in three different sets 
where $\gamma_n$ behave in a characteristic way \cite{Landau-65, bonami-21}.
For numerical purposes, 
we have used that $\gamma_n \approx 0$ when $n \geqslant 2(n_0+2)$.

\subsection{Entanglement entropies}
\label{sec_sub_ee}

The eigenvalues of the spectral problems \eqref{k1} provide the entanglement entropies of an interval $A = [0,R] \subset \RR^+$. For Neumann and Dirichlet b.c., they are given respectively by
\be
\label{renyi-entropies-def-sums}
S_{A,+}^{(\alpha)} = \sum_{n = 0}^{\infty} s_\alpha(\gamma_n^+) 
\;\;\;\;\qquad\;\;\;\;
S_{A,-}^{(\alpha)} = \sum_{n = 0}^{\infty} s_\alpha(\gamma_n^-) 
\ee
where 
\be
\label{def-s-alpha}
s_\alpha(x) \equiv\, \frac{1}{1-\alpha}\, \log\!\big[ x^\alpha +(1-x)^\alpha\big]\,.
\ee
In the limits $\alpha \to 1$ and $\alpha \to \infty$, this function becomes respectively
\be
\label{s-ee-s-infty-def}
s(x) \equiv -\,x \log(x) - (1-x) \log(1-x) 
\;\;\qquad\;\;
s_\infty(x) \equiv 
\left\{\begin{array}{ll}
\, - \log(1-x) 
\hspace{.5cm}
& x\in [0,1/2]
\\
\rule{0pt}{.5cm}
- \log(x)
& x\in (1/2,1]
\end{array}\right. 
\ee
that are employed in \eqref{renyi-entropies-def-sums} to evaluate 
the entanglement entropy $S_{A,\pm}$ and the single copy entanglement 
$S_{A,\pm}^{(\infty)}$ as follows 
\be
\label{replica limit-sec}
\lim_{\alpha \to 1} S_{A,\pm}^{(\alpha)} = S_{A,\pm}
\;\;\;\;\qquad\;\;\;\;
\lim_{\alpha \to \infty} S_{A,\pm}^{(\alpha)} = S_{A,\pm}^{(\infty)}\,.
\ee

\begin{figure}[t!]
\vspace{-.2cm}
\hspace{-2cm}
\includegraphics[width=1.15\textwidth]{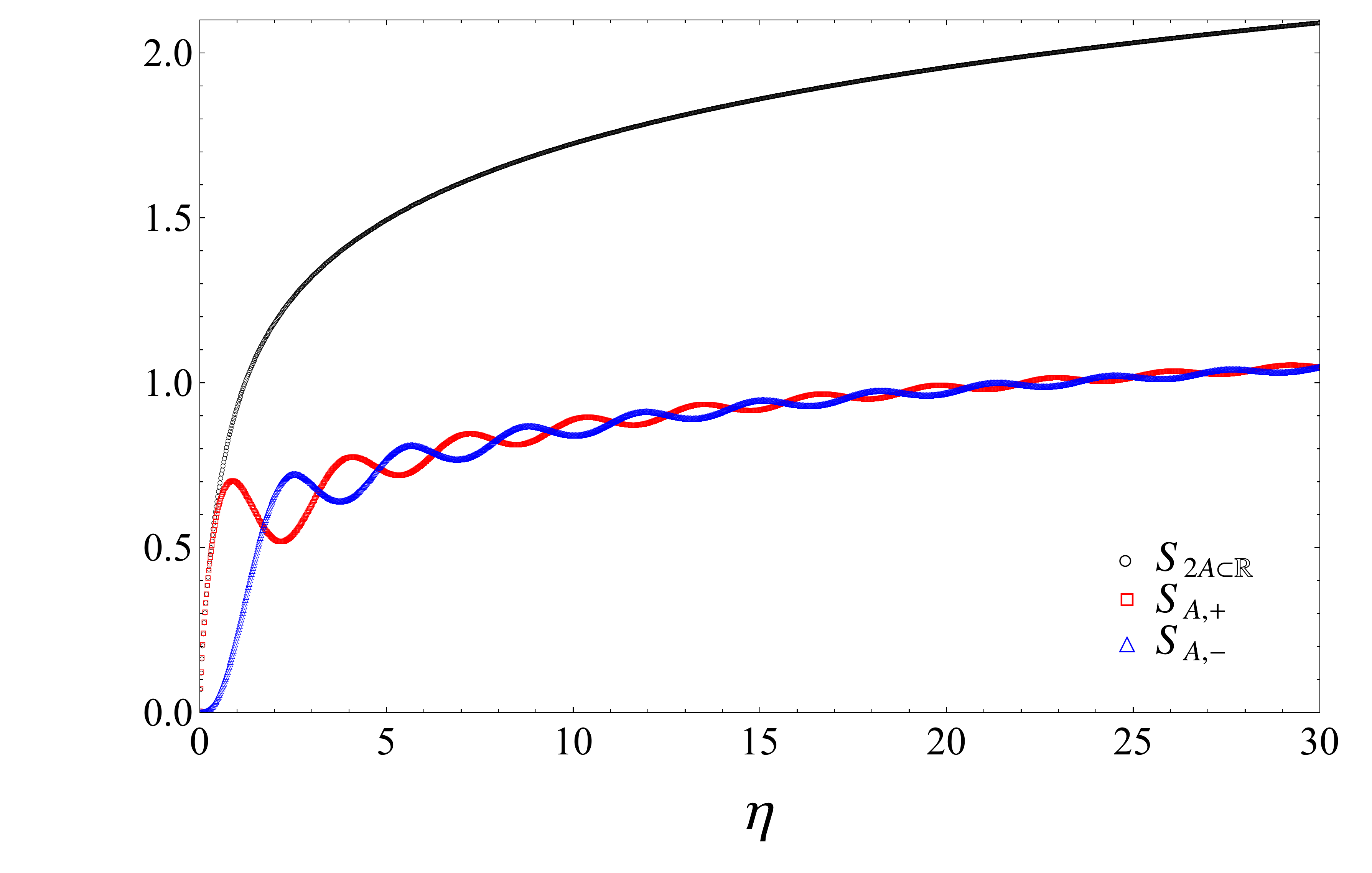}
\vspace{-.7cm}
\caption{
Entanglement entropy of an interval $A =[0,R]$ adjacent to the boundary of the half-line, for Neumann b.c. (red curve $S_{A, +}$) and Dirichlet b.c. (blue curve $S_{A, -}$), obtained numerically from \eqref{renyi-entropies-def-sums}. 
The black curve corresponds to the entanglement entropy 
$S_{2A \subset \RR}$
of an interval of length $2R$ on the line. The relation  \eqref{sum-entropies-line-pm} occurs among these quantities.
}
\label{fig:entropies-half-line}
\end{figure}

Summing up the two expressions for the 
entanglement entropies in \eqref{renyi-entropies-def-sums}
corresponding to the two different boundary conditions, one obtains
\be
\label{sum-entropies-line-pm}
S_{2A \,\subset \,\mathbb{R}}^{(\alpha)}  
\,=\,
S_{A, +}^{(\alpha)}  + S_{A, -}^{(\alpha)} 
\ee
where $S_{2A \,\subset \,\mathbb{R}}^{(\alpha)}$ are the entanglement entropies of an interval of length $2R$
on the line for the Schr\"odinger field theory at zero temperature and finite density, 
which have been studied in \cite{Mintchev:2022xqh}. 
Since $S_{A, \pm}^{(\alpha)} $ are positive functions of $\eta$
and $S_{2A \,\subset \,\mathbb{R}}^{(\alpha)}$ is finite for any given $\eta$ 
(the proof has been reported in Sec.\,4 of \cite{Mintchev:2022xqh}),
also $S_{A, \pm}^{(\alpha)} $ are finite functions of $\eta$.

In Fig.\,\ref{fig:entropies-half-line} we show $S_{A, \pm}$, 
evaluated numerically from \eqref{renyi-entropies-def-sums}, 
and compare them with 
the entanglement entropy $S_{2A \,\subset \,\mathbb{R}}$
of an interval of length $2R$ on the line. 
These three quantities are related through \eqref{sum-entropies-line-pm}.
The numerical analysis has been performed as explained in \cite{Mintchev:2022xqh},
by employing an optimised Fortran code
provided to us by Vladimir Rokhlin.
In particular, the infinite sums \eqref{renyi-entropies-def-sums} have been truncated to $n \leqslant 2(n_0 + 2)$, where $n_0$ is the critical index \eqref{n0-critical}.
We checked numerically that the entanglement entropies do not change significantly 
by including more terms. 
This truncation criterion, which is the one adopted in \cite{Mintchev:2022xqh}, 
has been applied to evaluate numerically all the quantities in this manuscript that involve a sum over the spectra \eqref{soln-spectral-problem-neumann} and \eqref{soln-spectral-problem-dirichlet}.

\begin{figure}[t!]
\subfigure
{\hspace{-1.8cm} \includegraphics[width=.6\textwidth]{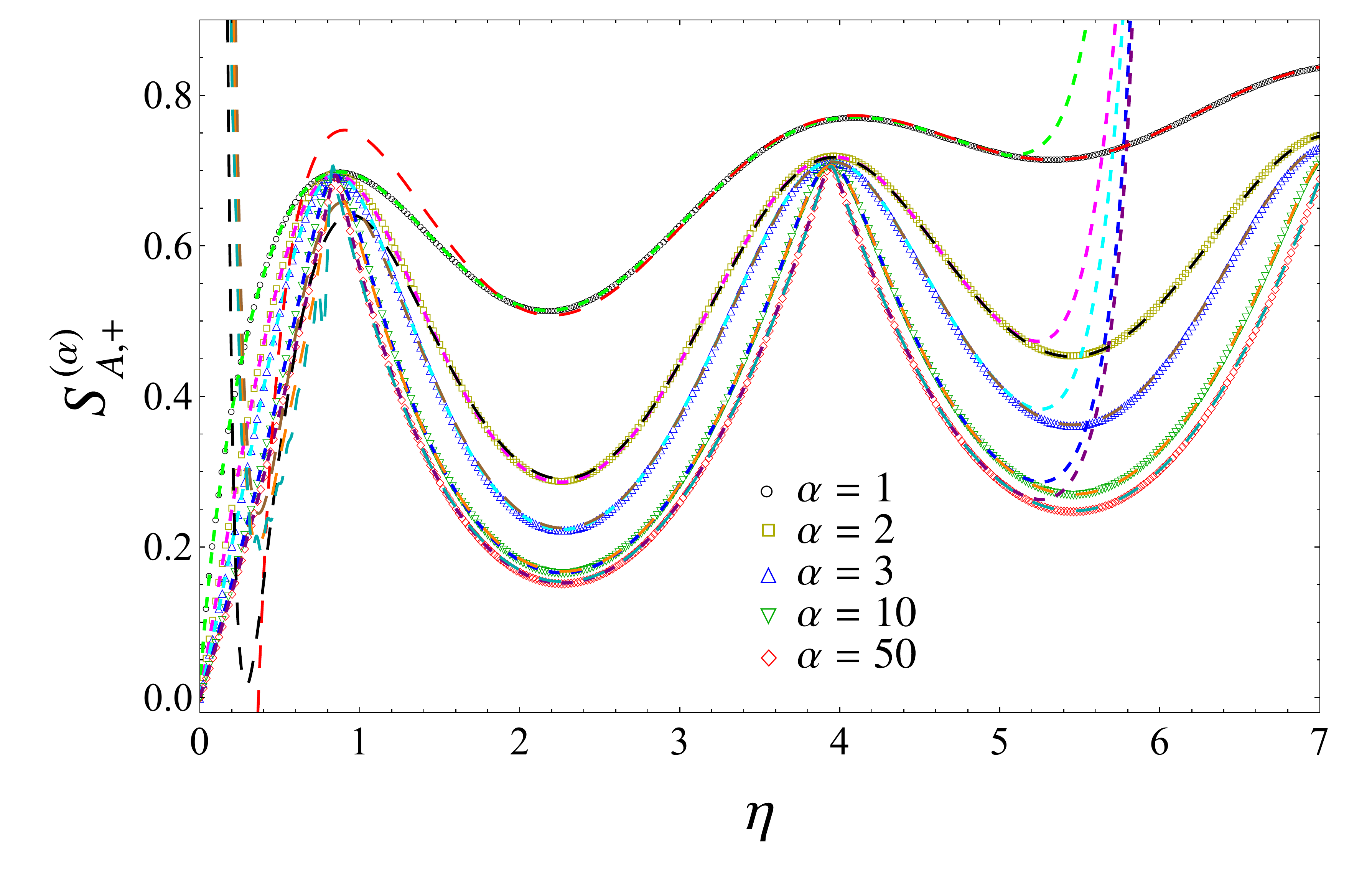}}
\subfigure
{\hspace{-.2cm}\includegraphics[width=.6\textwidth]{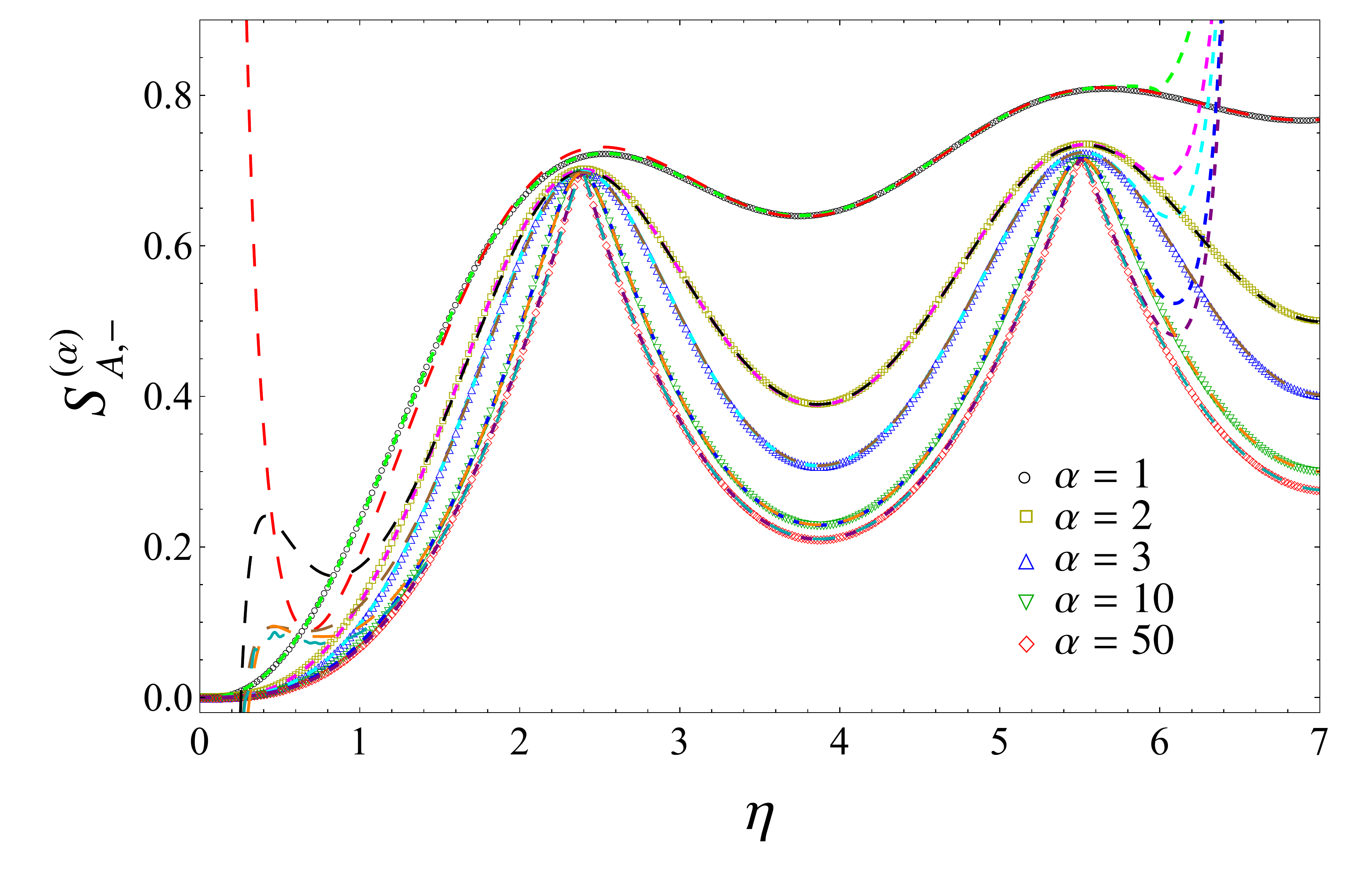}}
\vspace{-0.8cm}
\caption{
Entanglement entropies for either Neumann (left panel) or Dirichlet (right panel) 
boundary conditions.
The data points have been obtained numerically from \eqref{renyi-entropies-def-sums}, while the dashed lines correspond to the small (see Sec.\,\ref{sec_small_eta}) 
and large  (see Sec.\,\ref{sec_large_eta}) interval expansion.
} 
\vspace{0.4cm}
\label{fig:small-eta-renyi-rough}
\end{figure}

The main feature to highlight in Fig.\,\ref{fig:entropies-half-line} is the fact that 
$S_{A, \pm}$ are not monotonic functions.
Instead, $S_{2A \,\subset \,\mathbb{R}}$ is a monotonic function
as proved in \cite{Mintchev:2022xqh}.
The proof of this feature of $S_{2A \,\subset \,\mathbb{R}}$ exploits 
the invariance under translations, which does not hold for the model on the half line. 

\begin{figure}[t!]
\vspace{-.2cm}
\hspace{-.8cm}
\includegraphics[width=1.05\textwidth]{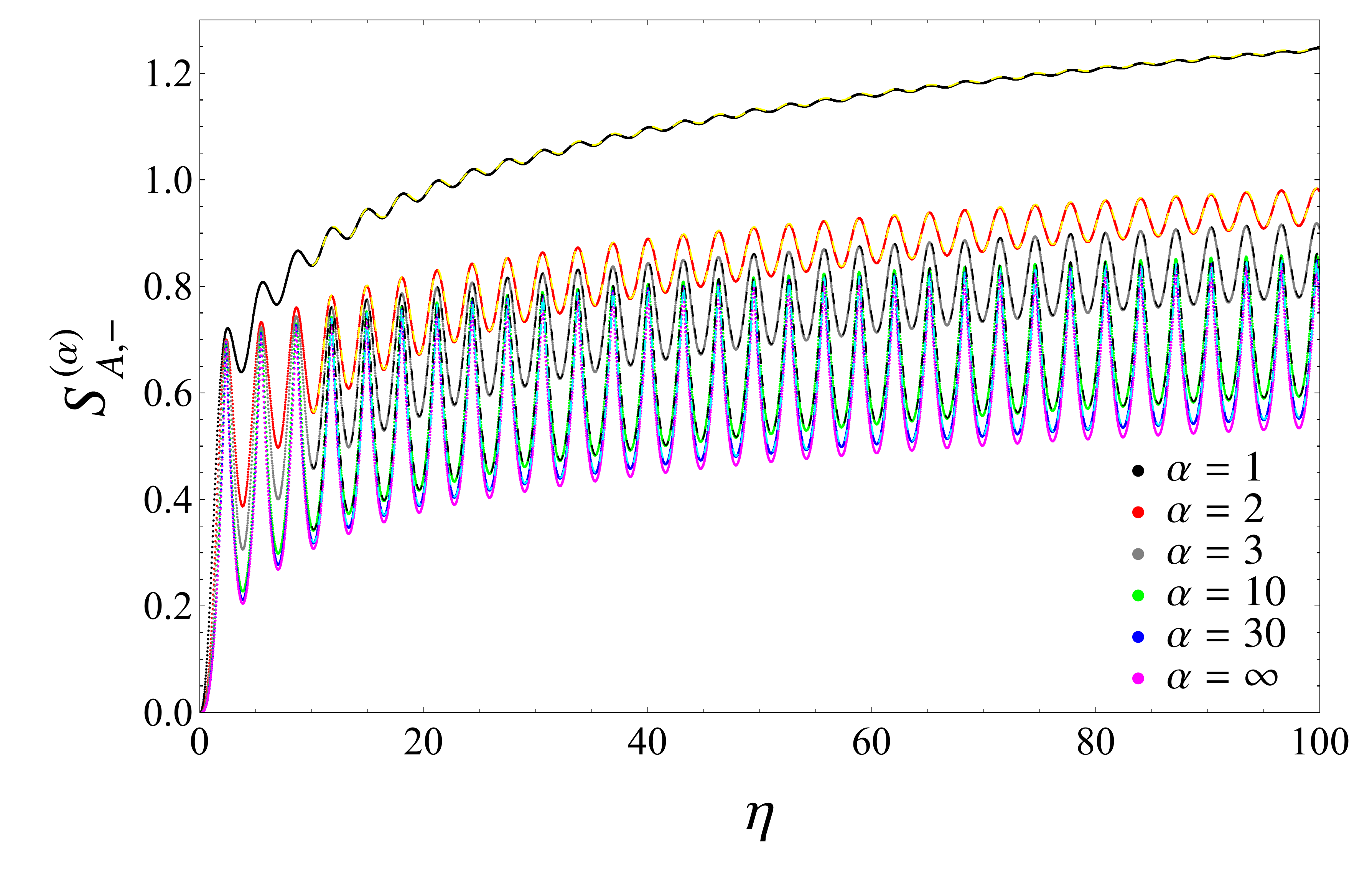}
\vspace{-.7cm}
\caption{
Entanglement entropies $S_{A,-}^{(\alpha)}$ 
for Dirichlet boundary conditions. 
The dashed lines (shown only for $\eta \geqslant 10$) correspond to the large $\eta$ expansion discussed in Sec.\,\ref{sec_large_eta}.
}
\label{fig:ent-3-large-eta}
\end{figure}

In Fig.\,\ref{fig:small-eta-renyi-rough} and Fig.\,\ref{fig:ent-3-large-eta}
we also show the entanglement entropies $S_{A, \pm}^{(\alpha)}$ 
for different values of $\alpha$, 
obtained numerically from \eqref{renyi-entropies-def-sums}.

Since $\gamma_n^\pm \in (0,1)$ for any $n \in \mathbb{N}$ and $\eta >0$ in \eqref{renyi-entropies-def-sums},
we can adapt to our case the procedure employed in \cite{Jin_2004, Keating_04} 
to evaluate the entanglement entropies in some spin chains, 
as done in \cite{Mintchev:2022xqh} for the entanglement entropies 
of an interval on the line for the Schr\"odinger model at finite density and zero temperature. 
This allows to write \eqref{renyi-entropies-def-sums} as the following contour integral in the complex plane
\be
\label{entropy-tau-pm}
S_{A, \pm }^{(\alpha)} 
\,=\,
  \lim_{\epsilon, \delta \to 0}\,
   \frac{1}{2\pi \textrm{i}} 
  \oint_{\mathfrak{C}}
 s_\alpha(z) \;
 \partial_z  \log( \tau_\pm )\, \textrm{d}z
\ee
where $s_\alpha(z)$ is the holomorphic function obtained from \eqref{def-s-alpha}.
The closed path $\mathfrak{C}$ encircles the interval $[0,1] \subset \mathbb{R}$ 
and is parameterised by the infinitesimal parameters $\epsilon$ and $\delta$
through its decomposition $\mathfrak{C} = \mathfrak{C}_0 \cup \mathfrak{C}_- \cup \mathfrak{C}_1 \cup \mathfrak{C}_+ $, where $\mathfrak{C}_0$ and $\mathfrak{C}_1$ are two arcs of radius $\epsilon/2$ centered in $0$ and $1$ respectively, while $\mathfrak{C}_\pm$ are the segments belonging to the horizontal lines $ x \pm \textrm{i} \delta$ with $x\in \mathbb{R}$ and intersecting $\mathfrak{C}_0$ and $\mathfrak{C}_1$ (see e.g. Fig.\,1 of \cite{Jin_2004}, where a similar path is shown);
hence $\epsilon \to 0$ implies $\delta \to 0$.
The functions $\tau_\pm$ in the integrand of \eqref{entropy-tau-pm} are the tau functions associated to the kernels \eqref{kernel-pm-def}
\be
\label{tau-function-pm}
\tau_\pm
\,\equiv\,  
\textrm{det}\big( I - z^{-1} K_\pm \big) 
=
\prod_{n=0}^{\infty} \! \big( 1 - z^{-1} \,\gamma_n^\pm \big) 
\ee
i.e. the Fredholm determinants of the corresponding kernels,
where $I$ denotes the identity operator, $z\in\CC$ and $\gamma_n^\pm$ are the eigenvalues  of $K_\pm$,
which are obtained from the eigenvalues of the sine kernel $K_{\textrm{\tiny sine}}$ 
(see (\ref{soln-spectral-problem-neumann}) and (\ref{soln-spectral-problem-dirichlet})).
From this relation, it is straightforward to observe that 
the tau function 
$\tau_{\textrm{\tiny sine}} \equiv  \textrm{det}( I - z^{-1} K_{\textrm{\tiny sine}})$ 
associated to the sine kernel
can be written in terms of the tau functions in \eqref{tau-function-pm} as follows\footnote{The relation (\ref{sine-tau-function-factorisation})
implies $\sigma_{\textrm{\tiny sine}} = \sigma_+ +  \sigma_-$ for the corresponding auxiliary functions (see \cite{Gamayun:2013auu} and also Appendix\;\ref{app_bessel}).} 
\be
\label{sine-tau-function-factorisation}
\tau_{\textrm{\tiny sine}}
\,=\,
\tau_+\, \tau_-
\ee
(see also Proposition\;1 in \cite{Witte-04} with $a_{\textrm{\tiny there}} = 0$).

The relation (\ref{sine-tau-function-factorisation}), combined with (\ref{entropy-tau-pm}),
provides (\ref{sum-entropies-line-pm}) in a straightforward way. 
This observation can be extended to a class of quantities 
having the form $G_A = \sum_{n\geqslant 0} g(\gamma_n)$, 
where $g(0)=0$.
Indeed, by writing these quantities like in (\ref{entropy-tau-pm})
and exploiting (\ref{sine-tau-function-factorisation}),
one finds the relation $ G_{2A \,\subset \,\mathbb{R}} = G_{A, +} + G_{A, -}\,$,
where $G_{2A \,\subset \,\mathbb{R}}$ corresponds to $G_A$
for the interval $[-R, R]$ on the line, 
and $G_{A, \pm}$ to $G_A$ for the interval $[0,R]$
adjacent to the boundary of the half line where
either Neumann ($+$) or Dirichlet ($-$) are imposed at the origin. 
Also the Schatten norms (\ref{schatten}) 
(see Sec.\,\ref{sec_cumulants_ee}) belong to this class of quantities
and (\ref{blow-up-for-schatten}) gives the above mentioned relation for them.

Fredholm determinants of integrable kernels occur in many interesting problems in physics and mathematics.
In particular, the Fredholm determinants \eqref{tau-function-pm} 
are related to some
probability distribution of the level spacings for random matrices
\cite{GAUDIN1961447} 
and to the inverse scattering problem \cite{Dyson-76}. 

In our analyses we exploit in a crucial way the fact that
the Fredholm determinants \eqref{tau-function-pm} are related to the 
solutions of a particular Painlevé $\textrm{III}$ differential equation
\cite{Tracy:1993xj, Jimbo-82, forrester-book}.
In particular, 
the kernels (\ref{kernel-pm-def}) can be obtained as special cases of a Bessel kernel 
(see \eqref{K-pm-from-K-B})
and the relation between the corresponding spectral problems is discussed in the Appendix\;\ref{app_bessel}.
This allows to write the tau functions $\tau_\pm$ in \eqref{entropy-tau-pm}
as special cases of the tau function of this Bessel kernel (see \eqref{tau-pm-from-tau-B}).
The auxiliary sigma function associated to this Bessel kernel tau function 
satisfies a particular Painlevé $\textrm{III}_1$ differential equation
(see (\ref{sigma-bessel-def}) and (\ref{P3-TW})).
Combining these observations, 
the relation between the Painlevé $\textrm{III}_1$ and $\textrm{III}'_1$ 
(see \eqref{tau_3_3p}) 
and the small $\eta$ expansion of $\tau_\pm$ given in \cite{Tracy:1993xj} (see Appendix\;\ref{app_tau_small_eta}),
we obtain 
\be
\label{fd_t3p}
\tau_{\pm}(\eta)=\frac{2^{1/8}}{\pi^{\pm 1/2}\; \textrm{e}^{\eta^{2}/8}\,\eta^{1/8}}
\; \tau_{\textrm{\tiny III}'}(\eta^{2}/4)\big|_{\theta_{\ast}=\,\theta_{\star}=\,\pm 1/4}
\ee
where the tau function $\tau_{\textrm{\tiny III}'}(t)$ in \cite{Gamayun:2013auu} 
is employed. 
The explicit expressions of
the tau functions occurring in the r.h.s. of (\ref{fd_t3p})
are discussed below (see (\ref{tau-GIL-sec}) and (\ref{t_pm_final})).

Analytic expressions for the expansions of $\tau_{\pm}$ as $\eta \to 0$ and $\eta \to \infty$ have been obtained in \cite{Gamayun:2013auu, Bonelli:2016qwg}.
In Sec.\,\ref{sec-small-eta-tau} and Sec.\,\ref{sec_large_eta}, 
we have employed them into \eqref{entropy-tau-pm} to get analytic results for the corresponding expansions of $S_{A, \pm}^{(\alpha)}$.
In Fig.\,\ref{fig:small-eta-renyi-rough} and Fig.\,\ref{fig:ent-3-large-eta}, 
the dashed curves have been found from these analytic expansions,
while the curves identified by the empty markers have been obtained numerically through \eqref{renyi-entropies-def-sums}, in the same way described for Fig.\,\ref{fig:entropies-half-line}.
In particular, 
we have used (\ref{approx-entropies-small-eta}) for small $\eta$
and the expressions discussed in Sec.\,\ref{sec_large_eta} 
(see e.g. \eqref{ee-large-eta-dec}, \eqref{S-infty-leading-sec} and \eqref{S_larg_exp})
for large $\eta$.
We emphasise that $S_{A, \pm}^{(\alpha)}$ are oscillating functions of $\eta$ for any value of $\alpha>0$.
A remarkable agreement 
between the numerical results and the analytic expressions for the small and large $\eta$ expansions
is observed.
Furthermore, in Fig.\,\ref{fig:small-eta-renyi-rough} 
an intermediate regime of $\eta$ can be identified where 
the curves corresponding to the small and large $\eta$ expansions overlap.
The size of this crossover regime depends both on the boundary condition and on the value of $\alpha$.

While in Fig.\,\ref{fig:small-eta-renyi-rough} 
the entanglement entropies $S_{A, \pm}^{(\alpha)}$ are shown only for $\eta \in [0,7]$,
a larger domain has been considered in Fig.\,\ref{fig:ent-3-large-eta},
where only $S_{A, -}^{(\alpha)}$ are reported 
because the curves for $S_{A, +}^{(\alpha)}$ are qualitatively very similar.
In Fig.\,\ref{fig:ent-3-large-eta} one observes the logarithmic growth of the entanglement entropies
(in particular, from (\ref{ee-large-eta-dec}) and (\ref{S-infty-leading-sec}) we have that 
$S^{(\alpha)}_{A,\pm}=\tfrac{1}{12}\big(1+\tfrac{1}{\alpha}\big)\log(\eta)+O(1)$ as $\eta \rightarrow \infty$)
and also their oscillatory behaviour.
For a given value of $\alpha$, 
the amplitude of the oscillations vanishes as $\eta \to \infty$.
Instead, this amplitude
increases with $\alpha$ for a given value of $\eta$.

We find it worth introducing the following combinations of entanglement entropies
\be
\label{b-comb-def-pm}
B_{A, \pm}^{(\alpha)} 
\,\equiv\,
S_{A, \pm}^{(\alpha)}  - \frac{1}{2}\, S_{2A \,\subset \,\mathbb{R}}^{(\alpha)}  
\,=\,
\pm \, 
\frac{S_{A, +}^{(\alpha)} - S_{A, -}^{(\alpha)}}{2}
\ee
where (\ref{sum-entropies-line-pm}) has been employed. 
For boundary conformal field theories in $d=1$,
the r.h.s. of (\ref{b-comb-def-pm}) 
can be defined for any conformally invariant boundary condition:
the resulting combination,
which depends on the boundary condition 
imposed at the origin of the half line,
is UV finite \cite{Calabrese:2004eu}.
In the Schr\"odinger models that we are considering, 
both $S_{A, \pm}^{(\alpha)} $ and $ S_{2A \,\subset \,\mathbb{R}}^{(\alpha)}  $ 
in the r.h.s. of (\ref{b-comb-def-pm}) are finite functions of $\eta$;
hence this property holds for any linear combination of these two quantities.
In the closing paragraph of Sec.\,\ref{sec_large_eta}
an interesting feature of the special combination (\ref{b-comb-def-pm})
is highlighted.

\begin{figure}[t!]
\vspace{-.2cm}
\hspace{-2cm}
\includegraphics[width=1.15\textwidth]{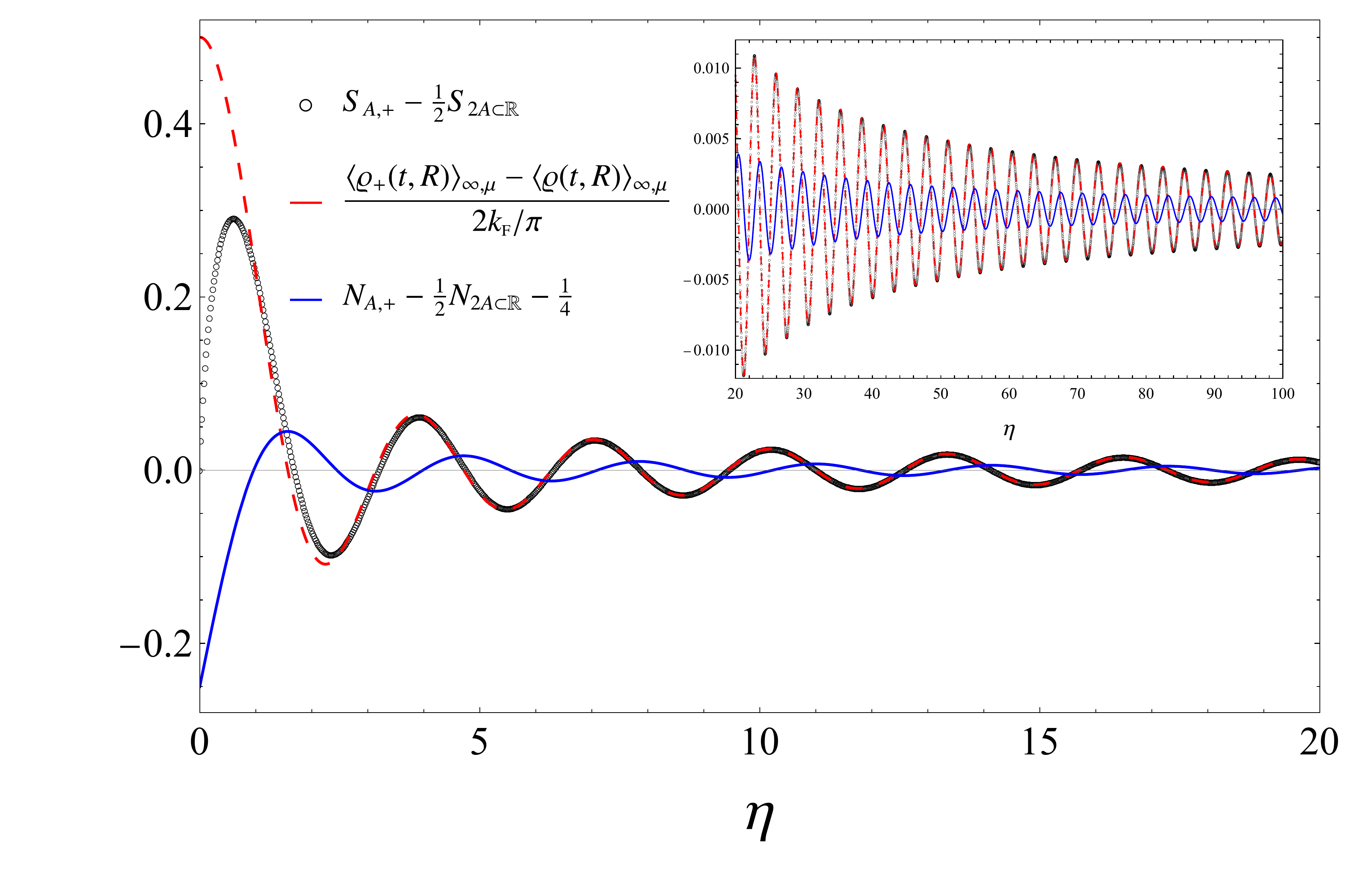}
\vspace{-.7cm}
\caption{
Entanglement entropy, 
normalised density at the entangling point (\ref{normalised-densities-at-R})
and mean particle number in the interval (\ref{mpn1})
in the case of Neumann b.c.,
where half of the corresponding quantity on the line has been subtracted 
(see \eqref{b-comb-def-pm}  for $\alpha =1$, (\ref{mpn2}) and (\ref{mpn3}))
to highlight the oscillatory behaviour.
\label{fig:friedel-ee}
}
\end{figure}

Since $S_{2A \,\subset \,\mathbb{R}} $ is a monotonic function \cite{Mintchev:2022xqh}
while $S_{A, \pm} $ display an oscillatory behaviour
(see Fig.\,\ref{fig:entropies-half-line}), 
the combinations \eqref{b-comb-def-pm}  in the special case of $\alpha =1$ 
are oscillating functions of $\eta$ for both Neumann and Dirichlet boundary conditions. 
In Fig.\,\ref{fig:friedel-ee} we focus on Neumann b.c. as prototypical case, 
and compare the combination \eqref{b-comb-def-pm}  for $\alpha =1$ 
with the density at the entangling point (\ref{normalised-densities-at-R}),
which exhibits the Friedel oscillations,
and the mean particle number in the interval (\ref{mpn1}).
The oscillatory behaviours are highlighted when 
half of the corresponding quantities on the line are subtracted 
(see \eqref{b-comb-def-pm}  for $\alpha =1$ and (\ref{mpn2})).
The density at the entangling point (\ref{normalised-densities-at-R})
can be rewritten as follows
\be
\label{density-sub-at-R}
\frac{1}{2k_{\textrm{\tiny F}}/\pi }
\left[\,
\langle \varrho_\pm (t,R)\rangle_{\infty,\mu} 
- 
\frac{1}{2} \big(\langle \varrho (t,R)\rangle_{\infty,\mu} + \langle \varrho (t,-R)\rangle_{\infty,\mu} \big) 
\right]
= \,
\pm \,\frac{\sin(2 \eta)}{4 \, \eta}
\ee
where the r.h.s. provides the Friedel oscillations
and the quantity within the square brackets 
has a form similar to the one of the other quantities displayed in Fig.\,\ref{fig:friedel-ee}.
The functions in (\ref{density-sub-at-R})
oscillate around zero with decreasing amplitudes and 
their zeros correspond to $2\eta = n \pi$, where $n \in \mathbb{N}$ and $\eta >0$.
These zeros correspond to the values of $\eta$ where the critical index \eqref{n0-critical} has jump discontinuities. 
The parity of the critical index is a $\pi$-periodic function of $\eta$ 
taking values $+1$ for $\eta\in\left[k \pi,(k +1/2)\pi \right)$ and $-1$ for $\eta\in\left[(k +1/2)\pi, (k+1) \pi \right)$, where $k\in \NN_0$. 
When $n_0$ is even, 
the eigenvalue $\gamma_{n_0}$ contributes to $S_{A,+}$;
hence one expects $S_{A,+} \gtrsim S_{A,-}$ (i.e. $B_{A,+} \gtrsim 0$),
as observed for this curve in Fig.\,\ref{fig:friedel-ee} for $\eta \gtrsim 5$.
Analogously, $B_{A,+} \lesssim 0$ when $n_0$ is odd.

In Fig.\,\ref{fig:friedel-ee},
a remarkable agreement between $B_{A,+}$ and \eqref{density-sub-at-R} 
is observed for $\eta > 4$.
This can be explained 
by anticipating some results discussed in Sec.\,\ref{sec_large_eta}
about the expansion of 
$S_{A,\pm} $ at large $\eta$ 
(see (\ref{S-infty-leading-sec}), (\ref{S-0-large-eta-sub}) 
and (\ref{normalised-densities-at-R})),
which allow to write
\be
\label{ee_pm_friedel}
S_{A,\pm} = 
\frac{1}{6} \log(4 \eta) 
+ \frac{E_1}{2} 
+ \frac{\langle\varrho_\pm(t,R) \rangle_{\infty,\mu} - \langle\varrho(t,R) \rangle_{\infty,\mu}}{2 k_{\textrm{\tiny F}} / \pi}
+O\big(1/\eta^2\big) 
\ee
where $E_1$ is the constant (\ref{E_alpha-def}) 
in the special case of $\alpha = 1$. 
The expansion (\ref{ee_pm_friedel}) tells us that
the Friedel oscillations occurring in the 
normalised density at the entangling point 
provide the first subleading correction of the entanglement entropy 
in the regime of large $\eta$, 
which vanish as $\eta \to \infty$.
As for the mean particle number in the interval
(see the blue solid curve in Fig.\,\ref{fig:friedel-ee},
obtained from \eqref{mpn2} and \eqref{mpn3}),
its oscillations have the same frequency and 
they are shifted by $\pi/4$.

The entanglement entropies of the interval adjacent to the boundary
of the half line display oscillations 
(see e.g. Fig.\,\ref{fig:entropies-half-line}, Fig.\,\ref{fig:small-eta-renyi-rough}
and Fig.\,\ref{fig:ent-3-large-eta});
hence it is worth asking whether a monotonically increasing function of $\eta$
can be constructed.
Since $S^{(\alpha)}_{A, \pm} \geqslant 0$, it is natural to consider
\be
\label{integ-entropy}
\mathcal{S}^{(\alpha)}_{A, \pm}
\equiv
\int_0^\eta S^{(\alpha)}_{A, \pm}(\xi)\,\textrm{d}\xi \,.
\ee
Let us investigate the class of functions of $\eta$
whose generic element is
$G(\eta) = \sum_{n \geqslant 0} g(\gamma_n)$,
where $g(x) \to 0$ as $x \to 0$ in a proper way 
to guarantee the convergence of the series that defines $G$.
Since the spectrum of the sine kernel satisfies the following property
(see Eq.\,(3.51) in \cite{Rokhlin-book})
\be
\label{gamma-n-prime}
\gamma'_n = \frac{2}{\eta}\, \gamma_n \,f_n(\eta;1)^2 
\ee
we have that $G'(\eta) = \sum_{n \geqslant 0} g'(\gamma_n)\, \gamma'_n$
with $\gamma'_n \geqslant 0$;
hence the condition  $g'(x)  \geqslant 0$ for $x \in (0,1)$ implies $G'(\eta) \geqslant 0 $.
The expressions in (\ref{integ-entropy}) correspond to
the particular choice given by $g'(x) = s_\alpha(x)$ 
and to the restriction to the eigenvalues of the sine kernel spectral problem
labelled by either even or odd  values of $n$.

\section{Integer Lifshitz exponents}
\label{sec_lifshitz}

In this section we study a hierarchy of two component Lifshitz fermion fields $\psi(t,x)$ 
whose time evolution on the half line $x\geqslant 0$ is given by 
\begin{equation}
\left[ \,\ri \sigma_0\,\partial_t -\frac{1}{(2m)^{z-1}} \,\big(\! - \! \ri \sigma_3\, \partial_x\big)^z \, \right]\psi (t,x) = 0 
\;\;\;\;\;\qquad \;\;\;\;\;
z \in {\mathbb N} 
\label{e1L}
\end{equation} 
where 
\begin{equation} 
\psi (t,x)=
\Bigg(\,
\begin{array}{c}\psi_1(t,x) \\ \rule{0pt}{.5cm} \psi_2(t,x) \\ \end{array}
\Bigg)
\;\;\qquad \;\;
\sigma_0 = \left(\begin{array}{cc}\; 1 \; & \;0\; \\  \rule{0pt}{.5cm} 0 &  1 \\ \end{array} \right) 
\;\; \qquad \;\;
\sigma_3 = \left(\begin{array}{cc}\; 1 \; & \;0\; \\  \rule{0pt}{.5cm} 0 &  -1 \\ \end{array} \right) .
\label{e2L}
\end{equation} 
We assume in addition that $\psi_i(t,x)$ satisfy the equal-time anti-commutation relations \eqref{e2a-b} and introduce the generalised Fermi momentum 
\be 
k_{\textrm{\tiny F},z} \equiv (2m)^{1-1/z} \mu^{1/z} \,.
\label{eL3}
\ee
Notice that $k_{\textrm{\tiny F},1} = \mu$, $k_{\textrm{\tiny F},2} = k_{\textrm{\tiny F}}$, and $k_{\textrm{\tiny F},z} \to 2m$ for $z\to \infty$.

From \eqref{e1L} it follows that for $z=1$ the fields $\psi_i(t,x)$ are the left and right moving components of a massless Dirac fermion $\psi(t,x)$ on the half line. 
For $z=2$ one has instead two independent  Schr\"odinger fields \eqref{e1}. 
It is useful to consider first these two cases because the ones corresponding 
to $z=2n-1$ and $z=2n$ for any $n \in  {\mathbb N}$ 
can be studied as direct extensions of the models 
having $z=1$ and $z=2$ respectively. 

The Dirac case $z=1$ has been considered in detail in \cite{Mintchev:2020uom}.
It has two types of boundary conditions at $x=0$
that ensure energy conservation. 
The boundary condition
\be
\label{bc5}
\psi_1 (t, 0) = \e^{\ri \alpha_{\textrm{\tiny v}}}\, \psi_2 (t,0)
\;\;\;\;\qquad\;\;\;\;
\alpha_{\textrm{\tiny v}} \in [0,2\pi)
\;\;\;\;\qquad\;\;\;\;
t \in \RR 
\ee
preserves the electric charge but not the helicity,
while the opposite holds for the boundary condition 
\be
\label{bc6}
\psi_1 (t, 0) = \e^{-\ri \alpha_{\textrm{\tiny a}}} \,\psi^*_2 (t,0) 
\;\;\;\;\qquad\;\;\;\;
\alpha_{\textrm{\tiny a}} \in [0,2\pi) 
\;\;\;\;\qquad\;\;\;\;
t \in \RR \,.
\ee
The boundary conditions (\ref{bc5}) and (\ref{bc6}) define respectively the vector and axial phases 
of the massless Dirac fermion on the half line. 
Notice that both \eqref{bc5} and \eqref{bc6} provide a scale invariant coupling of components with different chirality at $x=0$. 
For this reason the kernels of the spectral problem for the entanglement entropies in the two phases have off-diagonal elements. 
In fact, by imposing (\ref{bc5}) or (\ref{bc6}) for any $z=2n-1$, one finds 
\be
\label{vcm}
\boldsymbol{K}_{2n-1}(x,y;\alpha ) 
=
\Bigg(\,
\begin{array}{cc} 
K(k_{\textrm{\tiny F},2n-1};x-y) \;\; & \e^{\ri \alpha } \,K(k_{\textrm{\tiny F},2n-1};x+y)
\\ 
\rule{0pt}{.5cm}
\e^{-\ri \alpha }\, K(k_{\textrm{\tiny F},2n-1};-x-y)  \; \;& K(k_{\textrm{\tiny F},2n-1};-x+y) 
\end{array} 
\Bigg)  
\;\;\qquad\;\;
\alpha \in \big\{ \alpha_{\textrm{\tiny v}}\, , \alpha_{\textrm{\tiny a}} \big\}
\ee
where 
\be
\label{not2}
K(\mu;\zeta) 
\,\equiv\, \frac{\e^{-\ri \zeta \mu}}{2\pi \ri \, (\zeta - \ri \varepsilon)} 
\;\;\qquad\;\;
  \varepsilon \rightarrow 0^+ 
\ee
hence $\boldsymbol{K}_{2n-1}$ can be written in terms of the correlators as discussed in 
\cite{Mintchev:2020uom}.

For $z=2n$ one finds instead two fully decoupled Schr\"odinger fields
and either the Neumann or the Dirichlet boundary conditions (see (\ref{h12}))
can be imposed for each of them
(the case $z=2$ has been discussed in Sec.\,\ref{sec_model}).
Accordingly, 
\be
\label{vcm}
\boldsymbol{K}_{2n}(x,y; \kappa_1, \kappa_2) 
=
\Bigg(\,
\begin{array}{cc} 
K_{\kappa_1}(k_{\textrm{\tiny F},2n};x,y) \; & 0
\\ 
\rule{0pt}{.5cm}
0  \; & K_{\kappa_2}(k_{\textrm{\tiny F},2n};x,y) 
\end{array} \Bigg) 
\;\; \qquad \;\; 
\kappa_{1}\,, \kappa_2 \in \big\{+\,, - \big\}
\ee
where $K_\pm$ are given by (\ref{kernel-pm-def}). 

When $z=2n-1$ the entanglement entropies for the interval $A=[0,R]$ can be expressed in terms of the eigenvalues $\zeta_s$ of the spectral problem 
\be 
\int_0^R \boldsymbol{K}_{2n-1}(x,y;\alpha ) \;\Phi (y,s)\,\rd y 
\,=\, 
\zeta_s \, \Phi(x,s) 
\;\;\qquad \;\;
\Phi(x,s) \equiv 
\Bigg(\,
\begin{array}{cc} 
\Phi_1(x,s)
\\ 
\rule{0pt}{.5cm}
\Phi_2(x,s)
\end{array} \Bigg)  
\;\; \qquad \;\;
s \in {\mathbb R} \,.
\label{sp1}
\ee
The solution of (\ref{sp1}) is given by 
\be 
\zeta_s = \frac{1-\tanh(\pi s)}{2} 
\label{sp2a}
\ee
and
\be
\Phi_1(x,s) = \e^{\ri k_{\textrm{\tiny F},2n-1}x} \phi_s(x) 
\;\;\;\qquad \;\;\; 
\Phi_2(x,s) = \e^{-\ri k_{\textrm{\tiny F},2n-1}x} \e^{-\ri \alpha }\phi_s(-x)
\label{sp2b}
\ee
for $x\in [0,R]$, and $\zeta_s$ and $\phi_s(x)$ satisfy the simpler and well known 
\cite{Musk-book, Casini:2009vk, Arias:2016nip} spectral problem
\be 
\int_{-R}^R \frac{1}{2\pi \ri\,(x-y -\ri \varepsilon )} \; \phi_s (y)\,\rd y 
\,=\, 
\zeta_s \,\phi_s(x) 
\;\;\;\;\qquad \;\;\;\;
x \in [-R,R]\,.
\label{sp3}
\ee 
We remark that the dependence on $k_{\textrm{\tiny F},2n-1}$ 
and $\alpha$ is carried by the eigenfunctions, 
while the eigenvalues $\zeta_s$ are independent of these parameters. 
The explicit form \cite{Musk-book, Casini:2009vk, Arias:2016nip} of $\phi_s(x)$ 
is not needed because the entanglement entropies are 
fully expressed in terms of the eigenvalues. 
Thus, all the Lifshitz fermions with odd $z$ 
have the entanglement entropies of the relativistic 
massless Dirac fermion, i.e. 
\be 
S_A^{(\alpha)} 
\,=\, 
\frac{1}{12}  \left(1+ \frac{1}{\alpha}\right)
\log (2R/\epsilon) + O(1) 
\;\;\; \qquad \;\;\;
z= 2n-1\,.
\label{Eodd}
\ee
It is worth mentioning that the independence of the spectrum on 
$k_{\textrm{\tiny F},2n-1}$ leads to a well known logarithmic ultraviolet divergency, which induces the presence of the UV cutoff $\epsilon$ in \eqref{Eodd}.

When  $z=2n$ we have two independent Schr\"odinger fields
and each of them satisfy either the Neumann or the Dirichlet 
boundary condition. Therefore 
\be 
S_{A; \kappa_1,\kappa_2}^{(\alpha)} 
\,=\, S_{A,\kappa_1}^{(\alpha)} + S_{A,\kappa_2}^{(\alpha)}  
\;\; \qquad \;\; 
\kappa_{1}\,, \kappa_2 \in \big\{+\,, - \big\}
\;\;\; \qquad \;\;\;
z=2n 
\label{Eeven}
\ee 
where $S_{A,\kappa}^{(\alpha)}$ is given by \eqref{renyi-entropies-def-sums} 
with the substitution $k_{\textrm{\tiny F}} \mapsto k_{\textrm{\tiny F},2n}$.

The mean particle density \eqref{dnd} 
of the Schr\"odinger fermion exhibits Friedel oscillations. 
It turns out that such oscillations are absent for the massless Dirac fermion 
$z=1$ on the half line.
In fact,  in this case,  for both the vector and axial phases,
one finds \cite{Mintchev:2020uom}
\be 
\langle \varrho(t,x) \rangle_{\infty , \mu} 
\equiv 
\langle :\!\psi_1^* \, \psi_1 \! : \!(t,x) \rangle_{\infty , \mu} 
+  
\langle  :\!\psi_2^* \, \psi_2\!: \!(t,x) \rangle_{\infty , \mu} 
= 
\frac{|\mu|}{\pi}
\label{d1}
\ee
where $: \cdots :$ denotes the normal product. 
The mixed correlation functions in the vector and axial phases 
are given by
\be 
\langle  :\!\psi_1^*\, \psi_2\!:\!(t,x) \rangle_{\infty , \mu} 
= 
\e^{-\ri \alpha_{\textrm{\tiny v}} }\,\frac{\e^{\ri \mu x}}{2\pi x}\, \sin (\mu x)
\qquad
\langle :\!\psi_1 \,\psi _2\!:\!(t,x) \rangle_{\infty , \mu} 
= 
\e^{-\ri \alpha_{\textrm{\tiny a}} }\,\frac{\e^{\ri \mu x}}{2\pi x} \,\sin (\mu x)
\ee
respectively, which display an oscillatory behaviour. 

\section{Small $\eta$ expansion}
\label{sec_small_eta}

In this section we investigate the expansion of the entanglement entropies as 
$\eta \to 0$ by employing two different methods. 
The first approach (Sec.\,\ref{sec-small-eta-prolate})
is based on the expansion of the PSWF,
while the second one (Sec.\,\ref{sec-small-eta-tau})
exploits the expansion of the tau functions (\ref{fd_t3p}).

\subsection{PSWF approach}
\label{sec-small-eta-prolate}

The asymptotic behaviour of the PSWF leads to
the following small $\eta$ expansion for the eigenvalues \cite{Rokhlin-book,Slepian-expansions}
\be
\label{eigen-small-1}
\gamma_{n} \,=\, 
\tilde{g}_{n} \, \eta^{2n+1}\, \big[1+\tilde{a}_{n}\,\eta^{2}+O\big(\eta^{4} \big) \big]
\ee
where
\be
\label{eigen-small-3}
\tilde{g}_{n} = \frac{2}{\pi}\left(\frac{2^{2n}(n!)^{3}}{(2n)!\,(2n+1)!}\right)^{2}
\;\;\;\;\qquad\;\;\;\;
\tilde{a}_{n} = -\frac{2n+1}{(2n-1)^{2}\,(2n+3)^{2}}\,.
\ee
By using \eqref{eigen-small-1} combined with either \eqref{soln-spectral-problem-neumann} for Neumann b.c. or \eqref{soln-spectral-problem-dirichlet} for Dirichlet b.c. into \eqref{renyi-entropies-def-sums}, we obtain the expansions  of the entanglement entropies reported below.

\begin{figure}[t!]
\vspace{.4cm}
\subfigure
{\hspace{-1.6cm} \includegraphics[width=.58\textwidth]{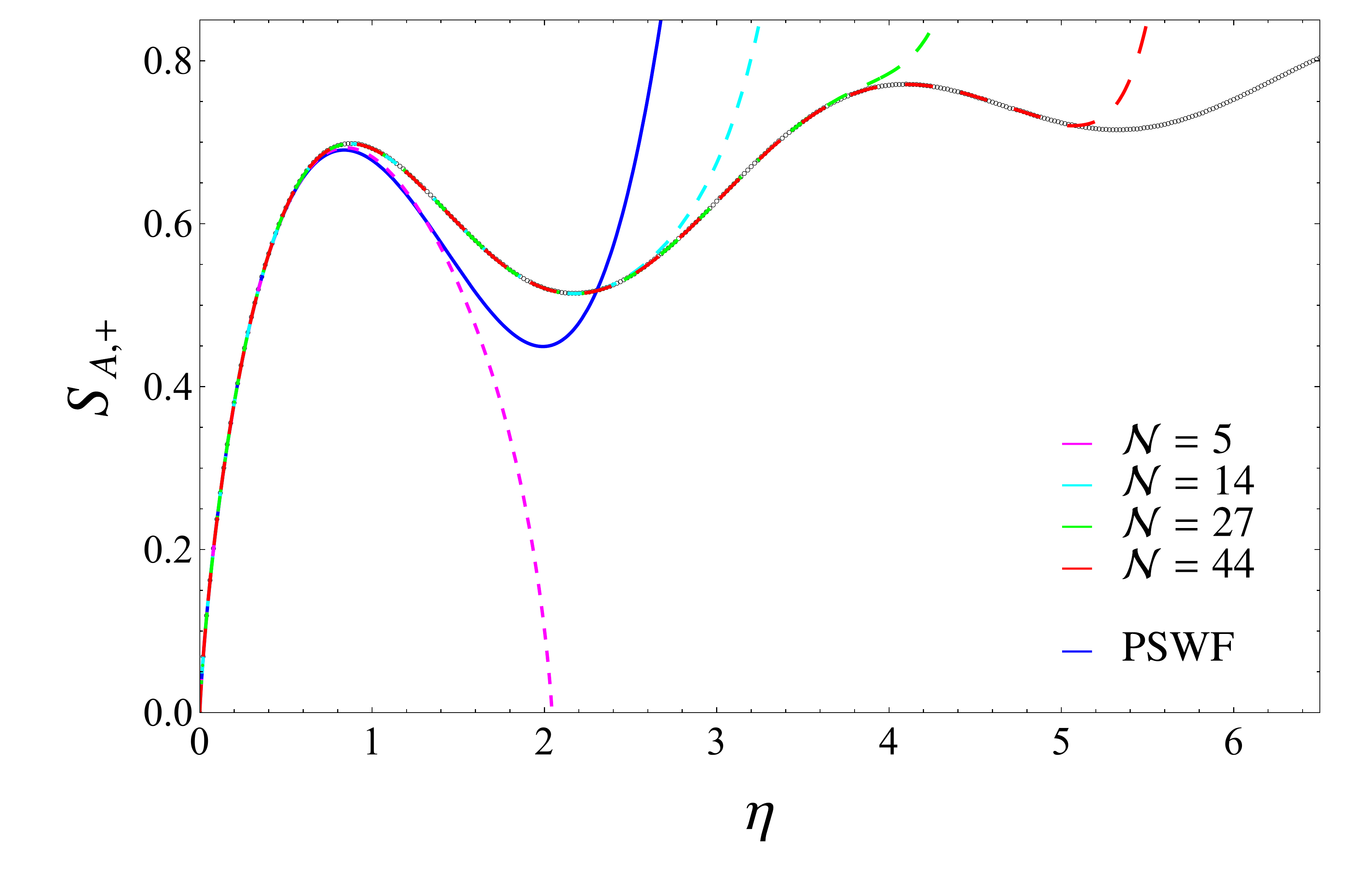}}
\subfigure
{\hspace{.1cm}\includegraphics[width=.58\textwidth]{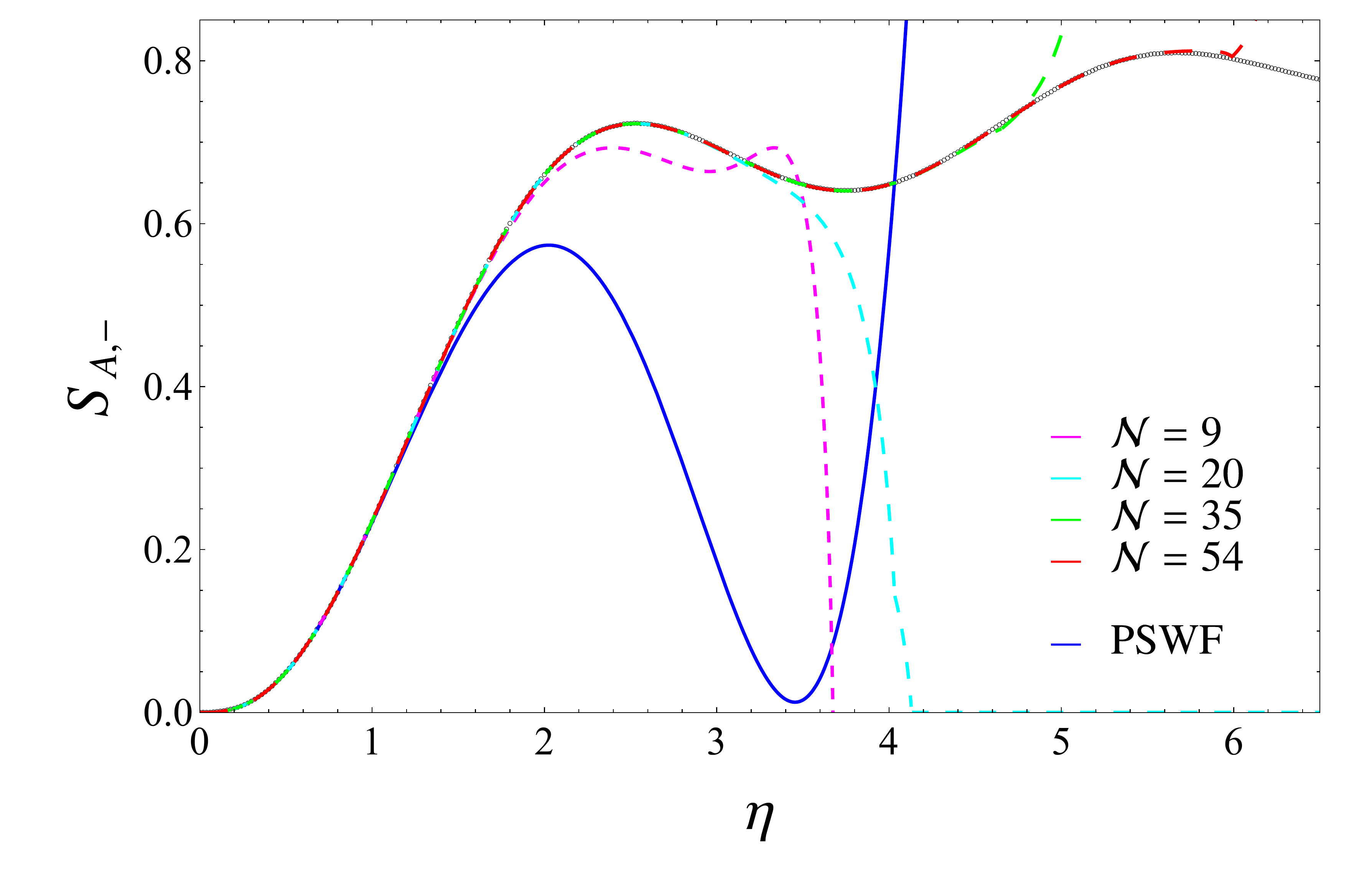}}
\subfigure
{\hspace{-1.45cm}\includegraphics[width=.58\textwidth]{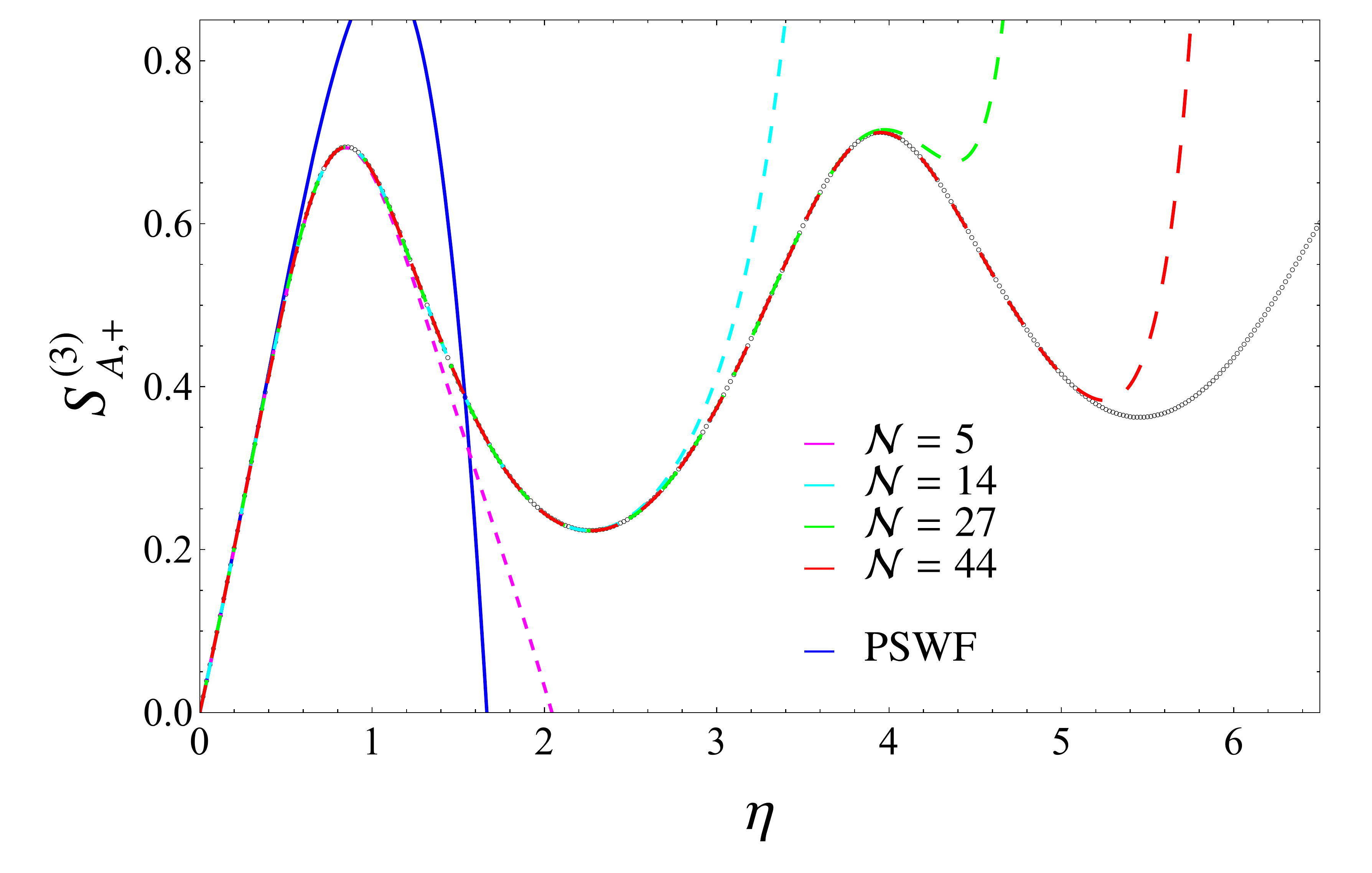}}
\subfigure
{\hspace{-.05cm} \includegraphics[width=.58\textwidth]{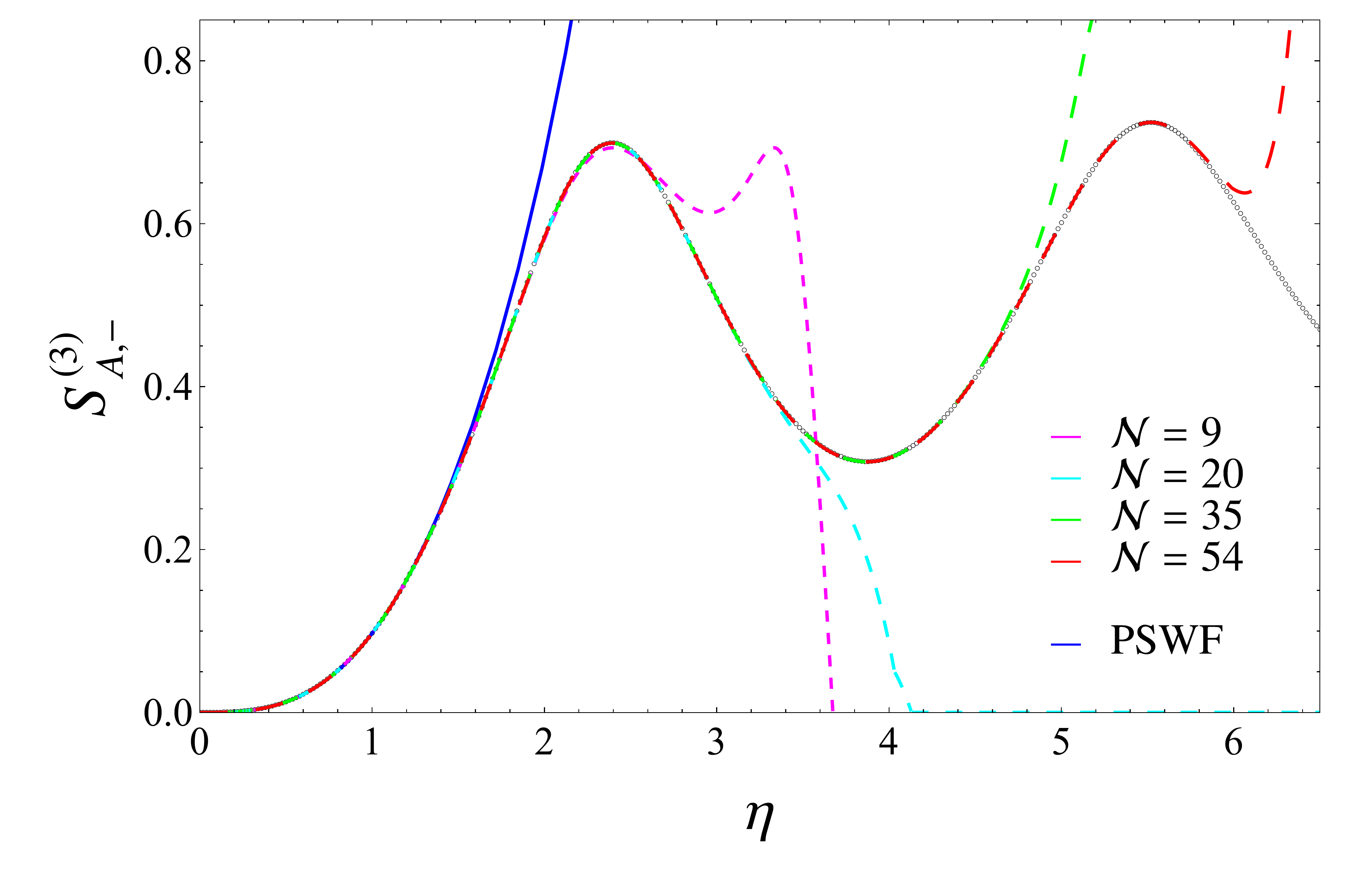}}
\subfigure
{\hspace{-1.55cm} \includegraphics[width=.58\textwidth]{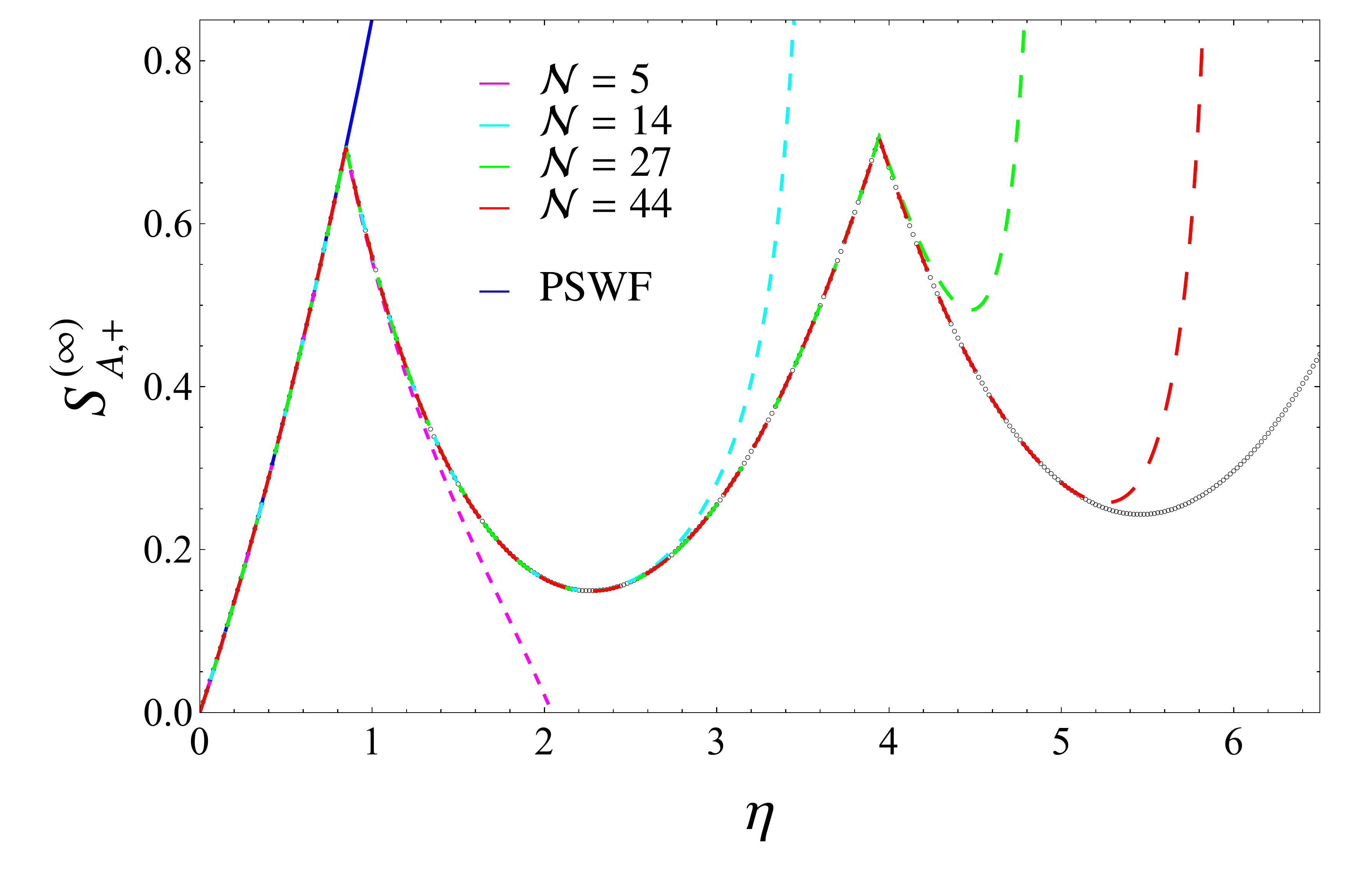}}
\subfigure
{\hspace{-.08cm} \includegraphics[width=.58\textwidth]{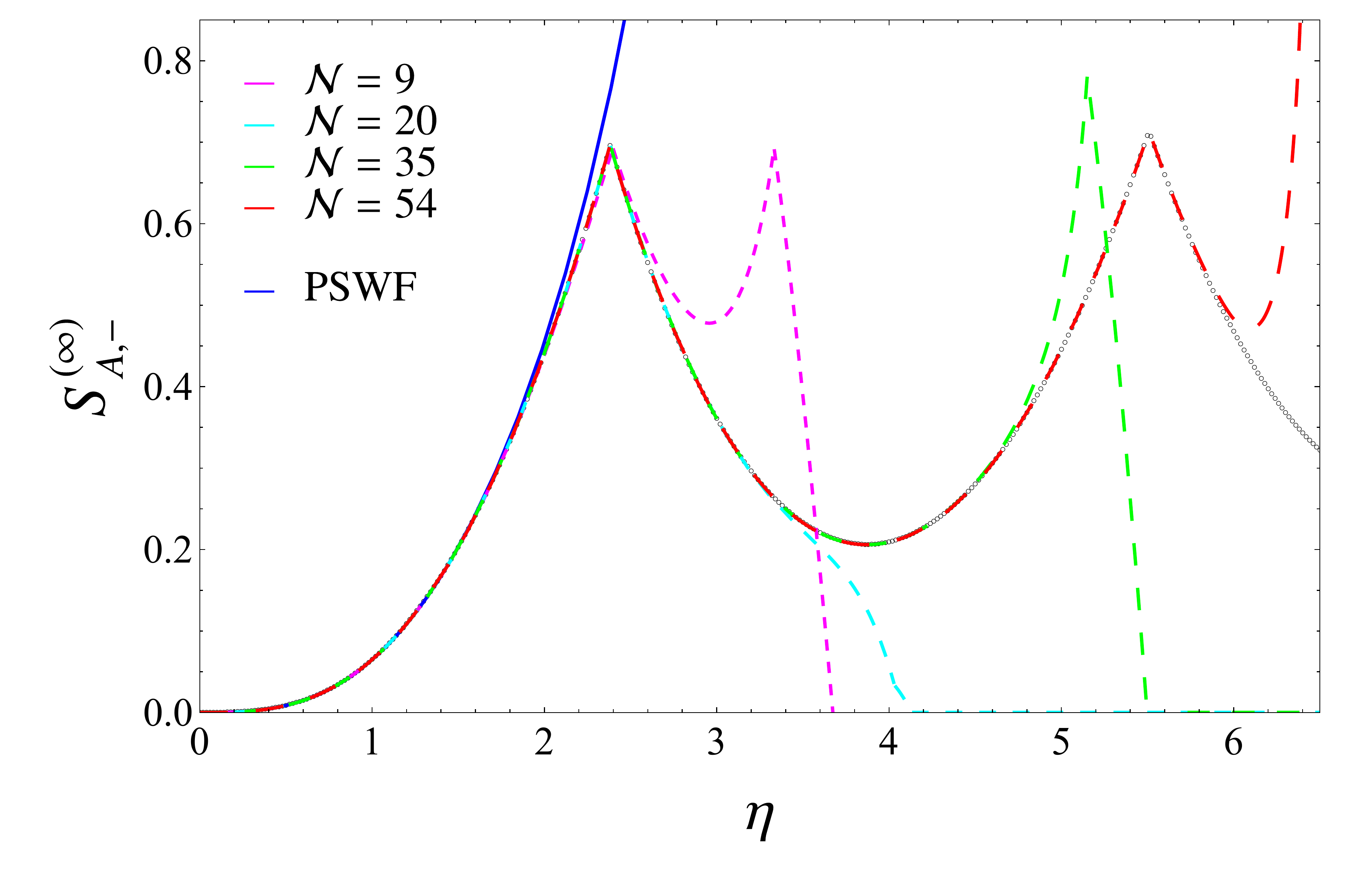}}
\caption{
Entanglement entropies $S_A^{(\alpha)}$ 
for $\alpha =1$ (top panels), $\alpha =3$ (middle panels) and $\alpha \to \infty$ (bottom panels) 
in the small $\eta$ regime.
The blue solid curves correspond to the expansions obtained through the PSWF
(see Sec.\,\ref{sec-small-eta-prolate}),
while the coloured dashed curves have been found through the expansion 
\eqref{tau-GIL-sec} of the tau function
(see Sec.\,\ref{sec-small-eta-tau} and \eqref{approx-entropies-small-eta}).
} 
\vspace{0.4cm}
\label{fig:small-eta-panels}
\end{figure}

As for the entanglement entropy, for Neumann b.c.  we find 
\bea
 \label{ee-small-eta-plus}
S_{A,+} 
& = & 
-\,\frac{2}{\pi}\, \eta \log(\eta)
+ \frac{2}{\pi} \big[1-\log(2/\pi)\big] \, \eta
-  \frac{2}{\pi^{2}}\, \eta^{2}
+  \frac{2}{9\pi}\,\eta^{3} \log(\eta) 
 \nonumber
\\
\rule{0pt}{.7cm}
 &  & 
 + \, \frac{2}{3\pi}
 \left(\frac{1}{3}\log(2/\pi )
 -\frac{2}{\pi^2 }\right)\eta^{3}
 + \frac{4}{3\pi^2}
 \left(\frac{1}{3}-\frac{1}{\pi^2}\right) \eta^{4}
 +O\big(\eta^{5}\log(\eta)\big)
 \eea
 which comes only from $\gamma_0$
because  $s(\gamma_{2n})= O\big(\eta^{5}\log(\eta)\big)$ when $n \geqslant 1$.
Instead, for Dirichlet b.c. we obtain
 \bea
 \label{ee-small-eta-minus}
S_{A,-} & = & 
-\,\frac{2}{3\pi}\,\eta^{3}\log(\eta)
+\frac{2}{9\pi}\big[ 1-\log\!\big(2/(9\pi)\big)\big] \,\eta^{3}
+\frac{2}{25\pi} \,\eta^{5}\log(\eta) 
 \nonumber
\\
\rule{0pt}{.7cm}
 &  & 
 +\,\frac{2}{75\pi} \log\!\big(2/(9\pi)\big)\, \eta^{5}
 -\frac{2}{81\pi^{2}}\, \eta^{6}
  +O\big(\eta^{7}\log(\eta)\big)
\eea
where only $\gamma_1$ has been employed because $s(\gamma_{2n+1})= O\big(\eta^{7}\log(\eta)\big)$ when $n \geqslant 1$.
Comparing the leading terms of \eqref{ee-small-eta-plus} and \eqref{ee-small-eta-minus}, we have that $S_{A,+} > S_{A,-}$ when $\eta \to 0$, which can be observed from Fig.\,\ref{fig:entropies-half-line} and from the top panels of Fig.\,\ref{fig:small-eta-panels}, where the solid blue lines correspond to the expressions in \eqref{ee-small-eta-plus} and \eqref{ee-small-eta-minus}.

The analysis performed for the entanglement entropy can be adapted to find the expansion of the R\'enyi entropies (\ref{renyi-entropies-def-sums}) with $\alpha \neq 1$
as $\eta \to 0$.
We focus on the cases where $\alpha>1$ and finite.
For Neumann b.c. we find
\bea
\label{ren-small-eta-plus}
S_{A,+}^{(\alpha)} 
& = & 
\frac{\alpha}{\alpha-1}\;
\Bigg\{
\bigg[\,
1+\frac{\eta}{\pi} 
+\left(\frac{4}{3\pi^2}-\frac{1}{3}\right)\eta^2
+ \frac{2}{\pi} \left(\frac{1}{\pi^2}-\frac{1}{\pi}\right)\eta^3
\,\bigg]\, \frac{2\eta}{\pi}
 \\
 \rule{0pt}{.7cm}
 &  & \hspace{1.5cm}
 -\,
\bigg[\,
\frac{1}{\alpha}+\frac{2\eta}{\pi} + \left(\frac{2(\alpha+1)}{\pi^{2}}-\frac{1}{9}\right)\eta^{2}+\frac{2(\alpha+1)}{9\pi}\left(\frac{6(\alpha+2)}{\pi^{2}}-1\right)\eta^{3}
\, \bigg]
 \left(\frac{2\eta}{\pi}\right)^{\alpha}
 \nonumber 
 \\
  \rule{0pt}{.7cm}
 &  & \hspace{1.5cm}
 +\,
\bigg[\,\frac{1}{2\alpha}+\frac{2\eta}{\pi}
 +\left(\frac{2(2\alpha+1)}{\pi^{2}}-\frac{1}{9}\right)\eta^{2}
\,\bigg]
\left(\frac{2\eta}{\pi}\right)^{2\alpha}
 \nonumber 
 \\
  \rule{0pt}{.7cm}
 &  & \hspace{1.5cm}
-\,
\left(\frac{1}{3\alpha}+\frac{2\eta}{\pi} \right)
\left(\frac{2\eta}{\pi}\right)^{3\alpha}
+ \frac{1}{4\alpha} \left(\frac{2\eta}{\pi}\right)^{4\alpha}
\Bigg\} 
+O\big(\eta^{5} \big)
\nonumber
\eea
while for Dirichlet b.c. we obtain 
\bea
\label{ren-small-eta-min}
S_{A,-}^{(\alpha)} 
& = & 
\frac{\alpha}{\alpha-1}\;
\Bigg\{
\left( 1 - \frac{9 }{75} \, \eta^{2}+ \frac{1}{9\pi}\, \eta^3 \right)
\frac{2\eta^{3}}{9\pi}
 \\
   \rule{0pt}{.5cm}
 &  &
 \hspace{1.5cm}
 - \left(\frac{1}{\alpha}-\frac{3}{25}\,\eta^{2}+\frac{2}{9\pi}\,\eta^{3}\right)
 \left(\frac{2\eta^{3}}{9\pi}\right)^{\alpha}
 + \frac{1}{2\alpha} \left(\frac{2\eta^{3}}{9\pi}\right)^{2\alpha}
 \Bigg\}
 +O\big(\eta^{7}\big)\,.
 \nonumber
\eea
The derivation of the expansions \eqref{ren-small-eta-plus} and \eqref{ren-small-eta-min}
is reported in Appendix\;\ref{app_small_eta_pswf}.
Notice that the relevance of the various terms
in these expansions as $\eta \to 0$ depends on $\alpha$.

The expansion of the single copy entanglement as $\eta \to 0$ 
can be studied in a similar way, by employing the function $s_\infty(x)$ in \eqref{s-ee-s-infty-def}. When Neumann b.c. are imposed, we find
\be
\label{exp-pswf-small-inf-plus}
S_{A,+}^{(\infty)} 
\, =\, 
\frac{2}{\pi}\, \eta
+\frac{2}{\pi^{2}} \,\eta^{2}
+ \frac{2}{3\pi} \left(\frac{4}{\pi^2}-\frac{1}{3}\right)\eta^{3}
+\frac{4}{\pi^{2}}\left(\frac{1}{\pi^{2}}-\frac{1}{9}\right)\eta^{4}
+O\big(\eta^{5}\big)
\ee
while for Dirichlet b.c. the expansion of the single copy entanglement reads
\be
\label{exp-pswf-small-inf-minus}
S_{A,-}^{(\infty)} 
\,=\, 
\frac{2}{9\pi} \, \eta^{3}
-\frac{2}{75\pi}\,\eta^{5}
+\frac{2}{81\pi^{2}}\,\eta^{6}
+O\big(\eta^{7}\big)\,.
\ee

The relation \eqref{sum-entropies-line-pm} can be employed to check the above expansions. Indeed, by summing up  
either \eqref{ee-small-eta-plus} and \eqref{ee-small-eta-minus},
or \eqref{ren-small-eta-plus} and \eqref{ren-small-eta-min}, 
or \eqref{exp-pswf-small-inf-plus} and \eqref{exp-pswf-small-inf-minus}, 
we recover the expansions found in Sec.\,7.2 of \cite{Mintchev:2022xqh}.

In Fig.\,\ref{fig:small-eta-panels} we compare the small $\eta$ expansions reported above (blue solid lines)
with the corresponding exact curves, obtained numerically 
through \eqref{renyi-entropies-def-sums} (black empty circles).
The cases $\alpha=1$, $\alpha=3$ and $\alpha \to \infty$ are considered, for both boundary conditions. 
Notice that these small $\eta$ expansions do not capture the first local maximum of the corresponding curves.

\subsection{Tau function approach}
\label{sec-small-eta-tau}

The expansion of the entanglement entropies as $\eta \to 0$
can be found by plugging into (\ref{entropy-tau-pm})
the expansion of $\tau_\pm$ in this regime.
In the Appendix\;\ref{app_tau_small_eta}
the latter expansion is obtained as a special case 
of the expansion given in the Conjecture\;4 
of \cite{Gamayun:2013auu}.
The result reads
\be
\label{tau-GIL-sec}
\tau_\pm
\,=\,
\frac{1}{\pi^{\pm1/2}\, \textrm{e}^{\eta^2/4} } 
\sum_{n\,=\,0}^{\infty}
C_{\textrm{\tiny III}'}\big(\! \pm\! \tfrac{1}{4} , \pm \tfrac{1}{4} , \tfrac{1}{4} \mp n\big)\,
\frac{(\eta / 2 )^{n(2n\mp 1)} }{z^n}\;
 \mathcal{B}_{\pm}(n; \eta^2/4) 
\ee
where the coefficients are written in terms of the Barnes $G$-function $G(z)$ as follows
\be
\label{C3-theta-theta-def}
C_{\textrm{\tiny III}'}(\theta , \theta , \hat{\sigma})
\equiv
\frac{\big[ G(1+ \theta + \hat{\sigma})\, G(1+ \theta - \hat{\sigma})\big]^2  }{G(1+2\hat{\sigma})\, G(1-2\hat{\sigma})}\,.
\ee
The functions $\mathcal{B}_{\pm}(n; t)$ are defined in \eqref{B-pm-def} and the first terms of their Taylor expansions for small $t$ are reported in \eqref{B-n-plus-small-t} and \eqref{B-n-minus-small-t}.
Thus, the expressions in \eqref{tau-GIL-sec} are double expansions; 
both in positive powers of $\eta$ and in negative powers of $z$.

Approximate analytic expressions for the entanglement entropies 
in the regime of small $\eta$ can be found 
by adapting the analysis performed in \cite{Mintchev:2022xqh}
for the interval on the line
to the cases we are considering on the half line.

Given a positive integer $\mathcal{N} \geqslant 1$,
let us truncate \eqref{tau-GIL-sec} 
by discarding the terms of order $o(\eta^{\mathcal{N}})$.
This condition provides also a truncation of the series in $n$ occurring in \eqref{tau-GIL-sec} to $n \leqslant N_\pm$, where $N_\pm = N_\pm(\mathcal{N})$.
 The functional form of $N_\pm(\mathcal{N})$ depends on the boundary condition. 
Since $o(1/z^{N_\pm})$ terms in \eqref{tau-GIL-sec} have been neglected,
denoting by $\tilde{\tau}_{\mathcal{N},N_\pm}$  the resulting finite sum,
we have that 
$\tilde{\tau}_{\mathcal{N},N_\pm} =z^{-N_\pm} P_{N_\pm,\,\mathcal{N}}(z)$,
where $P_{N_\pm,\,\mathcal{N}}(z)$ is a polynomial of degree $N_\pm$ 
whose coefficients are polynomials in $\eta$ of different degrees 
that are smaller than or equal to $\mathcal{N}$.
Thus
\be
\label{der-log-tau-approx}
\partial_z \log( \tilde{\tau}_{\mathcal{N},N_\pm} )
\,=\,
\sum_{i \,=\, 1}^{N_\pm} \frac{1}{z-z_i}
- \frac{N_\pm}{z}
\ee
where $z_i \in \mathcal{P}_{N_\pm,\,\mathcal{N}}$ are the zeros of 
$P_{N_\pm,\,\mathcal{N}}(z)$, 
that are highly non trivial functions of $\eta$.
According to the Abel-Ruffini theorem,
the roots of a polynomial of degree five or higher cannot be written through radicals.
This fundamental algebraic obstruction tells us that analytic results
can be found only for $N_\pm \leqslant 4$.

Plugging the finite sum \eqref{der-log-tau-approx} into \eqref{entropy-tau-pm} 
and exploiting that $s_\alpha(0) = 0$, one obtains
\be
\label{approx-entropies-small-eta}
\widetilde{S}_{A;\,\mathcal{N}}^{(\alpha)} 
\,\equiv\,
\sum_{j} s_\alpha (\tilde{z}_j)
\;\;\;\qquad\;\;\;
\tilde{z}_j \in \mathcal{P}_{N_\pm,\,\mathcal{N}} \cap [0,1]
\ee
where only the zeros of $P_{N_\pm,\,\mathcal{N}}(z)$ belonging to $[0,1]$ contribute.
The finite sums (\ref{approx-entropies-small-eta}) approximate the entanglement entropies in the small $\eta$ regime. 
The analytic expressions for \eqref{approx-entropies-small-eta} 
have been obtained by combining the procedure discussed in the Appendix\;E.2 of \cite{Mintchev:2022xqh} with the results in the Appendix\;\ref{app_tau_small_eta}.
The results are quite lengthy and not very instructive; 
hence they have not been reported here.

Some curves obtained from \eqref{approx-entropies-small-eta} 
are shown in Fig.\,\ref{fig:small-eta-panels} and Fig.\,\ref{fig:small-eta-renyi-rough},
for either Neumann (left panels) or Dirichlet (right panels) boundary conditions,
and they correspond to coloured dashed lines.
The exact curves obtained numerically are indicated through empty markers. 
In Fig.\,\ref{fig:small-eta-panels} we have considered $N_+, N_- \in \{1,2,3,4\}$
and the largest values of $\mathcal{N}$ providing them.
In particular, we have that
 $N_+(5)=1$, $N_+(14)=2$, $N_+(27)=3$, and $N_+(44)=4$ 
 for Neumann boundary condition and
 $N_-(9)=1$, $N_-(20)=2$, $N_-(35)=3$, and $N_-(54)=4$ 
 for Dirichlet boundary condition.
 In Fig.\,\ref{fig:small-eta-renyi-rough} we have reported the curves 
 corresponding to $N_+(44)=4$ and $N_-(54)=4$, 
 for various values of $\alpha$.
It is evident in Fig.\,\ref{fig:small-eta-panels}  
that the agreement between the approximations obtained from \eqref{approx-entropies-small-eta}
and the corresponding exact curves improve as $N_\pm$ increases. 
The best approximations for small $\eta$ obtained from \eqref{approx-entropies-small-eta}
(see Fig.\,\ref{fig:small-eta-renyi-rough} and the red curves in Fig.\,\ref{fig:small-eta-panels})
nicely reproduce the first two oscillations of the entanglement entropies.
%

\section{Large $\eta$ expansion}
\label{sec_large_eta}

In this section we study the expansions of the entanglement entropies as $\eta \to \infty$,
which are obtained by plugging 
the expansions of the tau functions $\tau_\pm$ in this regime
into (\ref{entropy-tau-pm}).

The expansion of the Painlevé $\textrm{III}'_1$ tau function 
given in Eq.\,(A.30) of \cite{Bonelli:2016qwg},
properly specialised to the cases we are considering,
provides the large $\eta$ expansions of 
the tau functions $\tau_\pm$ in (\ref{fd_t3p}).
This analysis is described in the Appendix\;\ref{app_large_eta_tau} 
and the results read
\be
\label{t_pm_final}
\tau_{\pm}
=\!
\sum_{n\,\in\,\mathbb{Z}}
\mathrm{e}^{\pm \textrm{i} \frac{\pi}{2}(\nu+n)}\,
\frac{ \mathrm{e}^{\textrm{i} 2\eta(n+\nu)} }{(4\eta)^{(\nu+n)^2}}
\; G(1+\nu+n)\,G(1-\nu-n)\,
\sum_{p\,=\,0}^{\infty}\frac{\mathcal{D}_{p}(\nu+n)}{\big(2\eta \big)^{p}}
\ee
where
\be
\label{nu-z-def}
\nu =\frac{1}{2\pi \textrm{i}\,}\log(1-1/z)
\ee
and
\be
 \label{D_hl}
\mathcal{D}_{0}(\nu)=1
\;\;\qquad\;\;
\mathcal{D}_{1}(\nu)=-\,\textrm{i}\,\nu^{3}
\;\;\qquad\;\;
\mathcal{D}_{2}(\nu)=-\frac{1}{4}\,\nu^{2}\big(2\nu^{4} +5\nu^{2} + 1\big)\,.
\ee
We find it worth remarking that 
the expansions \eqref{t_pm_final} is an asymptotic series in $1/\eta$ \cite{Bonelli:2016qwg};
hence also the corresponding expansions of the entanglement entropies derived from them 
are asymptotic. 
The expansion \eqref{t_pm_final} is not valid for $z\in [0,1]$, where
a different expansion of $\tau_\pm$ is expected
(see e.g. \cite{Ehrhardt_07} for $z=1$).

In order to study the expansion of the entanglement entropies (\ref{entropy-tau-pm}) as $\eta \to \infty$,
we find it convenient to write \eqref{t_pm_final} as follows
\begin{equation}
\label{tau-pm-prod-dec}
\tau_{\pm} \,=\, \tilde{\tau}_{\pm,\infty}\, \mathcal{T}_{\pm,\infty}
\end{equation}
where $\tilde{\tau}_{\pm,\infty}$  is the term corresponding to $n=0$ and $p=0$ in \eqref{t_pm_final}, namely
\be
\label{tau_lead}
\tilde{\tau}_{\pm,\infty}
\,=\,  
\frac{ \mathrm{e}^{\mp \textrm{i}\frac{\pi}{2}\nu}\, \mathrm{e}^{\textrm{i} 2\nu \eta} \, G(1+\nu)\,G(1-\nu) }{  (4\eta)^{\nu^{2}} }
\ee
while $\mathcal{T}_{\pm,\infty} $ reads
\be
\label{tau_sublead}
\mathcal{T}_{\pm,\infty} 
=
 \sum_{n\,\in\,\mathbb{Z}}( \pm\textrm{i} )^{n}\,
 \frac{\mathrm{e}^{ \textrm{i} 2n\eta} }{ (4\eta)^{n(n+2\nu)} }\;
 \frac{G(1+\nu+n)\,G(1-\nu-n)}{G(1+\nu)\,G(1-\nu)}\,
\sum_{p\,=\,0}^{\infty}\frac{\mathcal{D}_{p}(\nu+n)}{\left(2\eta\right)^{p}}\,.
\ee
The term (\ref{tau_lead}) 
corresponds to Eq.\,(1.35) of  \cite{Bothner_2019} 
specialised to our cases
(see Appendix\;\ref{app_large_eta_tau}).

Plugging (\ref{tau-pm-prod-dec}) into (\ref{entropy-tau-pm}), 
the following decomposition for $S_{A,\pm}^{(\alpha)}$ is obtained 
\be
\label{ee-large-eta-dec}
S_{A,\pm}^{(\alpha)}
=
S_{A,\pm,\infty}^{(\alpha)}
+
\widetilde{S}_{A,\pm,\infty}^{(\alpha)}
\ee
where $S_{A,\pm,\infty}^{(\alpha)}$,
which originates from $\log(\tilde{\tau}_{\pm,\infty})$,
provides the leading contributions to the entanglement entropies for large $\eta$; 
while $\widetilde{S}_{A,\pm,\infty}^{(\alpha)}$ 
comes  from $\log(\mathcal{T}_{\pm,\infty})$
and gives the subleading corrections that vanish when $\eta \to \infty$.

The leading terms occurring in $S_{A,\pm,\infty}^{(\alpha)}$ 
can be found by first taking the logarithm of \eqref{tau_lead}, i.e.
\be
\label{log-tau-infty}
\log\! \big(\tilde{\tau}_{\pm,\infty}\big)  
=\,
\textrm{i} \, 2\,\nu\, \eta 
-\nu^{2}\log(4\eta)
\pm \, \textrm{i}\, \frac{\pi}{2}\, \nu
+\log \! \big[ G(1-\nu)\,G(\nu+1)\big]
\ee
and then plugging the resulting expression into \eqref{entropy-tau-pm}.
While the integrals corresponding to the linear terms in $\nu$ in the r.h.s. of 
\eqref{log-tau-infty} vanish, the remaining terms provide non vanishing contributions
to the entanglement entropies and the result reads
\be
\label{S-infty-leading-sec}
S_{A,\pm,\infty}^{(\alpha)}
=
\frac{1}{12}
\bigg(1+\frac{1}{\alpha}\bigg)
\log(4\eta)
+\frac{E_{\alpha}}{2}
\ee
where the  constant term $E_\alpha$ is 
\be
\label{E_alpha-def}
E_\alpha \equiv
\left(1 + \frac{1}{\alpha} \right)
\int_0^\infty
\!\left( 
\frac{\alpha\, \textrm{csch}(t)}{\alpha^2 - 1} \, 
\big( \textrm{csch}(t/\alpha) -\alpha\, \textrm{csch}(t)  \big)- \frac{\textrm{e}^{-2t}}{6} 
\,\right)
\frac{\textrm{d} t}{t}\,.
\ee
We remark that \eqref{S-infty-leading-sec} is independent of the boundary condition.
Furthermore, \eqref{S-infty-leading-sec} is equal to half of the corresponding terms 
in the large $\eta$ expansion of $S_{2A \,\subset \,\mathbb{R}}^{(\alpha)}$ 
(see Eq.\,(8.16) of \cite{Mintchev:2022xqh}),
which has been previously found in the lattice \cite{Jin_2004, Keating_04, Calabrese-Essler-10}  
by using the Fisher-Hartwig conjecture, and also in the continuum
by employing a result of Slepian \cite{Slepian-expansions},
as shown in \cite{EislerPeschelProlate, Susstrunk_2012} 
(see also \cite{Spitzer-14} for a rigorous derivation of the logarithm term).
The fact that the  leading terms in the  large $\eta$ expansions 
of $S_{A, \pm}^{(\alpha)}$ are half of the corresponding ones 
for $S_{2A \,\subset \,\mathbb{R}}^{(\alpha)}$ 
can be observed in Fig.\,\ref{fig:entropies-half-line}.

The term $\widetilde{S}_{A,\pm,\infty}^{(\alpha)}$ in \eqref{ee-large-eta-dec}
is obtained from \eqref{tau_sublead}
and contains all the subleading contributions in the large $\eta$ expansion of $S_{A, \pm}^{(\alpha)}$, which vanish for $\eta \to \infty$.
It can be written as follows
\be
\label{S_larg_exp}
\widetilde{S}_{A,\pm,\infty}^{(\alpha)}
=
\sum_{N=0}^{\infty}\frac{\widetilde{S}_{A,\pm,\infty,N}^{(\alpha)}}{(4\eta)^{N}} \,.
\ee
In the Appendix\;\ref{app_large_eta_entropies} 
the derivation of the coefficients $\widetilde{S}_{A,\pm,\infty,N}^{(\alpha)}$ 
for $N \in \{0,1,2\}$ is described.
These are non trivial functions of $\eta$ that vanish when $\eta \to \infty$ 
and their expressions are reported in the following.

When $N=0$, for the entanglement entropy 
and the R\'enyi entropies with $\alpha \neq 1$
we find respectively (see the Appendix\;\ref{app-large-eta-N-0})
\be
\label{S-0-large-eta-sub}
\widetilde{S}_{A,\pm,\infty,0}
\,=\,
\pm\sin(2\eta)\sum_{k=1}^{\infty}
(-1)^{k+1} \, \frac{(2k-1)\, [(k-1)!]^{2}}{(4\eta)^{2k-1}} 
\,=\,\pm\, \frac{\sin(2\eta)}{4\eta} \left( 1 - \frac{3}{16 \eta^2}   \right) + O\big(1/\eta^5\big)
\ee
and
\be
\label{Ren-0-large-eta-sub}
\widetilde{S}_{A,\pm,\infty,0}^{(\alpha)}
\,=\,
\frac{2}{\alpha-1}
\sum_{j=1}^{\infty}\frac{1}{j}\,
\cos\!\big[(2\eta-\tfrac{\pi}{2})j\big] \,
\sum_{k=1}^{\infty}
\bigg( \frac{\pm\,\Omega(\hat{y}_{k} / \alpha)}{(4\eta)^{(2k-1) /\alpha}} \bigg)^j
\ee
where $\hat{y}_{k}\equiv  \ri (k- 1/2 )$ 
and $\Omega(\hat{y}_{k} / \alpha) \equiv
\tfrac{\Gamma( 1/2 - 1/(2\alpha) +k/\alpha)}{\Gamma( 1/2 + 1/(2\alpha) - k/\alpha)}$ 
(see \eqref{hat-y-def} and \eqref{Omega_eva} respectively).

The coefficient $\widetilde{S}_{A,\pm,\infty,1}^{(\alpha)}$
is derived in the Appendix\;\ref{app-large-eta-N-1}.
For $\alpha=1$ and $\alpha \neq 1$
we obtain respectively 
\be
\label{S-1-large-eta-sub}
\widetilde{S}_{A,\pm,\infty,1}
\,=\,
\pm\cos(2\eta)\sum_{k=1}^{\infty}(-1)^{k+1}\,
\frac{(2k-1)\, [(k-1)!]^{2} \,\mathcal{P}_{1}(\hat{y}_{k})}{(4\eta)^{2k-1}} 
\,=\,
 \mp\, \frac{\cos(2\eta)}{2 \eta}+O\big(1/\eta^3\big)
\ee
with $\mathcal{P}_{1}(\hat{y}_{k})=-6k^{2}+6k-2$ coming from \eqref{P1_def}, 
and
\be
\label{Ren-1-large-eta-sub}
\widetilde{S}_{A,\pm,\infty,1}^{(\alpha)}
\,=\,
- \frac{2}{\alpha-1}
\sum_{j=1}^{\infty}  \,\sin\!\big[(2\eta-\tfrac{\pi}{2})j\big] 
\sum_{k=1}^{\infty} 
\mathcal{P}_{1}(y_{k}/\alpha)\,
\bigg( \frac{\pm\,\Omega(\hat{y}_{k} / \alpha)}{(4\eta)^{(2k-1) /\alpha}} \bigg)^j
\ee
where 
$\mathcal{P}_{1}(y_{k} / \alpha)=-\frac{3}{2} \big[ (2k-1)/\alpha \big]^2-\frac{1}{2}\,.$

\begin{figure}[t!]
\vspace{-.2cm}
\hspace{-.8cm}
\includegraphics[width=1.05\textwidth]{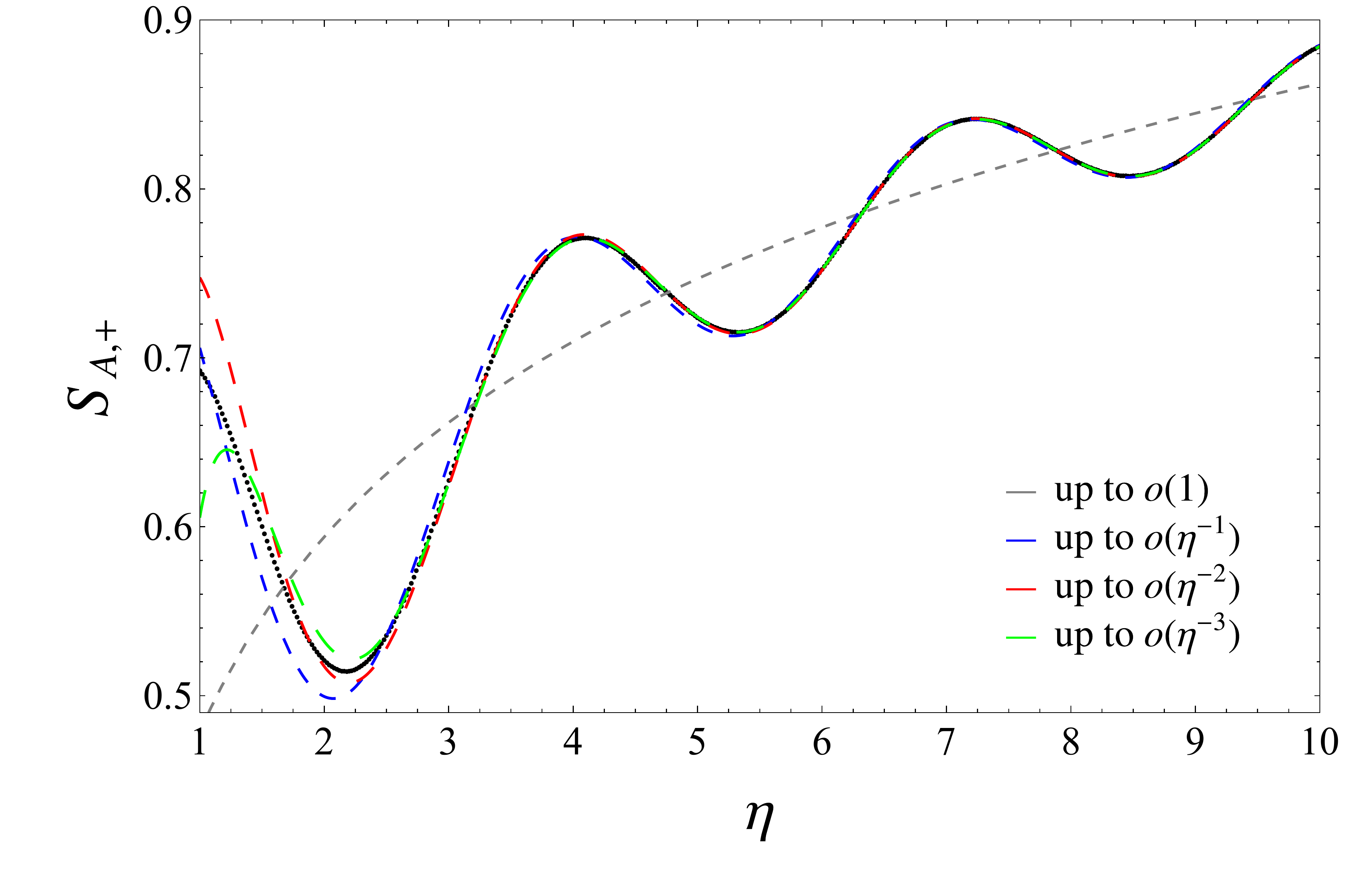}
\vspace{-.7cm}
\caption{
Entanglement entropy for Neumann b.c. in a range where the large $\eta$ expansion 
(see \eqref{ee-large-eta-dec}, \eqref{S-infty-leading-sec}, and \eqref{S_larg_exp}) 
deviates from the curve obtained numerically (black points).
}
\label{fig:ee-large-eta-orders}
\end{figure}

Finally, for $\widetilde{S}_{A,\pm,\infty,1}^{(\alpha)}$ with either 
$\alpha=1$ or $\alpha \neq 1$ we find 
(see the Appendix\;\ref{app-large-eta-N-2})
\bea
\label{S-2-large-eta-sub}
\widetilde{S}_{A,\pm,\infty,2}
&=&
-\,\frac{1}{6}
\pm \sin(2\eta)\sum_{k=1}^{\infty}(-1)^{k+1}\,
\frac{(2k-1)\, [(k-1)!]^{2} 
\big[\mathcal{P}_{2,c}(\hat{y}_{k}) -\mathcal{P}_{2,a}(\hat{y}_{k})\big]}{(4\eta)^{2k-1}} 
\nn
\\
\rule{0pt}{.7cm}
&=& -\,\frac{1}{6} \mp \frac{7\sin(2\eta)}{4 \eta} + O\big(1/\eta^3\big)
\eea
where $\mathcal{P}_{2,c}(\hat{y}_{k}) -\mathcal{P}_{2,a}(\hat{y}_{k}) =-18k^{4}+16k^{3}-8k+3$ is obtained from
\eqref{P2-a-def} and \eqref{P2-c-def},  and
\be
\label{Ren-2-large-eta-sub}
\widetilde{S}_{A,\pm,\infty,2}^{(\alpha)} 
\,=\,
\frac{(\alpha+1)(3\,\alpha^{2}-7)}{48\,\alpha^{3}}
 - 
 \frac{2}{\alpha-1}\,
 \sum_{j=1}^{\infty}\,  \cos\!\big[(2\eta-\tfrac{\pi}{2})j\big] 
 \sum_{k=1}^{\infty}
\widetilde{\mathcal{P}}_{2}(j ; y_{k}/ \alpha)\,
\bigg( \frac{\pm\,\Omega(\hat{y}_{k} / \alpha)}{(4\eta)^{(2k-1) /\alpha}} \bigg)^j
\ee
with $\widetilde{\mathcal{P}}_{2}(j;y) $ defined as  (from (\ref{P2-c-def}), (\ref{P2-d-def}) and (\ref{P2-e-def}))
\be
\widetilde{\mathcal{P}}_{2}(j;y) 
\equiv
-\,\mathcal{P}_{2,c}(y)+\mathcal{P}_{2,d}(y)+(1-j)\,\mathcal{P}_{2,e}(y)
=
18j\, y^{4}-3j \,y^{2}+\frac{j}{8}+20 \ri \,y^{3} - 5\ri \, y\,.
\ee

\begin{figure}[t!]
\vspace{-.2cm}
\hspace{-.8cm}
\includegraphics[width=1.05\textwidth]{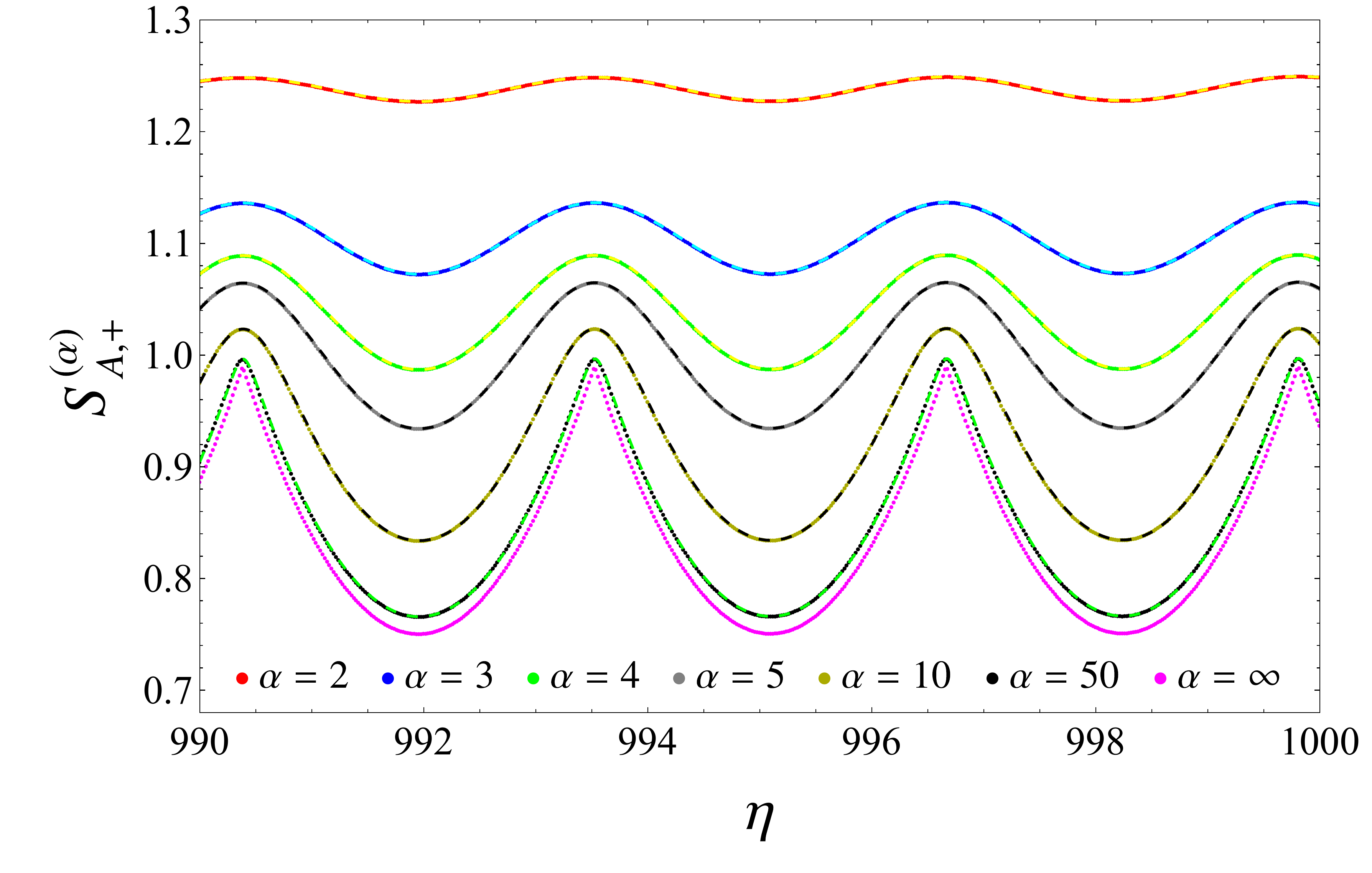}
\vspace{-.7cm}
\caption{
Oscillatory behaviour of the entanglement entropies with $\alpha > 1$  for Neumann b.c.
in the large $\eta$ regime.
The dashed curves are obtained from \eqref{ee-large-eta-dec}, 
\eqref{S-infty-leading-sec} and \eqref{S_larg_exp} with $N \leqslant 2$.
}
\label{fig:ren-very-large-eta}
\end{figure}

In the Appendix\;\ref{app-large-eta-checks} 
some consistency checks for the analytic expressions of $\widetilde{S}_{A,\pm,\infty,N}^{(\alpha)} $ reported above have been discussed. 
In particular, we have considered the limit $\alpha \to 1$, the relation \eqref{sum-entropies-line-pm} and the double scaling limit of the lattice results obtained in \cite{Fagotti:2010cc}.

\begin{figure}[t!]
\vspace{-.2cm}
\hspace{-.8cm}
\includegraphics[width=1.05\textwidth]{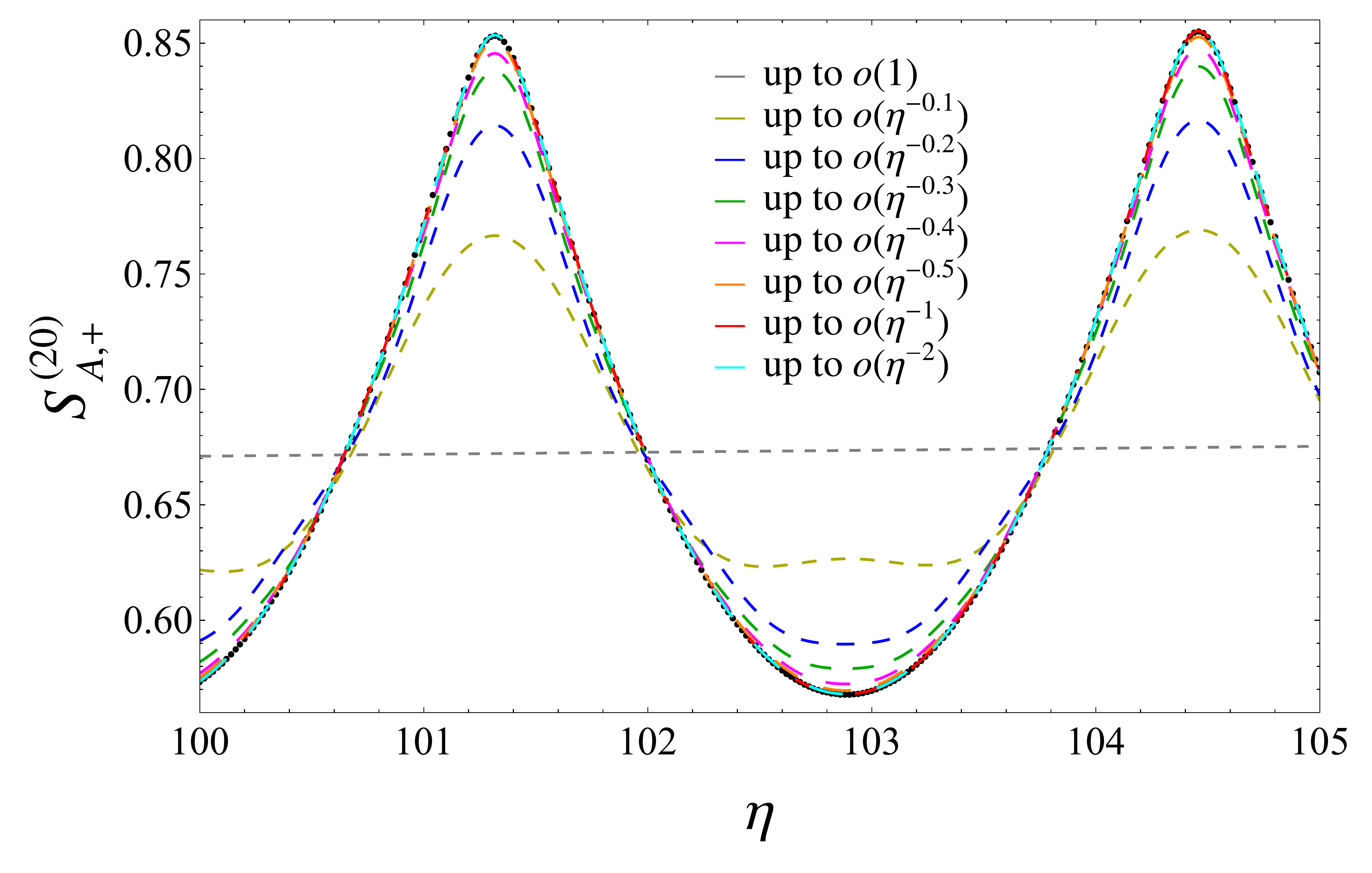}
\vspace{-.7cm}
\caption{
The R\'enyi entropy of order $\alpha=20$ for Neumann b.c. in the regime of large $\eta$. 
}
\label{fig:ren20-orders}
\end{figure}

In Fig.\,\ref{fig:ee-large-eta-orders}, Fig.\,\ref{fig:ren-very-large-eta} and Fig.\,\ref{fig:ren20-orders}, the curves for the entanglement entropies found numerically are compared with the curves corresponding to the analytic expressions 
valid at larg $\eta$ obtained from \eqref{ee-large-eta-dec}, \eqref{S-infty-leading-sec} and \eqref{S_larg_exp},
where $\widetilde{S}_{A,\pm,\infty,N}$ and $\widetilde{S}_{A,\pm,\infty,N}^{(\alpha)}$ are respectively \eqref{S-0-large-eta-sub} and \eqref{Ren-0-large-eta-sub} for $N=0$, \eqref{S-1-large-eta-sub} and \eqref{Ren-1-large-eta-sub} for $N=1$,
and \eqref{S-2-large-eta-sub} and \eqref{Ren-2-large-eta-sub} for $N=2$.
We show these results only for Neumann b.c. because the ones for Dirichlet b.c. are qualitatively very similar.

In Fig.\,\ref{fig:ee-large-eta-orders} the entanglement entropy $S_{A, +}$ is considered:
the exact curve is labelled by the black circles,
the dashed grey line correspond only to the leading terms \eqref{S-infty-leading-sec}
and the dashed coloured lines are obtained by including 
also the three subleading terms having $N \in \{0,1,2\}$ in \eqref{S_larg_exp}.
The analytic approximation that takes into account also the subleading terms 
nicely reproduce the exact curve for $\eta \gtrsim 6.5\,$.
For smaller values of $\eta$, 
the subleading term corresponding to $N=0$ is not enough to capture the exact curve. 
When $\eta < 2.5$ all the approximate curves obtained from an analytic expression
that we have considered deviate from the one of $S_{A, +}$.

In Fig.\,\ref{fig:ren-very-large-eta} we focus on the range $\eta \in[990,1000]$, 
where $\eta$ takes large values and the amplitude of the oscillations is small.
The curves for the R\'enyi entropies found numerically (coloured circles)
are nicely captured by the corresponding approximate analytic expressions 
(dashed lines) found by including all the subleading terms that we have evaluated
(i.e. the ones having $N \in \{0,1,2\}$ in (\ref{S_larg_exp})),  
truncated at order $O(1/\eta^3)$.

In Fig.\,\ref{fig:ren20-orders} we consider $S_{A, +}^{(20)}$ in a regime where $\eta$ is large enough and show how the agreement between the exact curve (black circles) and the ones obtained from the analytic expressions of the large $\eta$ expansion truncated at some order improves as the number of subleading terms in this truncated sum increases, i.e. when higher orders in $1/\eta$ are included. 
The dashed grey line corresponds to the leading terms \eqref{S-infty-leading-sec}.

We find it worth remarking that the oscillating terms in $\widetilde{S}_{A,+,\infty,N}$ 
and $\widetilde{S}_{A,-,\infty,N}$ have opposite signs (see \eqref{S-0-large-eta-sub}, \eqref{S-1-large-eta-sub} and \eqref{S-2-large-eta-sub}), while the non oscillating ones (only the constant in \eqref{S-2-large-eta-sub} for the terms we are considering) are equal to half of the corresponding ones in the large $\eta$ expansion of $S_{2A \,\subset \,\mathbb{R}}^{(\alpha)}$ \cite{Mintchev:2022xqh}.
This implies that the oscillating terms cancel in the r.h.s. of \eqref{sum-entropies-line-pm}
(see also \eqref{check-app-full-line}) and therefore 
$S_{2A \,\subset \,\mathbb{R}}$ does not oscillate, as found in \cite{Mintchev:2022xqh}.
Notice that, instead, this exact cancellation of the oscillating terms in the r.h.s. of (\ref{sum-entropies-line-pm}) does not occur when $\alpha \neq 1$.

By applying the above observations to the combination (\ref{b-comb-def-pm}),
we find that the non oscillating terms simplify, for any $\alpha >0$.
This cancellation is due to the relative factor of $-1/2$ 
between $S_{A, \pm}^{(\alpha)}$ and $S_{2A \,\subset \,\mathbb{R}}^{(\alpha)}  $ in (\ref{b-comb-def-pm}).

\section{Cumulants expansion}
\label{sec_cumulants_ee}

In this section we discuss the relation, 
found in \cite{Klich-Levitov-09, Klich-Song-11, Klich-Laflorencie-12},
between entanglement entropy and the charge cumulants  
for the models that we are considering.

The cumulants of 
the time independent local charge operator
$Q_{A, \pm}=\int_{A} \varrho_\pm(t=0, x) \, \rd x$
are
$\mathcal{C}_A^{(k)} \equiv \big[\partial^k_{\ri \zeta} 
\log \!\big(\langle \e^{\textrm{i} \zeta Q_{A, \pm}} \rangle \big)
\big]\big|_{\zeta=0}\,$, where $k \geqslant 1$.
Their generating function  can be expressed in terms of the tau function 
\eqref{tau-function-pm} as follows
\cite{Klich-Levitov-09,Klich-Laflorencie-12,Ivanov-2013,Mintchev:2022xqh}
\be
\label{QA-pm-def}
\log\! \big[  \langle \e^{\textrm{i} \zeta Q_{A, \pm}} \rangle \big]
=
\textrm{Tr} \big[ \log  \big( I + (\e^{\textrm{i} \zeta} - 1) \, K_\pm \big) \,\big]
=
\textrm{Tr} \big[ \log  \big( I - z^{-1}\, K_\pm \big) \,\big]
=\,
\log(\tau_\pm)
\ee
where $I$ is the identity operator, $K_\pm$ are the kernels \eqref{kernel-pm-def}
and  $\zeta=2\pi\nu$, with $\nu=\nu(z)$ being defined in \eqref{nu-z-def}.

From \eqref{QA-pm-def}, one finds that  the first cumulants are given by 
\be
\label{Cum_01234}
\mathcal{C}_{A,\pm}^{(1)}=\mathrm{Tr}\big(K_{\pm}\big)
\;\;\qquad\;\;
\mathcal{C}_{A,\pm}^{(2)}=\mathrm{Tr}\big(K_{\pm}-K_{\pm}^{2}\big)
\;\;\qquad\;\;
\mathcal{C}_{A,\pm}^{(3)}=\mathrm{Tr}\big(K_{\pm}-3K_{\pm}^{2}+2K_{\pm}^{3}\big)
\ee
where the Schatten $p$-norm 
of the kernels \eqref{kernel-pm-def} is defined as
\be
\label{schatten}
\mathrm{Tr}\big(K_{\pm}^{p}\big)
=
\sum_{n=0}^{\infty}\left(\gamma_{n}^{\pm}\right)^{p} 
\;\;\;\qquad\;\;\;
p\geqslant 1\,.
\ee
The $n$-th cumulant $\mathcal{C}_{A,\pm}^{(n)}$ is a finite linear combinations of the Schatten norms \eqref{schatten} with $p \in \mathbb{N}$ and $p=1,\dots,n$;
hence we can evaluate numerically the cumulants
by computing the Schatten norms from \eqref{schatten},
as done for the entanglement entropies.

In  \cite{Klich-Levitov-09, Klich-Song-11, Klich-Laflorencie-12} a remarkable relation between the entanglement entropies and the charge cumulants has been studied 
(the final form has been reported in \cite{Klich-Song-11}). 
A similar relation has been  found also for the R\'enyi entropies with integer index \cite{Klich-Laflorencie-12}. Focussing only on the entanglement entropy for simplicity, for the models we are considering this relation reads
\be
\label{ent_cum}
S_{A,\pm}
=
\lim_{q \to \infty}
\sum_{n\,=\,1}^{q+1}a_{n}(q) \, \mathcal{C}_{A,\pm}^{(n)}
\;\;\;\qquad\;\;\;
a_{n}(q)
\equiv
\left\{
\begin{array}{ll}
\;0  
& 
\textrm{odd}\;\; n
\\
\displaystyle
\; 2\! \sum_{k=n-1}^{q} \! \frac{S_{1}(k,n-1)}{k! \, k}
\hspace{.7cm}& 
\textrm{even}\;\; n
\end{array}
\right.
\ee
where $S_{1}(k,n)$ are the unsigned Stirling numbers of the first kind. 
We remark that the coefficients in \eqref{ent_cum} are independent of $\eta$;
hence the dependence on $\eta$ of $S_{A, \pm}$
is encoded only in the cumulants.

\begin{figure}[t!]
\vspace{-.2cm}
\hspace{-.8cm}
\includegraphics[width=1.05\textwidth]{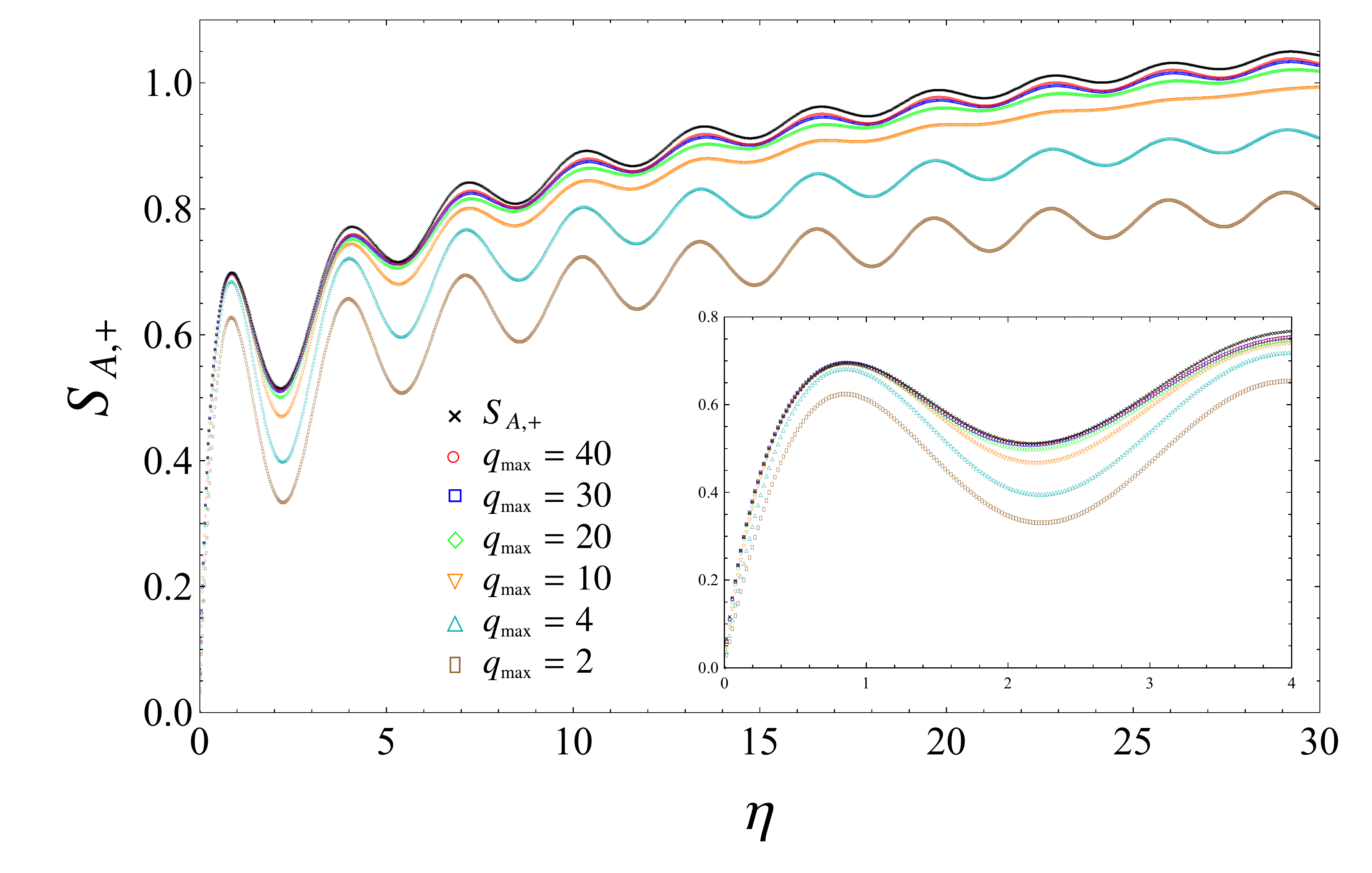}
\vspace{-.7cm}
\caption{Cumulant expansion of the entanglement entropy for Neumann b.c. (see \eqref{ent_cum}). 
The inset zooms in on a range of small values of $\eta$, 
where a complete overlap is observed when $q_{\textrm{\tiny max}} $ is large enough. 
\label{fig:cumulants-ee}
}
\end{figure}

In Fig.\,\ref{fig:cumulants-ee} we show some results about the cumulant expansion 
(\ref{ent_cum}) in the case of Neumann b.c.
(the curves for $S_{A, -}$ are very similar).
The black crosses correspond to the curve obtained numerically, 
while the curves identified by the empty markers 
are given by the finite sums  obtained by restricting \eqref{ent_cum} 
to $n \leqslant q_{\textrm{\tiny max}} + 1 $,
for different values of $q_{\textrm{\tiny max}} $.
The approximation of the exact curve 
improves as $q_{\textrm{\tiny max}}$ increases.
Notice that a complete agreement is obtained 
for small values of $\eta$, as highlighted in the inset of Fig.\,\ref{fig:cumulants-ee}.

In \cite{Klich-Levitov-09} the cumulant expansion 
$S_{A,\pm}=\lim_{q\rightarrow \infty}\sum_{n\,=\,1}^{q}\tilde{a}_{n} \, \mathcal{C}_{A,\pm}^{(n)}$
has been first proposed,
where $\tilde{a}_{n}=\lim_{q\rightarrow \infty}a_{n}(q)=2\zeta(n)$ and $\zeta(z)$ is the Riemann zeta function.
This expansion is divergent \cite{Klich-Song-11}. 
Indeed, testing numerically this expansion, we found 
that its deviation from the entanglement entropy increases with $q$.

The Schatten norms, and consequently also the cumulants,
can be expressed in terms of the tau function \eqref{tau-function-pm}
by adapting to \eqref{schatten} the procedure leading 
to the contour integral \eqref{entropy-tau-pm} 
for the entanglement entropies.
The result reads
\be
\label{int_sn}
\mathrm{Tr}\big(K_{\pm}^{p}\big)
\,=\,
\frac{1}{2\pi\ri }\oint_{\mathfrak{C}}z^{p}\,\partial_{z}\log(\tau_{\pm})\,\rd z
\;\;\; \qquad\;\;\;
 p\in\mathbb{N}\,.
\ee
This expression allows us to write the expansions of the Schatten norms 
in the regimes of small and large $\eta$
by adapting to these quantities the analyses  
discussed in Sec.\,\ref{sec_small_eta} and Sec.\,\ref{sec_large_eta}  
for the entanglement entropies, as done in 
\cite{Mintchev:2022xqh} for the interval on the line.

\begin{figure}[t!]
\vspace{-.2cm}
\hspace{-.8cm}
\includegraphics[width=1.05\textwidth]{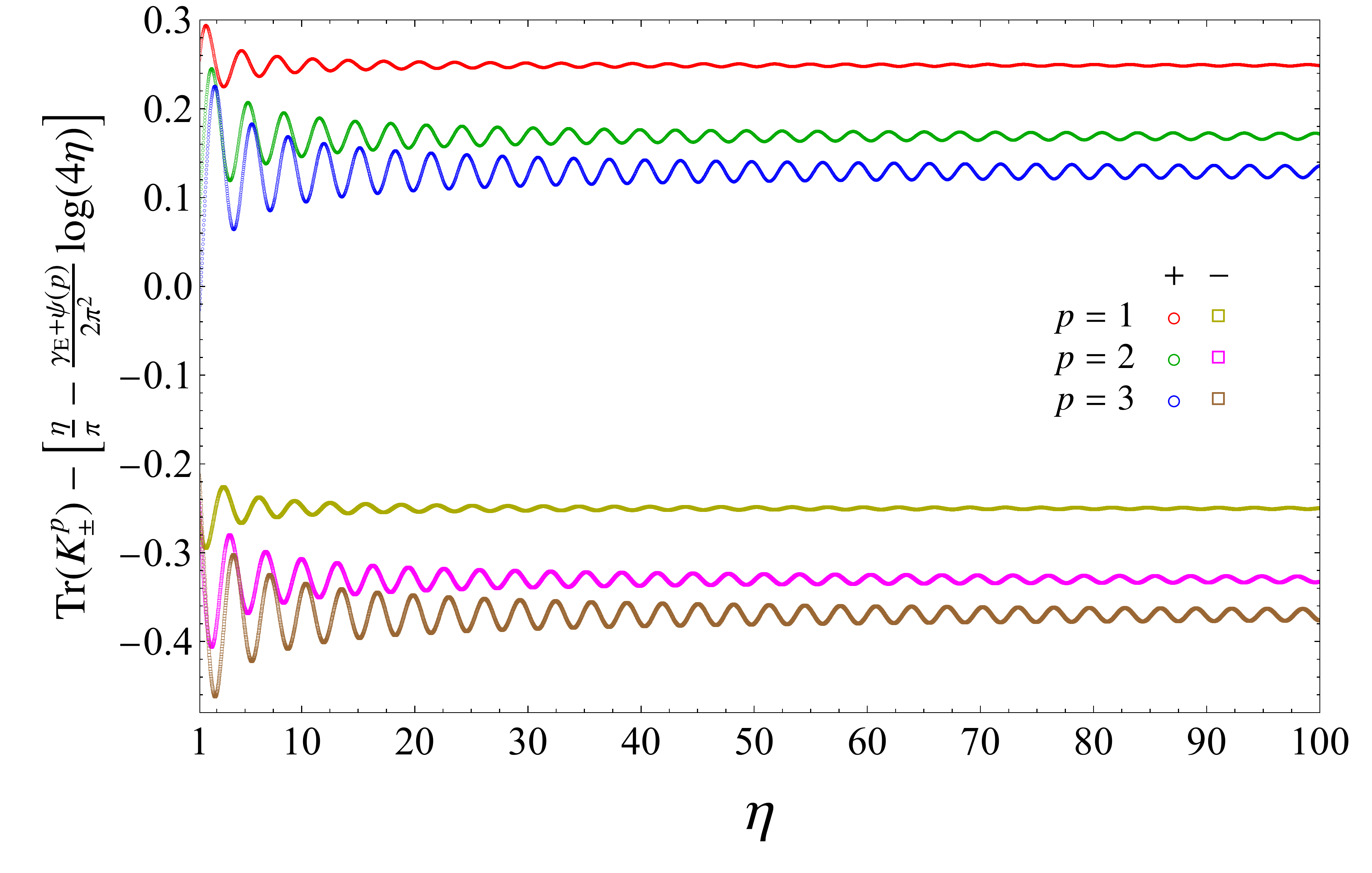}
\vspace{-.7cm}
\caption{
Schatten norms (\ref{schatten}) for some values of $p$,
when either Neumann ($+$) or Dirichlet ($-$) b.c. are imposed,
in the large $\eta$ regime (see (\ref{sn_large})). 
\label{fig:schatten-norms}
}
\end{figure}

From (\ref{sine-tau-function-factorisation}) 
and (\ref{int_sn}), it is straightforward to observe that
\be
\label{blow-up-for-schatten}
\mathrm{Tr}\big(K_{\textrm{\tiny sine}}^{p}\big)
\,=\,
\mathrm{Tr}\big(K_{+}^{p}\big)
+
\mathrm{Tr}\big(K_{-}^{p}\big)
\ee
where $\mathrm{Tr}\big(K_{\textrm{\tiny sine}}^{p}\big)$
are the Schatten norms for the interval $[-R,R]$ on the line 
considered in \cite{Mintchev:2022xqh}.
The case $p=1$ of (\ref{blow-up-for-schatten}) is interesting 
because the l.h.s. is known analytically.
In particular, by employing Eq.\,(3.55) in \cite{Rokhlin-book}, 
we have
\be
\label{blow-up-for-schatten=p=1}
\frac{2\eta}{\pi}
\,=\,
\mathrm{Tr}\big(K_{+}\big)
+
\mathrm{Tr}\big(K_{-}\big)\,.
\ee

\begin{figure}[t!]
\vspace{-.2cm}
\hspace{-1.5cm}
\includegraphics[width=1.1\textwidth]{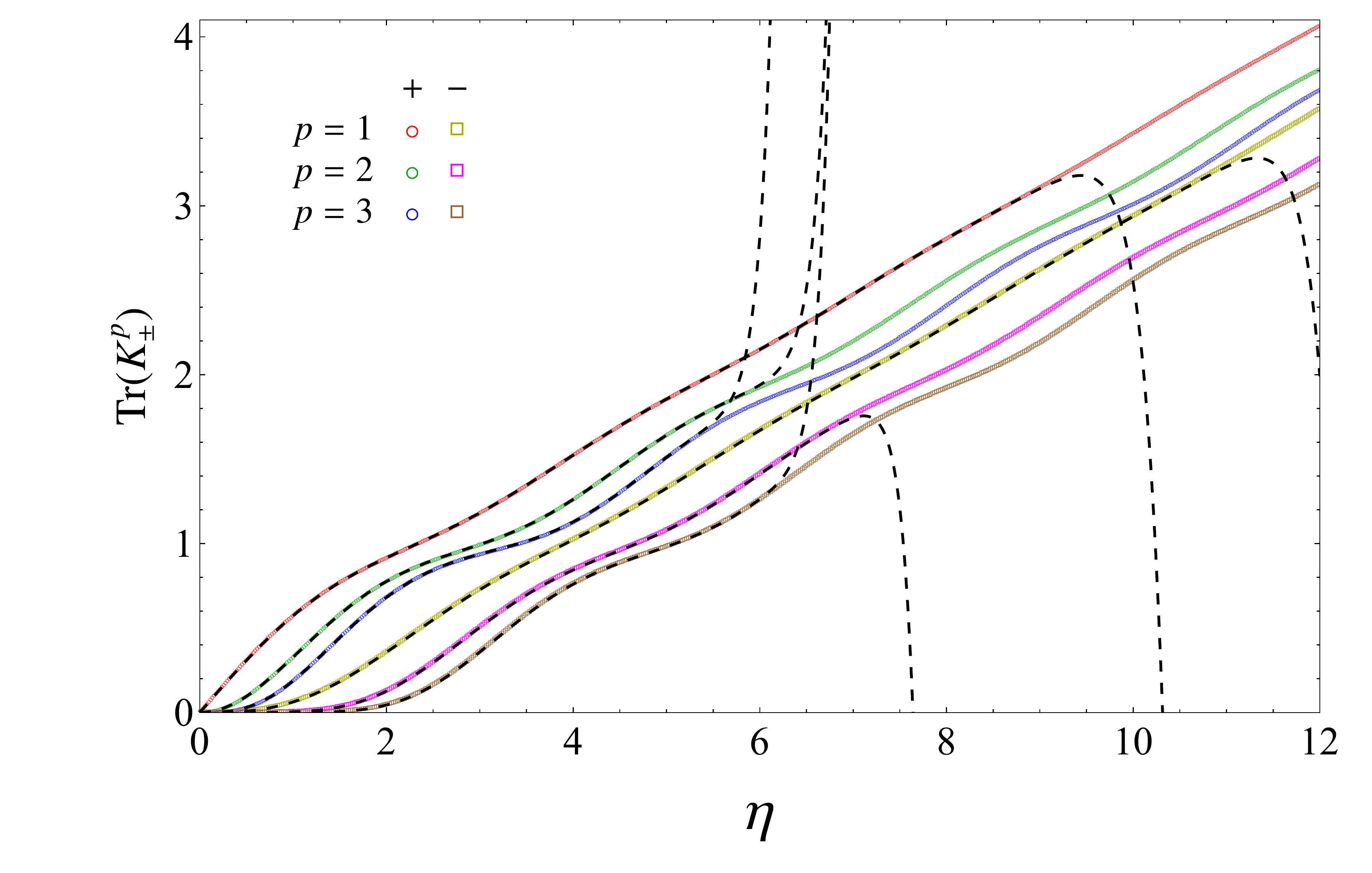}
\vspace{-.7cm}
\caption{
Schatten norms (\ref{schatten}) for some values of $p$ in the small $\eta$ regime,
when either Neumann ($+$) or Dirichlet ($-$) b.c. are imposed.
The dashed black curves have been obtained through 
(\ref{approx-sn-small-eta}) in the best 
approximation where analytic expressions can be found (see Sec.\,\ref{sec-small-eta-tau}).
\label{fig:schatten-norms-small-eta}
}
\end{figure}

In the regime of large $\eta$, 
by using (\ref{logtBIP}) into (\ref{int_sn})
we find that
\be
\label{sn_large}
\mathrm{Tr}(K_{_\pm}^{p}) 
\,=\,
\frac{\eta}{\pi} 
- \frac{\gamma_{\textrm{\tiny E}} +\psi(p)}{2\pi^{2}} \, \log(4\eta)
\pm \frac{1}{4}
+ C_0(p)
+ o(1)
\ee
where $\gamma_{\textrm{\tiny E}} \simeq 0.577$ is the Euler-Mascheroni constant,
$\psi(x)$ is the digamma function
(notice that $\psi(1) = - \,\gamma_{\textrm{\tiny E}}$)
and $C_0(p)$ is independent of the boundary conditions
because it comes from the term $\log[G(1+\nu)\,G(1-\nu)]$ in (\ref{logtBIP}).
The linear and the logarithmic terms in the expansion (\ref{sn_large}) 
are independent of the boundary conditions and
they are equal to $1/2$ of the corresponding terms in the expansion
of the Schatten norms of the two-point function for the interval
on the line (see Eq.\,(9.17) of \cite{Mintchev:2022xqh}). 
The dependence on the boundary conditions occurs 
in the constant term of \eqref{sn_large}.

The proper combination of the expansions \eqref{sn_large}
provide the corresponding expansions of the cumulants (see \eqref{Cum_01234}). 
In particular, 
one finds that the linear terms of \eqref{sn_large} simplify in the combinations \eqref{Cum_01234}
and therefore $\mathcal{C}_{A,\pm}^{(n)}=\beta_{\pm}^{(n)}\log(4\eta)+O(1)$ as $\eta \to \infty$.
By employing this observation in \eqref{ent_cum},
we conclude that all the cumulants contribute to the leading logarithmic term of 
$S_{A, \pm}$ in \eqref{S-infty-leading-sec}.

In Fig.\,\ref{fig:schatten-norms} 
we show the numerical results of the Schatten norms 
for different orders $p$, evaluated numerically through \eqref{schatten}
in the range $\eta\in[1,100]$. 
Subtracting the leading divergent terms for large values of $\eta$,
the resulting expression tends to a constant as $\eta \to \infty$
which depend both on $p$ and on the boundary condition.
The dependence on the boundary conditions of the $O(1)$ 
term in \eqref{sn_large} occurs only through the term $\pm 1/4$;
indeed, by using the curves reported in Fig.\,\ref{fig:schatten-norms},
we have checked that the curves for 
$\mathrm{Tr}(K_{_\pm}^{p}) 
- \big[ \frac{\eta}{\pi} - \frac{\gamma_{\textrm{\tiny E}} +\psi(p)}{2\pi^{2}} \, \log(4\eta)
\pm \frac{1}{4} \big]$
associated to different boundary conditions oscillate 
around the same constant value, which corresponds to $C_0(p)$.

In the regime of small $\eta$,
the analysis discussed in Sec.\,\ref{sec-small-eta-tau}
for the entanglement entropies
can be applied to the Schatten norms in a straightforward way.
The approximate analytic expressions for 
$\mathrm{Tr}(K_{\pm}^{p})$ as $\eta \to 0$ read
\be
\label{approx-sn-small-eta}
\sum_{j} \tilde{z}_j^p
\;\;\;\qquad\;\;\;
\tilde{z}_j \in \mathcal{P}_{N_\pm,\,\mathcal{N}} \cap [0,1]
\ee
where $\mathcal{P}_{N_\pm,\,\mathcal{N}}$ and $\tilde{z}_j $
have been introduced in Sec.\,\ref{sec-small-eta-tau}.
In Fig.\,\ref{fig:schatten-norms-small-eta}
we compare the numerical results for $\mathrm{Tr}(K_{\pm}^{p})$ 
when $\eta \in [0, 12]$ with the best analytic expressions
(dashed black curves) obtained from (\ref{approx-sn-small-eta})
as discussed in Sec.\,\ref{sec-small-eta-tau} for the
entanglement entropies. 
The agreement is remarkable in a range of $\eta$
whose width decreases with $p$.
%

\section{Conclusions}
\label{sec_conclusions}

We studied the entanglement entropies $S_{A,\pm}^{(\alpha)}$
of an interval adjacent to the boundary of the half line 
for the free fermionic spinless Schr\"odinger field theory
at finite density $\mu$ and zero temperature,
along the lines of the analysis made
 in \cite{Mintchev:2022xqh} of these quantities 
 for an interval on the line.
We have considered the models characterised 
by scale invariant boundary conditions (\ref{h12})
at the origin of the half line,
which are of either Neumann ($+$) 
or Dirichlet ($-$) type.

The spectral problems \eqref{k1}
can be solved (see Sec.\,\ref{sec_spectral})
through the sine kernel spectral problem
in the interval on the line, 
whose solution has been found by Slepian, Pollak and Landau in the seminal papers
\cite{Slepian-part-1, Slepian-part-2, Slepian-part-3, Slepian-part-4, Slepian-83}.
The corresponding eigenvalues $\gamma_n^\pm$,
which are functions of the 
dimensionless parameter $\eta$ introduced in \eqref{normalised-densities-at-R},
can be written in terms of the PSWF, as shown in 
\eqref{soln-spectral-problem-neumann}, \eqref{soln-spectral-problem-dirichlet} 
and \eqref{eigenvalues}. The relation \eqref{sum-entropies-line-pm} and the results of \cite{Mintchev:2022xqh}
allow to prove that $S_{A,\pm}^{(\alpha)}$ are finite functions of $\eta$
(see e.g. in Fig.\,\ref{fig:entropies-half-line}, Fig.\,\ref{fig:small-eta-renyi-rough} 
and Fig.\,\ref{fig:ent-3-large-eta}).
We remark that $S_{A,\pm}$ display an oscillatory behaviour,
differently from $S_{2A \,\subset \,\mathbb{R}}$,
as shown in Fig.\,\ref{fig:entropies-half-line}.

The numerical evaluation of $S_{A,\pm}^{(\alpha)}$
has been performed as described in \cite{Mintchev:2022xqh}.
In this manuscript  we have obtained
analytic expressions for the expansions of $S_{A,\pm}$,
both in the small $\eta$ regime
(see \eqref{approx-entropies-small-eta} and Fig.\,\ref{fig:small-eta-panels})
and in the large $\eta$ regime
(see \eqref{ee-large-eta-dec},
\eqref{S-infty-leading-sec}, \eqref{S_larg_exp}
with the coefficients corresponding to $N \in \{0,1,2\}$ written explicitly 
in Sec.\,\ref{sec_large_eta}, and Fig.\,\ref{fig:ent-3-large-eta}).
These analytic results are based on the expansions of the 
Fredholm determinants $\tau_\pm$ in \eqref{tau-function-pm},
that are given by \eqref{tau-GIL-sec} and \eqref{t_pm_final}.
Since $\tau_\pm$ are special cases of the Bessel kernel tau function 
\cite{Tracy:1993xj, Jimbo-82, forrester-book}
(see (\ref{tau-pm-from-tau-B})),
these expansions have been obtained from the
expansions found in \cite{Gamayun:2013auu} and \cite{Bonelli:2016qwg}
for the Painlev\'e $\textrm{III}_1'$ tau functions
as $\eta \to 0$ and $\eta \to \infty$ respectively
(see the Appendices\;\ref{app_tau_small_eta} and \ref{app_large_eta_tau}).
Some terms of the small $\eta$ expansions of $S_{A,\pm}^{(\alpha)}$ 
have been obtained also through the properties of the PSWF 
(see Sec.\,\ref{sec-small-eta-prolate}).

The large $\eta$ expansion of $S_{A,\pm}$
can be also written as in (\ref{ee_pm_friedel})
and this form highlights the fact that the first correction 
vanishing as $\eta \to \infty$ can be expressed 
through the Friedel oscillations occurring in the 
normalised density at the entangling point 
(\ref{normalised-densities-at-R}) (see Fig.\,\ref{fig:friedel-ee}).

In Sec.\,\ref{sec_lifshitz} we have shown that
the results obtained for the Schr\"odinger model,
whose Lifshitz exponent is $z=2$,
can be employed to obtain the corresponding ones 
for a hierarchy of two component Lifshitz fermion fields on the half line 
having even $z$.
In the same section, we have considered also a hierarchy of 
Lifshitz fermion fields on the half line with odd values of $z$,
whose first model ($z=1$) corresponds to 
the massless Dirac fermion on the half line 
\cite{Mintchev:2020uom}.

Finally, we have discussed the expansion (\ref{ent_cum}),
found in \cite{Klich-Levitov-09, Klich-Song-11, Klich-Laflorencie-12},
of $S_A$ in terms of charge cumulants (see Fig.\,\ref{fig:cumulants-ee}),
and the Schatten norms (\ref{schatten})
by adapting the procedure described for 
the entanglement entropies
(see Fig.\,\ref{fig:schatten-norms} and Fig.\,\ref{fig:schatten-norms-small-eta}).

Various directions can be explored to extend the results discussed in this manuscript. 
Considering free non-relativistic models, 
it would be interesting to study the entanglement entropies 
when the system is in a generic Gibbs state,
where both the density and the temperature are non vanishing \cite{Spitzer-16,Spitzer-17,Spitzer-22}.
Furthermore, it would be very insightful to investigate the same problems 
also for non-relativistic bosonic fields. 
Besides a physical boundary, a point-like defect provides
another way to break the invariance under translations
that would be interesting to explore for non-relativistic field theories
\cite{Sakai:2008tt,EP-defects,Calabrese:2011ru, 
EislerPeschel:2012def, Mintchev:2020jhc,Capizzi:2022xdt}.
It is worth considering also the entanglement entropies of
more complicated spatial bipartitions,
given e.g. by two disjoint intervals on the line
\cite{Calabrese:2009ez,Calabrese:2010he,Casini:2009vk,Coser:2013qda,Coser:2015dvp,DeNobili:2015dla,Arias:2018tmw, Grava:2021yjp}
or by a single interval not adjacent to the boundary of the half line
\cite{Fagotti:2010cc, Mintchev:2020uom}.
Interesting models where it is important to understand the properties of the
bipartite entanglement include 
the relativistic massive models and their non-relativistic limit
\cite{Casini:2005rm,Casini:2005zv, Daguerre:2020pte}.
The most important generalisations to study are 
the non-relativistic interacting models, like e.g.
the $d=1$ spinfull fermionic field with a quartic interaction  
\cite{Benfatto-book, Gallavotti_01, Gentile:2001gb}.

\vskip 20pt 
\centerline{\bf Acknowledgments} 
\vskip 5pt

It is our pleasure to thank 
Alexander Its and Oleg Lisovyy for insightful conversations and suggestions.
We are grateful to 
Giulio Bonelli, Pasquale Calabrese, Maurizio Fagotti, 
Davide Fioravanti, Pavlo Gavrylenko, Domenico Seminara, 
Germ\'an Sierra, Wolfgang Spitzer,
Luca Tagliacozzo and Alessandro Tanzini
for useful discussions. 
ET acknowledges the Instituto de F\'isica Te\'orica (Madrid)
for warm hospitality and support during part of this work. 
ET’s research has been conducted within the framework of the Trieste Institute for Theoretical Quantum Technologies (TQT).

\vskip 30pt

\appendix

\section{A Bessel kernel and a Painlev\'e III} 
\label{app_bessel}

Let us consider the Bessel kernel 
\cite{Tracy:1993xj, forrester-book,FORRESTER1993} 
\bea
\label{bessel-K-TW}
K_{\textrm{\tiny B}} (a ;x,y)
&\equiv&
\frac{ \sqrt{y}\; J_{a}(\sqrt{x}\,)\, J'_{a}(\sqrt{y}\,) -  \sqrt{x}\; J_{a}(\sqrt{y}\,)\, J'_{a}(\sqrt{x}\,) }{2\,(x - y)}
\\
\label{bessel-K-TW-v2}
\rule{0pt}{.8cm}
& = &
\frac{ \sqrt{x}\; J_{a+1}(\sqrt{x}\,)\, J_{a}(\sqrt{y}\,) -  \sqrt{y}\; J_{a+1}(\sqrt{y}\,)\, J_{a}(\sqrt{x}\,) }{2\,(x - y)}
\\
\label{bessel-K-TW-v3}
& = &
\frac{1}{4}\int_{0}^{1} J_{a}\big(\sqrt{xs}\,\big)\, J_{a}\big(\sqrt{ys}\,\big)\, \rd s \hspace{1.5cm} a>-1 
\eea
where $x>0$  and $y>0$ with $x\neq y$, 
$J_a'(\xi) \equiv \partial_\xi J_a(\xi)$
and some identities for Bessel functions have been employed\footnote{
The expressions in (\ref{bessel-K-TW-v2}) and (\ref{bessel-K-TW-v3}) have been obtained by using respectively $\xi J_a'(\xi)= a J_a(\xi)- \xi J_{a+1}(\xi)$ and 
\be
\int 
z\,J_{a}(\tilde{x}z)J_{a}(\tilde{y}z)\, \rd z
\,=\,
z\;
\frac{\tilde{x}J_{a+1}(\tilde{x}z)J_{a}(\tilde{y}z)-\tilde{y}J_{a}(\tilde{x}z)J_{a+1}(\tilde{y}z)}{\tilde{x}^{2}-\tilde{y}^{2}} \,.
\ee
}.

The kernels \eqref{kernel-pm-def} can be expressed in terms of the Bessel kernel \eqref{bessel-K-TW} as follows
\be
\label{K-pm-from-K-B}
K_\pm(\eta ;x,y) 
\,=\,
2\,\eta^2 \sqrt{x\, y}\; K_{\textrm{\tiny B}} \big(\! \mp \!\tfrac{1}{2}\, ; (\eta\, x)^2 , (\eta\, y)^2 \big)\,.
\ee
This identity leads us to write the spectral problems \eqref{k1} as
\be 
\label{sp-KB-pm-app0}
2\,\eta^2  \int_0^1 \! \rd y \, \sqrt{x\, y}\; K_{\textrm{\tiny B}} \big(\! \mp \!\tfrac{1}{2}\, ; (\eta\, x)^2 , (\eta\, y)^2 \big) \, f^\pm_n(\eta; y) 
\,=\, \gamma_n^\pm\, f^\pm_n(\eta ; x) \,.
\ee
In terms of $\tilde{x} \equiv (\eta\, x)^2$ and of the integration variable $\tilde{y} \equiv (\eta\, y)^2$, this becomes
\be 
\label{sp-KB-pm-app1}
 \int_0^{\eta^2} \!\! \rd \tilde{y} \; K_{\textrm{\tiny B}} \big(\! \mp \!\tfrac{1}{2}\, ; \tilde{x} , \tilde{y} \big) \, 
\frac{f^\pm_n\big(\eta; \sqrt{\tilde{y}}/\eta\, \big)}{\sqrt[4]{\tilde{y}}}
\,=\, 
\gamma_n^\pm\;
\frac{f^\pm_n\big(\eta; \sqrt{\tilde{x}}/\eta\, \big)}{\sqrt[4]{\tilde{x}}}\,.
\ee
Comparing this expression with (\ref{k1}), one realises that the spectral problem associated to 
$K_{\textrm{\tiny B}} \big(\! \mp \!\tfrac{1}{2}\, ; x , y \big)$ in the interval $[0, \eta^2] \in \RR^+$
and the one associated to $K_\pm (\eta; x,y) $ in the interval $[0, 1] \in \RR^+$ discussed in Sec.\,\ref{sec_entropies}
have the same spectrum; hence they share the same tau function.

The Fredholm determinant $\tau_{\textrm{\tiny B}}(t) \equiv  \textrm{det}( I - z^{-1} K_{\textrm{\tiny B}})$
associated to the integral operator $K_{\textrm{\tiny B}}$ acting on the interval $[0,t]$,
whose kernel is (\ref{bessel-K-TW}),
can be studied by introducing the auxiliary function 
\be
\label{sigma-bessel-def}
\sigma_{\textrm{\tiny B}}(t) 
\equiv \, -\ t\, \partial_t \log\! \big[\tau_{\textrm{\tiny B}}(t)\big]\,.
\ee
This function 
satisfies the following Painlev\'e III equation
\cite{Tracy:1993xj, Jimbo-82, forrester-book}
\be
\label{P3-TW}
\big(\,t\,\sigma_{\textrm{\tiny B}}'' \,\big)^2 
+ \big(4\, \sigma_{\textrm{\tiny B}}' - 1\big)\, \big(\sigma_{\textrm{\tiny B}} -t\, \sigma_{\textrm{\tiny B}}' \big)\, \sigma_{\textrm{\tiny B}}' 
- a^2 \big(\sigma_{\textrm{\tiny B}}' \big)^2
\,=\,0
\ee
with the boundary condition 
\be
\label{P3-TW-boundary-cond}
\sigma_{\textrm{\tiny B}}(t) \,=\, \frac{t^{1+a}}{z\,2^{2(1+a)} \, \Gamma(1+a)\, \Gamma(2+a)} + \dots
\;\;\;\;\qquad\;\;\;\;
t \to 0^+
\ee
where the dots correspond to subleading terms. 
Notice that, while the differential equation (\ref{P3-TW}) 
is not affected by the sign of $a$,
its boundary condition (\ref{P3-TW-boundary-cond}) depends on it.

Combining the observations collected above, we have that
\be
\label{tau-pm-from-tau-B}
\tau_\pm(\eta) \,=\, \tau_{\textrm{\tiny B}} \big(\eta^2\big) \big|_{a \,=\,\mp 1/2}\,.
\ee

We remark that two different versions of the 
Painlevé $\textrm{III}_1$ differential equation
have been introduced in the literature, which are usually denoted by 
Painlevé $\textrm{III}_1$ and Painlevé $\textrm{III}'_1$
\cite{Okamoto-P3}.
In the  $\sigma$-form of Jimbo, Miwa and Okamoto,
the Painlevé $\textrm{III}_1$ reads
\be
\label{P-III-sigma}
\big( t\, \sigma''_{\textrm{\tiny III}} \big)^2
+\big(4\, \sigma'_{\textrm{\tiny III}} - 1\big)\, \big(\sigma_{\textrm{\tiny III}}- t \, \sigma'_{\textrm{\tiny III}}\big)\,\sigma'_{\textrm{\tiny III}} 
-4\, \theta_\ast \theta_\star \big( \sigma'_{\textrm{\tiny III}}\big)^2
-\frac{(\theta_\ast - \theta_\star )^2}{4}
\,=\,0
\ee
while the Painlevé $\textrm{III}'_1$ is
\be
\label{P-III-prime-sigma}
\big( t\, \sigma''_{\textrm{\tiny III}'}\big)^2
- \big[4 \big( \sigma'_{\textrm{\tiny III}'}\big)^2 - 1\big]\, \big(  \sigma_{\textrm{\tiny III}'} - t \, \sigma'_{\textrm{\tiny III}'}  \big)
+ 4\, \theta_\ast \theta_\star \, \sigma'_{\textrm{\tiny III}'}
- \big(\theta_\ast^2 + \theta_\star^2 \big)
\,=\,0
\ee
where the notation of \cite{Gamayun:2013auu} has been adopted. 
These differential equations are invariant 
under $(\theta_{\ast},\theta_{\star}) \to (-\theta_{\ast},-\theta_{\star}) $.
The solutions of (\ref{P-III-sigma}) and (\ref{P-III-prime-sigma})  
are related as follows
(see e.g. remark\;2 in \cite{Forrester-Dai})
\be
\label{3_3p}
\sigma_{\textrm{\tiny III}}(t)
=
-\,\sigma_{\textrm{\tiny III}'}(t/4)+\frac{t}{8}+\theta_{\ast} \theta_{\star}\,.
\ee
The tau functions associated to the solutions of 
(\ref{P-III-sigma}) and (\ref{P-III-prime-sigma}) are defined respectively by 
\be
\sigma_{\textrm{\tiny III}}(t) \equiv -\,t\, \partial_t \log\! \big[  \tau_{\textrm{\tiny III}}(t) \big]
\;\;\;\; \qquad \;\;\;\;
\sigma_{\textrm{\tiny III}'}(t) 
\equiv \,t\, \partial_t \log\! \big[  \tau_{\textrm{\tiny III}'}(t) \big]\,.
\label{tau_p3}
\ee
From these definitions and the relation (\ref{3_3p}), one finds that
\be
\label{tau_3_3p}
\tau_{\textrm{\tiny III}}(t) 
\,\propto\,
\frac{\tau_{\textrm{\tiny III}'}(t/4) }{\mathrm{e}^{t/8} \, t^{\theta_{\ast}\theta_{\star}} }\,.
\ee

As for the Bessel kernel (\ref{bessel-K-TW}),
the differential equation (\ref{P3-TW}) satisfied by 
its auxiliary function (\ref{sigma-bessel-def})
corresponds to (\ref{P-III-sigma}) in the special case given by 
$\theta_{\ast}=\theta_{\star}= \pm a/2$.
In our analysis we set $\theta_{\ast}=\theta_{\star}= - a/2$.

The relations  \eqref{tau-pm-from-tau-B} and \eqref{tau_3_3p}
provide \eqref{fd_t3p} up to a proportionality constant
whose derivation is reported in the Appendix\;\ref{app_tau_small_eta}
(see \eqref{C0-pm-def}).

\section{On the small $\eta$ expansion} 
\label{app_small_eta}

In this appendix we describe the technical details underlying 
some results concerning the expansions of the entanglement entropies
as $\eta \to 0$ reported in Sec.\,\ref{sec_small_eta}.

\subsection{PSWF approach}
\label{app_small_eta_pswf}

In the following we discuss the derivation of 
the expansions of $S_{A, \pm}^{(\alpha)}$ given in 
(\ref{ren-small-eta-plus}) and (\ref{ren-small-eta-min}).

For finite $\alpha\neq 1$, the expression (\ref{def-s-alpha}) can be written as
$s_{\alpha}(x) =  s_{\alpha,1}(x)+s_{\alpha,2}(x)$, with
\bea
s_{\alpha,1}(x) 
& \equiv & \frac{\alpha}{1-\alpha}\log(1-x)=\frac{\alpha}{\alpha-1}\left[x+\frac{x^{2}}{2}+\frac{x^{3}}{3}+\frac{x^{4}}{4}+O(x^{5})\right]
\\
\rule{0pt}{.8cm}
s_{\alpha,2}(x) 
& \equiv & 
\frac{1}{1-\alpha}\, \log\big[1+\chi_{\alpha}(x)\big]
=
\frac{1}{1-\alpha}\sum_{j=1}^{\infty}\frac{(-1)^{j+1}}{j}
\; \chi_{j\alpha}(x)
\eea
where we have introduced $\chi_{\beta}(x)\equiv[x/(1-x)]^{\beta}$.

The contribution to (\ref{ren-small-eta-plus}) and (\ref{ren-small-eta-min})
coming from $s_{\alpha,1}(x)$ is given by
\bea
s_{\alpha,1}(\gamma_{0}) 
&=& 
\frac{\alpha}{\alpha-1}
\left[\,
\tilde{g}_{0}\eta
+\frac{\tilde{g}_{0}^{2}}{2}\,\eta^{2}
+\left(\frac{\tilde{g}_{0}^{2}}{3}+\tilde{a}_{0}\!\right)\tilde{g}_{0} \eta^{3}
+\left(\frac{\tilde{g}_{0}^{2}}{4}+\tilde{a}_{0}\!\right) \tilde{g}_{0}^{2}\eta^{4}
\,\right]+O(\eta^{5})
\\
\rule{0pt}{.7cm}
s_{\alpha,1}(\gamma_{1}) 
&=&
 \frac{\alpha}{\alpha-1}
 \left[\,\tilde{g}_{1}\eta^{3}+\tilde{g}_{1}\tilde{a}_{1}\eta^{5}+\frac{\tilde{g}_{1}^{2}}{2}\,\eta^{6} \,\right]
 +O(\eta^{7})
\eea
in terms of (\ref{eigen-small-3});
while $s_{\alpha,1}(\gamma_{2n}) = O(\eta^{5})$ and $s_{\alpha,1}(\gamma_{2n+1}) = O(\eta^{7})$
when $n\geqslant1$.
As for the contribution  to (\ref{ren-small-eta-plus}) and (\ref{ren-small-eta-min}) originating  from $s_{\alpha,2}(x)$, 
all the eigenvalues must be considered;
indeed for generic $\alpha$ we have 
\bea
\chi_{\beta}(\gamma_{0}) 
& = & 
\tilde{g}_{0}^{\beta}\eta^{\beta}
\bigg\{
1+\beta\tilde{g}_{0}\eta+\left(\beta\tilde{a}_{0}+\frac{\beta(\beta+1)}{2}\tilde{g}_{0}^{2}\right)\eta^{2}
 \\
 &  & \hspace{1.4cm}
 +\,\frac{\beta(\beta+1)}{6}\tilde{g}_{0}\left(6\tilde{a}_{0}+(\beta+2)\tilde{g}_{0}^{2}\right)\eta^{3}
 \bigg\}
 +O\big(\eta^{4+\beta}\big)
 \nonumber
 \\
 \rule{0pt}{.7cm}
\chi_{\beta}(\gamma_{2n}) 
& = & 
\tilde{g}_{2n}^{\beta}\eta^{\beta(4n+1)}
\Big\{
1+\beta\tilde{a}_{2n}\eta^{2}
\Big\} 
+O\big(\eta^{4+\beta(4n+1)}\big)
\hspace{2cm} n\geqslant1
\eea
and
\bea
\chi_{\beta}(\gamma_{1}) 
& = & 
\tilde{g}_{1}^{\beta}\eta^{3\beta}
\Big\{
1+\beta\tilde{a}_{1}\eta^{2}+\beta\tilde{g}_{1}\eta^{3}
\Big\}
+O\big(\eta^{4+3\beta}\big)
\\
 \rule{0pt}{.5cm}
\chi_{\beta}(\gamma_{2n+1}) 
& = & 
\tilde{g}_{2n+1}^{\beta}\eta^{\beta(4n+3)}
\Big\{
1+\beta\tilde{a}_{2n+1}\eta^{2}
\Big\}
+O\big(\eta^{4+\beta(4n+3)}\big)
\hspace{1.3cm} n\geqslant1\,. \;\;\;\;
\eea
%

Combining the above results, we arrive to
\bea
\label{S-plus-app-alpha-pswf}
S_{A,+}^{(\alpha)} & = & 
\frac{\alpha}{\alpha-1}\;
\Bigg\{ 
\left[\tilde{g}_{0}\eta+\frac{\tilde{g}_{0}^{2}}{2}\eta^{2}+\left(\frac{\tilde{g}_{0}^{3}}{3}+\tilde{g}_{0}\tilde{a}_{0}\right)\eta^{3}+\left(\frac{\tilde{g}_{0}^{4}}{4}+\tilde{g}_{0}^{2}\tilde{a}_{0}\right)\eta^{4}\right]
\nonumber \\
 &  & \hspace{-1cm}
 +\sum_{j=1}^{\infty}(-1)^{j}
 \left[\,\frac{1}{j\alpha}+\tilde{g}_{0}\eta+\left(\tilde{a}_{0}+(j\alpha+1)\frac{\tilde{g}_{0}^{2}}{2}\right)\eta^{2}+(j\alpha+1)\tilde{g}_{0}\left(\tilde{a}_{0}+(j\alpha+2)\frac{\tilde{g}_{0}^{2}}{6}\right)\eta^{3}\right]
  \tilde{g}_{0}^{j\alpha}\eta^{j\alpha}
 \nonumber \\
 &  & \hspace{-1cm}
 +\sum_{n=1}^{\infty}\sum_{j=1}^{\infty}
 (-1)^{j}
 \left(\,\frac{1}{j\alpha}+\tilde{a}_{2n}\eta^{2}\right)
\tilde{g}_{2n}^{j\alpha}\eta^{j\alpha(4n+1)} 
\Bigg\} 
 +O(\eta^{\min\{5,4+\alpha\}})
\eea
and
\bea
\label{S-min-app-alpha-pswf}
S_{A,-}^{(\alpha)} 
& = & 
\frac{\alpha}{\alpha-1}\;
\Bigg\{ 
\left(\tilde{g}_{1}\eta^{3}+\tilde{g}_{1}\tilde{a}_{1}\eta^{5}+\frac{\tilde{g}_{1}^{2}}{2}\eta^{6}\right)
+
\sum_{j=1}^{\infty}(-1)^{j}\tilde{g}_{1}^{j\alpha}\eta^{3j\alpha}\left(\frac{1}{j\alpha}+\tilde{a}_{1}\eta^{2}+\tilde{g}_{1}\eta^{3}\right)
\hspace{1cm}
 \\
 &  & \hspace{1.6cm}
 +\sum_{n=1}^{\infty}\sum_{j=1}^{\infty}(-1)^{j}\tilde{g}_{2n+1}^{j\alpha}\eta^{j\alpha(4n+3)}\left(\frac{1}{j\alpha}+\tilde{a}_{2n+1}\eta^{2}\right)\Bigg\}
 +O\big(\eta^{\min\{7,4+3\alpha\}}\big)\,.
 \nonumber
\eea
We remark that, despite (\ref{S-plus-app-alpha-pswf}) and (\ref{S-min-app-alpha-pswf})
contain an infinite number of terms,  
only a finite number of them are $O(\eta^{\min\{5,4+\alpha\}})$ or $O(\eta^{\min\{7,4+3\alpha\}})$
once $\alpha$ has been fixed.
In particular, when $\alpha>1$, 
from \eqref{S-plus-app-alpha-pswf} and \eqref{S-min-app-alpha-pswf} 
we obtain the expansions \eqref{ren-small-eta-plus} and \eqref{ren-small-eta-min} 
by discarding the terms of order $O(\eta^5)$ and $O(\eta^7)$ respectively.

\subsection{Tau function approach}
\label{app_tau_small_eta}

In this Appendix we derive (\ref{tau-GIL-sec}) as special cases of
the expansion of the Painlev\'e $\textrm{III}_1'$ tau function
found  in \cite{Gamayun:2013auu}.

The Conjecture\;4 of \cite{Gamayun:2013auu}, 
combined with (\ref{tau-pm-from-tau-B}) and (\ref{tau_3_3p}),
provides the following ansatz
\be
\label{tau-pm-small-eta-app}
\tau_{\pm}
\,=\,
\frac{\mathcal{C}_{0,\pm}}{\textrm{e}^{\eta^2/8} \, \eta^{1/8}} 
\sum_{n \in \mathbb{Z}}
C_{\textrm{\tiny III}'}\big( \pm\! \tfrac{1}{4} , \pm \tfrac{1}{4} , \hat{\sigma}_\pm +n \big) 
\,s_{\textrm{\tiny III}', \pm}^n\,
\bigg(\frac{\eta}{2}\bigg)^{\!2(\hat{\sigma}_\pm +n)^2}
\!\mathcal{B}_{\textrm{\tiny III}'}\big( \pm\! \tfrac{1}{4} , \pm \tfrac{1}{4} , \hat{\sigma}_\pm +n; \,\eta^2/4 \big) \;
\ee
where the explicit expressions of
$C_{\textrm{\tiny III}'}(\theta_\ast,\theta_{\star},\hat{\sigma})$ and $\mathcal{B}_{\textrm{\tiny III}'}(\theta_\ast,\theta_{\star},\hat{\sigma};t)$
are reported in \cite{Gamayun:2013auu}.\footnote{The parameter $\hat{\sigma}$ corresponds to the one denoted by $\sigma$ in \cite{Gamayun:2013auu}.}
In our cases we need the special case given by 
$C_{\textrm{\tiny III}'}(\theta , \theta , \hat{\sigma})$
(see (\ref{C3-theta-theta-def})).

The parameters $\mathcal{C}_{0,\pm}$, $\hat{\sigma}_\pm$ and $s_{\textrm{\tiny III}', \pm}$ in (\ref{tau-pm-small-eta-app})
are fixed by imposing the proper behaviour as $\eta \to 0$.
This behaviour is obtained by employing the small $\eta$ expansion of the eigenvalues
(see (\ref{eigen-small-1}) and (\ref{eigen-small-3}))
into the definition (\ref{tau-function-pm}) of the tau functions $\tau_\pm$,
finding 
\bea
\label{tau-N-exp}
\tau_{+} & = & 
1-\tilde{g}_{0}\,\frac{\eta}{z}-\tilde{g}_{0}\,\tilde{a}_{0}\, \frac{\eta^{3}}{z} + O\big(\eta^{5}\big)
\,=\,
1-\frac{2}{\pi}\;\frac{\eta}{z}+\frac{2}{9\pi}\;\frac{\eta^{3}}{z}+O\big(\eta^{5}\big)
\\
\rule{0pt}{.7cm}
\label{tau-D-exp}
\tau_{-} 
& = & 
1-\tilde{g}_{1}\,\frac{\eta^{3}}{z}-\tilde{g}_{1}\,\tilde{a}_{1}\, \frac{\eta^{5}}{z}+O\big(\eta^{7}\big)
\,=\,
1-\frac{2}{9\pi}\;\frac{\eta^{3}}{z} + \frac{2}{75\pi}\;\frac{\eta^{5}}{z}+O\big(\eta^{7}\big)
\eea
which agree with the expansions obtained from Eq.\,(1.22) of  \cite{Tracy:1993xj}.

In order to get consistency between  (\ref{tau-pm-small-eta-app}) and 
the expansions in (\ref{tau-N-exp}) and (\ref{tau-D-exp}),
first we observe that both $\tau_{\pm}$ and 
$\mathcal{B}_{\textrm{\tiny III}'} ( \pm \tfrac{1}{4} , \pm \tfrac{1}{4} , \hat{\sigma}_\pm +n; \eta^2/4)$ 
do not diverge as $\eta \to 0$. 
This tells us that the factor $1/\eta^{1/8}$ multiplying the series in the r.h.s. of (\ref{tau-pm-small-eta-app})
must simplify with the factor 
$(\eta/2)^{2\hat{\sigma}_\pm^2 + 4 \hat{\sigma}_\pm n +2n^2}$ 
in the summand of the series in (\ref{tau-pm-small-eta-app}); hence $\hat{\sigma}_\pm = 1/4$.
For this specific value of $\hat{\sigma}_\pm$, we have that
\be
C_{\textrm{\tiny III}'}\big(\! +\! \tfrac{1}{4} , + \tfrac{1}{4} , \tfrac{1}{4} +n\,\big) \big|_{n>0} =\, 0
\;\;\;\qquad\;\;\;
C_{\textrm{\tiny III}'}\big(\! -\! \tfrac{1}{4} , - \tfrac{1}{4} , \tfrac{1}{4} +n\,\big) \big|_{n<0} =\, 0
\ee
which simplify (\ref{tau-pm-small-eta-app}) in a significant way. 
For $n=0$ we have
$C_{\textrm{\tiny III}'} \big(\! \pm\! \tfrac{1}{4} , \pm \tfrac{1}{4} , \tfrac{1}{4}\big) 
= \pi^{\pm 1/2}$.

Then, by considering the terms corresponding to $n \in \{ -1,0,1\}$ 
in \eqref{tau-pm-small-eta-app},
which involve also 
$\mathcal{B}_{\textrm{\tiny III}'}( \pm \tfrac{1}{4} , \pm \tfrac{1}{4} , \tfrac{1}{2} +n; \,\eta^2/4)$ for these values of $n$,
agreement with (\ref{tau-N-exp}) and (\ref{tau-D-exp})  is obtained when
\be
\label{C0-pm-def}
\mathcal{C}_{0,\pm} = \frac{2^{1/8}}{\pi^{\pm 1/2}}
\;\;\;\;\qquad\;\;\;\;
s_{\textrm{\tiny III}', \pm} = z^{\pm 1}\,.
\ee

Combining the above results, we find that the expansion \eqref{tau-pm-small-eta-app} simplifies to \eqref{tau-GIL-sec}, with
\be
\label{B-pm-def}
\mathcal{B}_{\pm}(n;t)
=\!
\sum_{\lambda,\mu\in\mathbb{Y}}
\mathcal{B}_{\lambda,\mu}^{\,\textrm{\tiny III}'}
\big(\! \pm\! \tfrac{1}{4} , \pm \tfrac{1}{4} , \tfrac{1}{4} \mp n \big)\;
t^{|\lambda |+ |\mu |}
\ee
where $\mathbb{Y}$ is the set of all Young diagrams
(see Sec.\,3.1 of \cite{Gamayun:2013auu} for a more detailed explanation)
and the coefficients 
$\mathcal{B}_{\lambda,\mu}^{\,\textrm{\tiny III}'}(\theta_{*},\theta_{\star},\sigma)$
are given by Eq.\,(4.19) of \cite{Gamayun:2013auu}. 
From (\ref{B-pm-def}), we obtain 
the following expansions of $\mathcal{B}_{+}(n;t)$ for some values of $n$
\bea
\label{B-n-plus-small-t}
\mathcal{B}_{+}(0;t) & = & 
1 + t + \frac{t^2}{2} + \frac{t^3}{6} + \frac{t^4}{24} + \frac{t^5}{120} 
+ \frac{t^6}{720}  + \frac{t^7}{5040}  + \frac{t^8}{40320} + \frac{t^9}{362880}    
+\dots
\\
\rule{0pt}{.7cm}
\mathcal{B}_{+}(1;t) & = & 
1 + \frac{5}{9}\,t + \frac{121}{450}\,t^2 + \frac{733}{7350}\,t^3  + \frac{187483}{7144200}\,t^4 
 + \frac{4481899}{864448200}\,t^5  + \frac{241147213}{292183491600}\,t^6 
+\dots
\nonumber
\\
\rule{0pt}{.7cm}
\mathcal{B}_{+}(2;t) & = & 
1+ \frac{25}{49}\, t + \frac{1003}{7350}\, t^2 + \frac{67979}{2668050}\, t^3 
+ \frac{338600363}{88376488200}\, t^4 + \frac{74803}{146076840}\, t^5 
+\dots
\nonumber
\\
\rule{0pt}{.7cm}
\mathcal{B}_{+}(3;t) 
& = & 
1 + \frac{61}{121}\, t + \frac{5289}{40898}\, t^2 + \frac{1216997}{54108054}\, t^3 
+ \frac{22574889473}{7568418161304}\, t^4 
+\dots
\nonumber
\\
\rule{0pt}{.7cm}
\mathcal{B}_{+}(4;t) 
& = & 
1 + \frac{113}{225}\, t + \frac{2797321}{21978450}\, t^2 + \frac{6943216561}{320013558150}\, t^3 
+ \frac{34818356971}{12442127140872}\, t^4
+\dots
\nonumber
\eea
where the dots denote subleading terms;
while for the expansions of $\mathcal{B}_{-}(n;t)$ 
corresponding to the same values of  $n$ we find 
\bea
\label{B-n-minus-small-t}
\mathcal{B}_{-}(0;t) 
& = & 1 + t + \frac{t^2}{2} + \frac{t^3}{6} + \frac{t^4}{24} + \frac{t^5}{120} 
+ \frac{t^6}{720}  + \frac{t^7}{5040}  + \frac{t^8}{40320} + \frac{t^9}{362880}    
+\dots
\\
\rule{0pt}{.7cm}
\mathcal{B}_{-}(1;t) 
& = & 1+ \frac{13}{25}\, t + \frac{369}{2450}\, t^2 + \frac{6887}{198450}\, t^3 
+ \frac{680459}{96049800}\, t^4 + \frac{20582899}{16232416200}\, t^5 
+ \dots
\nonumber
\\
\rule{0pt}{.7cm}
\mathcal{B}_{-}(2;t) 
& = & 
1+ \frac{41}{81}\, t + \frac{126409}{960498}\, t^2 + \frac{1266497}{54108054}\, t^3 
+ \frac{4321645}{1348539192}\, t^4 
+ \frac{45832437061}{126661543608600}\, t^5
+ \dots
\nonumber
\\
\rule{0pt}{.7cm}
\mathcal{B}_{-}(3;t) 
& = & 
1+ \frac{85}{169}\, t + \frac{130929}{1022450}\, t^2 
+ \frac{526486511}{23934532050}\, t^3 + \frac{16747684673123}{5840887463353800}\, t^4 
+ \dots
\nonumber
\\
\rule{0pt}{.7cm}
\mathcal{B}_{-}(4;t) 
& = & 
1+ \frac{145}{289}\, t + \frac{38897}{306850}\, t^2 
+ \frac{2786241737}{129592267350}\, t^3 + \frac{8223548620739}{2986329051875400}\, t^4 
+ \dots
\nonumber
\eea
and $\mathcal{B}_{\pm}(n;t) = 1+ \dots $ when $n\geqslant5$.
We have access to more terms in these expansions, 
but they have not been written here to avoid lengthy expressions. 
In the case of $n=0$, 
these expansions lead us to conjecture that 
$\mathcal{B}_{\pm}(0;t)=\mathrm{e}^{t}$.

The best results reported in Sec.\,\ref{sec-small-eta-tau} 
for the small $\eta$ expansions of the entanglement entropies
involve a polynomial of fourth degree in $z$.
This is achieved by employing 
the expansion of $\mathcal{B}_{+}(n;t)$ and $\mathcal{B}_{-}(n;t)$
up to $O(t^p)$, where the pairs $(n,p)$ are given respectively by 
\bea
(n,p) &\in& \big\{(0,23), (1,22), (2,20), (3,15), (4,9), (5,0)\big\}
\\
(n,p) &\in& \big\{(0,28), (1,26), (2,23), (3,17), (4,10), (5,0)\big\}
\eea
and whose first terms are reported in (\ref{B-n-plus-small-t}) 
and (\ref{B-n-minus-small-t}).

\section{On the large $\eta$ expansion} 
\label{app_large_eta}

In this Appendix we discuss the derivation of some results 
o the expansion of the entanglement entropies
as $\eta \to \infty$ employed in Sec.\,\ref{sec_large_eta}.

\subsection{Tau functions}
\label{app_large_eta_tau}

Consider the expansion of the Painlevé $\textrm{III}'_1$ tau function 
given in Eq.\,(A.30) of \cite{Bonelli:2016qwg}.
In the cases we are exploring 
$\theta_{*}=\theta_{\star}=\pm1/4$ (see the Appendix\;\ref{app_bessel});
hence this expansion becomes
\be
\label{tau-3prime-Lys-Bon}
\tau_{\textrm{\tiny III}'}(t) 
\,=\,
s^{-\frac{1}{24}} \sum_{n\,\in\,\mathbb{Z}}\mathsf{P}^{n}
\,\mathscr{C}(\nu+n,s)\sum_{p\,=\,0}^{\infty}\frac{\mathcal{D}_{p}(\nu+n)}{s^p}
\;\;\;\qquad\;\;\;
t=\frac{s^{2}}{16}
\ee
where
\be
\mathscr{C}(\nu,s) 
\equiv\,
\mathrm{e}^{\frac{s^{2}}{32}+ \textrm{i} \nu s+  \textrm{i} \frac{\pi}{2}\nu^{2}}
\frac{G(1+\nu)^2}{(2\pi)^{\nu}\, 2^{\nu^{2}}\, s^{\nu^{2}-1/6}}
\ee
and
$\mathcal{D}_{p}(\nu)$ for $p \in \{0,1,2\} $ have been reported in (\ref{D_hl}).
The function (\ref{tau-3prime-Lys-Bon}) is parameterised by 
$\mathsf{P}$ and $\nu$.

Combining the expansion (\ref{tau-3prime-Lys-Bon}) with (\ref{tau-pm-from-tau-B}) and (\ref{tau_3_3p}), 
we find that the tau functions occurring in (\ref{entropy-tau-pm}) can be written as
\bea
\label{tau3_pm-0}
\tau_{\pm} 
&=&
\mathsf{N}_{\pm}\, \frac{\tau_{\textrm{\tiny III}'}(\eta^2/4) }{\mathrm{e}^{\eta^2/8} \, \eta^{1/8} }
\\
\label{tau3_pm}
&=&
 2^{\frac{1}{8}} \,\mathsf{N}_{\pm}
 \sum_{n\,\in\,\mathbb{Z}}
 \mathsf{P}_{\pm}^{n}\,
\mathrm{e}^{ \textrm{i} \frac{\pi}{2}(\nu_{\pm}+n)^{2}+  \textrm{i} 2\eta(\nu_{\pm}+n)}\,
\frac{G(1+\nu_{\pm}+n)^{2}}{ (2\pi)^{(\nu_{\pm}+n)} \, (4\eta)^{(\nu_{\pm}+n)^{2}}}
\sum_{p\,=\,0}^{\infty}\frac{\mathcal{D}_{p}(\nu_{\pm}+n)}{\left(2\eta\right)^{p}}
\eea
where the parameters $\mathsf{N}_{\pm}$, $\nu_{\pm}$ and $\mathsf{P}_{\pm}$
(which depend on $z$ but are independent of $\eta$) 
occur in the first terms of the large $\eta$ expansion of $\tau_{\pm} $.

The leading terms in the expansion of $\log{\tau_\pm}$ as $\eta \to \infty$
have been reported in Eq.\,(1.35) of  \cite{Bothner_2019} 
and, in our notation\footnote{Comparing with the notation in Eq.\,(1.35) of  \cite{Bothner_2019}, we have $t_{\textrm{\tiny there}} = \eta^2$, 
$v_{\textrm{\tiny there}} = - 2\pi \textrm{i} \,\nu$
and $\alpha_{\textrm{\tiny there}} = \mp 1/2$.},
they read (see also (\ref{log-tau-infty}))
\be
\label{logtBIP}
\log (\tau_\pm) 
\,=\,
\ri \,2\nu\,\eta
-\nu^{2}\log(4\eta)
\pm \ri \,\frac{\pi}{2}\,\nu
+\log\! \big[G(1+\nu)\,G(1-\nu)\big]
+O(1/\eta)
\ee
where $\nu$ is given by (\ref{nu-z-def}). 
Comparing (\ref{logtBIP}) with  the  term having $n=p=0$ in \eqref{tau3_pm},
we find 
\be
\label{Nbip}
\mathsf{N}_{\pm} 
= 
\frac{(2\pi)^{\nu}}{2^{1/8}} \;
\frac{G(1-\nu)}{G(1+\nu)} \;
\e^{\ri \frac{\pi}{2} (\pm \nu-\nu^{2})} 
\;\;\;\qquad\;\;\;
\nu_\pm = \nu 
\ee
which are independent of $\eta$, as expected. 
The constant $\mathsf{P}_\pm$ 
cannot be obtained from the terms reported in  \eqref{logtBIP}
because it  occurs in the subleading contributions of \eqref{tau3_pm} 
having $n \neq 0$.

We find the parameter $\mathsf{P}_{-}$
in terms of $z$ by imposing that $\tau_-$ in \eqref{tau3_pm} agrees with the proper limit of the corresponding lattice result obtained 
for the XX model in the semi-infinite chain with open boundary conditions
\cite{Basor-01,Basor2002,Basor2008,Deift-11,Fagotti:2010cc}.
In particular, 
from Eq.\,(39) of \cite{Fagotti:2010cc} and by employing the notation adopted there,
we have the following lattice result\footnote{The expression \eqref{D_from_FC_proposal} 
has been obtained by simply removing a factor
$\exp(-2\ri \beta k_{\textrm{\tiny F,there}} \,\ell)$
in Eq.\,(39) of \cite{Fagotti:2010cc}.
The proper limits of (\ref{D_from_FC_proposal}) 
agree with Eq.\,(36) of \cite{Fagotti:2010cc} and with the expansion (\ref{logtBIP}).}
\bea
\label{D_from_FC_proposal}
\frac{D_{\ell} ( \lambda ) }{ (\lambda +1)^{\ell} }
& = & 
\mathrm{\e}^{-2\ri \beta k_{\textrm{\tiny F,there}} \,\ell}
\sum_{n\,\in\,\mathbb{Z}}
\e^{\ri \frac{\pi}{2}(\beta+n)}
\big[\,
4(\ell+1/2)\,
\big|\sin(k_{\textrm{\tiny F,there}})\big|
\,\big]^{-(n+\beta)^{2}}
\\
 &  & \hspace{3cm}
 \times\,
  \e^{-\ri \beta k_{\textrm{\tiny F,there}}} \,
 \e^{-2\ri k_{\textrm{\tiny F,there}}n(\ell+1/2)} \,
 G(1+n+\beta) \, G(1-n-\beta)
 \hspace{.2cm}
 \nn
\eea
where $\ell$ is the number of consecutive sites of the block located 
at the beginning of the semi-infinite chain
and $\beta \equiv \frac{1}{2\pi \ri }\log\! \big(\frac{\lambda+1}{\lambda-1}\big)$.
The parameter $\lambda$ in (\ref{D_from_FC_proposal}) and $z$ 
are related by $\lambda = 2z-1$;
hence $\beta  = -\nu$, with $\nu$ being defined in \eqref{nu-z-def}.

In the double scaling limit given by 
$\ell\rightarrow+\infty$ and $k_{\textrm{\tiny F,there}}\rightarrow0^{+}$ 
with $\ell \,k_{\textrm{\tiny F,there}}=\eta$ kept fixed
(which implies $L_{k, \textrm{\tiny there}}=4(\ell+1/2)\left|\sin(k_{\textrm{\tiny F,there}})\right|\rightarrow4\eta$),
the expression (\ref{D_from_FC_proposal}) becomes
\be
\label{fd_FC_dsl}
\frac{D_{\ell} ( \lambda ) }{ (\lambda +1)^{\ell} }
\;\;\longrightarrow\;
\sum_{n\,\in\,\ZZ}
\e^{- \ri \frac{\pi}{2}(\nu+n)}\,
\e^{\textrm{i} 2\eta(\nu+n)}\;
\frac{G(1+\nu+n)\,G(1-\nu-n)}{(4\eta)^{(\nu+n)^{2}}}\,.
\ee
By imposing that this expansion coincides with 
the series in $n$ given by 
the truncation of $\tau_-$ in \eqref{tau3_pm} 
obtained by considering only the term $p=0$ in 
the second series,
we find\footnote{In this calculation we have used the identities
\be
\frac{G(1-\nu)\,G(1+\nu+n)}{G(\nu+1)} 
=  \textrm{i}^{(n-1)n}
\big[ \Gamma(\nu+1)\Gamma(-\nu)\big]^{n}G(1-n-\nu)
\;\;\qquad\;\;
\Gamma(\nu+1)\,\Gamma(-\nu) = -\frac{\pi}{\sin(\pi\nu)}
\ee
and that $1/\sin(\pi\nu) = - 2\textrm{i}\,z\,\textrm{e}^{\textrm{i}\pi\nu}$ (from (\ref{nu-z-def})).}
\be
\label{P-minus-explicit}
\mathsf{P}_{-}
\,=\,
\textrm{i}\,\frac{\e^{-\textrm{i}2\pi\nu}}{z}\,.
\ee

Finally,  plugging the expressions for 
$\nu_-$, $\mathsf{N}_{-}$ and $\mathsf{P}_{-}$ 
(see \eqref{nu-z-def}, \eqref{Nbip} and \eqref{P-minus-explicit})
into \eqref{tau3_pm},
the expression for $\tau_-$ reported in \eqref{t_pm_final} is obtained.

The ansatz for $\tau_{+}$ can be found by using the relation (\ref{sine-tau-function-factorisation}),
with \cite{Gamayun:2013auu}
\be
\label{tau_sine}
\tau_{\textrm{\tiny sine}}  
=
\sum_{n\,\in\,\mathbb{Z}}
\e^{ \textrm{i} 4\eta(\nu+n)}
\left[\,
\frac{G(1+\nu+n)\, G(1-\nu-n)}{(4\eta)^{(\nu+n)^{2}}}
\,\right]^2
\,\sum_{p\,=\,0}^{\infty}\frac{\mathcal{D}^{\textrm{\tiny sine}}_p(\nu+n)}{ (4\eta\, \textrm{i} )^{p}}
\ee
where $\nu$ and $\mathcal{D}^{\textrm{\tiny sine}}_k(\nu)$
are given respectively in (\ref{nu-z-def})
and in Eqs.\,(8.3)-(8.5) of \cite{Mintchev:2022xqh}\footnote{The expression (\ref{tau_sine}) coincides with Eq.\,(8.1) of \cite{Mintchev:2022xqh} after some manipulations.}.
In particular, by using (\ref{t_pm_final}),  \eqref{tau_sine} and \eqref{tau3_pm} 
for $\tau_{-}$, $\tau_{\textrm{\tiny sine}}$ and $\tau_{+}$ respectively into (\ref{sine-tau-function-factorisation}),
we obtain a relation that allows to determines the parameter
$\mathsf{P}_{+}$ as function of $z$.
Denoting by $n_+$, $n_-$ and $n_{\textrm{\tiny sine}}$
and by $p_+$, $p_-$ and $p_{\textrm{\tiny sine}}$ the labels $n$ and $p$ respectively in the corresponding expressions 
\eqref{tau3_pm}, \eqref{t_pm_final} and \eqref{tau_sine},
from the term having $n_{+}=n_{-}=n_{\textrm{\tiny sine}}\geqslant 1$ 
and $p_{+}=p_{-}=p_{\textrm{\tiny sine}}=0$ we arrive to
\be
\label{P-plus-explicit}
\mathsf{P}_{+}
\,=\,
-\,\textrm{i}\,\frac{\e^{-\textrm{i}2\pi\nu}}{z} \,.
\ee
Finally, the expression for $\tau_+$ in \eqref{t_pm_final}  is obtained 
by plugging \eqref{Nbip} and \eqref{P-plus-explicit} into \eqref{tau3_pm}.

Let us remark that it would be worth providing an alternative 
derivation of the parameters $\mathsf{N}_{\pm}$, $\nu_{\pm}$
and $\mathsf{P}_{\pm}$ in (\ref{tau3_pm}) 
through the connection formula for the Painlev\'e $\textrm{III}_1$, 
as done in \cite{Mintchev:2022xqh} for the interval on the line,
where the connection formula for the Painlev\'e $\textrm{V}$
given in \cite{Lisovyy:2018mnj} has been employed.

\subsection{A consistency check}
\label{app_cons_check}

In the final part of the Appendix\;\ref{app_large_eta_tau},
the ansatz for $\tau_+$ in (\ref{t_pm_final}) has been obtained 
by requiring the validity of \eqref{sine-tau-function-factorisation},
but only few terms of the resulting series have been employed
to fix the parameters occurring in (\ref{tau3_pm}) for $\tau_+$.
Hence, the relation \eqref{sine-tau-function-factorisation} 
can be used as consistency check of the expressions for 
$\tau_+$ and $\tau_{\textrm{\tiny sine}}  $ 
given in (\ref{t_pm_final}) and (\ref{tau_sine}) respectively, 
where free parameters do not occur. 
This analysis is performed by reorganising in the powers of $\eta$
the expansions involved in \eqref{sine-tau-function-factorisation}.

As for $\tau_{\textrm{\tiny sine}}$ (see \eqref{tau_sine}), 
let us first change $(n,k)$ into $(n,j)$, where $j=2n^{2}+k$;
hence  for each $n\in\mathbb{Z}$ we have $j\in\{2n^{2},2n^{2}+1,\cdots\}$.
This leads to write \eqref{tau_sine} as follows
\be
\tau_{\textrm{\tiny sine}}
\,= 
\sum_{n\,\in\,\mathbb{Z}}\;
\sum_{j=2n^{2}}^{\infty}\e^{\textrm{i} 4\eta(\nu+n)}\;
\frac{G(1+\nu+n)^{2}\,G(1-\nu-n)^{2}}{\textrm{i}^{j-2n^{2}} (4\eta)^{2\nu(\nu+2n)+j}}\;
\mathcal{D}^{\textrm{\tiny sine}}_{j-2n^{2}}(\nu+n)\,.
\ee
The condition $j\geqslant 2n^{2}$ for any $n\in\mathbb{Z}$ is equivalent to
$- \, n_{\textrm{\tiny s}}(j) \leqslant n \leqslant n_{\textrm{\tiny s}}(j)$
with $ j\in\mathbb{N}_{0}$, 
where we have introduced $n_{\textrm{\tiny s}}(j) \equiv \big\lfloor \sqrt{j/2} \big\rfloor$.
Thus, $\tau_{\textrm{\tiny sine}}$ can be written as 
\be
\label{t_sine_bis}
\tau_{\textrm{\tiny sine}}
\,= 
\sum_{j=0}^{\infty} \frac{1}{(4\eta)^{j}} \!
\sum_{n=-n_{\textrm{\tiny s}}(j)}^{n_{\textrm{\tiny s}}(j)}
\!\! \! \! 
(-1)^{n^{2}}
\e^{i4\eta(\nu+n)}\;
\frac{G(1+\nu+n)^{2}\,G(1-\nu-n)^{2}}{\textrm{i}^{j} (4\eta)^{2\nu(\nu+2n)}}\;
\mathcal{D}^{\textrm{\tiny sine}}_{j-2n^{2}}(\nu+n)\,.
\ee
Performing the same manipulations for $\tau_{\pm}$ in (\ref{t_pm_final}) as well,
one obtains the following expansions
\be
\label{t_pm_bis}
\tau_{\pm}\, 
=
\sum_{j=0}^{\infty} \frac{1}{(4\eta)^{j}} \!
\sum_{n=-n_{\textrm{\tiny h}}(j)}^{n_{\textrm{\tiny h}}(j)}
\!\! \! \! 
\e^{\pm \textrm{i} \frac{\pi}{2}(\nu+n)}\,
\e^{\textrm{i} 2\eta(n+\nu )}\,
\frac{G(1+\nu+n)\,G(1-\nu-n)}{(4\eta)^{\nu(\nu+2n)}} \;
2^{j-n^{2}} 
\mathcal{D}_{j-n^{2}}(\nu+n)
\ee
where $n_{\textrm{\tiny h}}(j)\equiv\left\lfloor \sqrt{j}\right\rfloor .$

By writing the expressions \eqref{t_sine_bis} and \eqref{t_pm_bis}
as $\tau_{\textrm{\tiny sine}} \equiv  \sum_{j=0}^{\infty}  \tau_{\textrm{\tiny sine},\, j} $ 
and $\tau_{\pm} \equiv \sum_{j=0}^{\infty} \tau_{\pm, \,j}$ respectively,
we find that \eqref{sine-tau-function-factorisation} is equivalent to
\be
\label{sine-tau-function-factorisation_j}
 \tau_{\textrm{\tiny sine},\, j} 
\, =
 \sum_{l=0}^{j}  \tau_{+,\, l}  \; \tau_{-,\, j - l} 
 \;\;\;\;\qquad\;\;\;
 \forall\, j \in \mathbb{N}_0\,.
\ee
We have checked the validity of this relation only for $j \in \{0,1,2\}$, finding agreement. 
In order to check (\ref{sine-tau-function-factorisation_j}) also for $j \geqslant 3$, 
one needs the explicit expressions for $\mathcal{D}_{k}(\nu)$ with $k\geqslant3$.

\subsection{Subleading terms of the entanglement entropies}
\label{app_large_eta_entropies}

In this subsection we discuss the evaluation of 
$\widetilde{S}_{A,\kappa,\infty}^{(\alpha)}$ in \eqref{ee-large-eta-dec},
where $\kappa \in \{+, -\}$, 
which contains the subleading terms of the entanglement entropies
that vanish as $\eta \to \infty$.
These terms are obtained from the expansion \eqref{tau_sublead}.

\subsubsection{Expansion of the vanishing term}

By using  \eqref{entropy-tau-pm} and (\ref{tau-pm-prod-dec}),
for the term $\widetilde{S}_{A,\kappa,\infty}^{(\alpha)}$
in \eqref{ee-large-eta-dec} we have
\be
\label{ent_sub_app}
\widetilde{S}_{A,\kappa,\infty}^{(\alpha)}
=
  \lim_{\epsilon, \delta \to 0}\,
\frac{1}{2\pi i}\oint_\mathfrak{C}
\rd z\,s_{\alpha}(z) \, \partial_z \log\left(\mathcal{T}_{\kappa,\infty} \right) 
= \, - \!
  \lim_{\epsilon, \delta \to 0}\,
\frac{1}{2\pi i}\oint_\mathfrak{C}
\rd z\,s'_{\alpha}(z) \, \log\left(\mathcal{T}_{\kappa,\infty} \right)
\ee
where an integration by parts has been performed
and the closed path $\mathfrak{C}$ in the complex plane,
which is parameterised by  $\epsilon>0$ and $\delta>0$,
has been  described in the text below \eqref{entropy-tau-pm}.
The integrals along $\mathfrak{C}_0$ and $\mathfrak{C}_1$ in (\ref{ent_sub_app})
vanish as $\epsilon \to 0$. 
In order to evaluated the remaining two terms, one needs
the limit of $\mathcal{T}_{\kappa,\infty}$ as  $ z\rightarrow x\pm \textrm{i} \,0^+ $, with $x\in[0,1]$. 
The dependence on $z$ in the expansion \eqref{tau_sublead}
occurs through $\nu = \nu(z)$ in \eqref{nu-z-def}, which gives
\be
\lim_{\delta\rightarrow0} \, \nu\big|_{z=x\pm \ri \delta}
\,=
\lim_{\delta\rightarrow0}\frac{1}{2\pi \ri} \, \log\! \big[ 1-(x\pm \ri\delta)^{-1}\big]
\,=\,
\pm\frac{1}{2} + \frac{1}{2\pi \ri}\log(1/x - 1)\,.
\ee
This suggests to adopt $y\equiv\frac{1}{2\pi}\log(1/x-1)$ as integration variable in the remaining two integrals;
hence \eqref{ent_sub_app} becomes
\be
\label{St_inf}
\widetilde{S}_{A,\kappa,\infty}^{(\alpha)}
=
\frac{1}{2\pi \ri}\int_{-\infty}^{+\infty}\! 
\hat{s}'_{\alpha}(y)
\left[\,
\log\!\left(\left.\mathcal{T}_{\kappa,\infty}\right|_{\nu=- \ri y-\frac{1}{2}}\right)
-\log\!\left(\left.\mathcal{T}_{\kappa,\infty}\right|_{\nu=-\ri y+\frac{1}{2}}\right)
\,\right]
\rd y
\ee
where,
for $\alpha =1 $ and $\alpha\in(0,1)\cup(1,\infty)$,
we have respectively 
\be
\label{s-alpha-der-app}
\hat{s}'_{1}(y) = -\frac{\pi^{2}y}{\left[\cosh(\pi y)\right]^{2}}
\;\;\;\qquad\;\;\;
\hat{s}'_{\alpha}(y) = \frac{\pi\alpha}{\alpha-1}\left[\tanh(\pi y)-\tanh(\alpha\pi y)\right] .
\ee
The change of integration variable $y\rightarrow-y$ in the second term of \eqref{St_inf} leads to
\be
 \label{S_sub_app_2}
\widetilde{S}_{A,\kappa,\infty}^{(\alpha)} 
=
 \frac{1}{2\pi \ri}\int_{-\infty}^{+\infty} \!
 \rd y\,\hat{s}'_{\alpha}(y)
\big[ 
\log\left(\mathcal{T}_{\kappa,\infty}^{-}\right)
 +\log\left(\mathcal{T}_{\kappa,\infty}^{+}\right)
\big]
\ee
where
\be
\label{nu_hat}
\mathcal{T}_{\kappa,\infty}^{\pm}\equiv\left.\mathcal{T}_{\kappa,\infty}\right|_{\nu=\pm\tilde{\nu}}
\;\;\;\; \qquad \;\;\;
\tilde{\nu}\equiv \frac{1}{2} + \ri \,y
\ee
which can be written explicitly by using \eqref{tau_sublead} and the result reads
\be
\label{app-Tk-pm-inf-v0}
\mathcal{T}_{\kappa,\infty}^{\pm} 
=
\sum_{n\in\mathbb{Z}} \; \sum_{k=0}^{\infty}
\frac{(\ri \kappa)^{n} \e^{\ri 2n\eta} \,2^k}{ (4\eta)^{\pm 2 \ri yn}(4\eta)^{n^{2}\pm n + k}}
\;
\frac{G(1\pm\tilde{\nu}+n)\,G(1\mp\tilde{\nu}-n)}{G(1+\tilde{\nu})\,G(1-\tilde{\nu})}\;
\mathcal{D}_{k}(\pm\tilde{\nu}+n)\,.
\ee

Let us first perform the change of variable $n\rightarrow -\,n$ 
only for $\mathcal{T}_{\kappa,\infty}^{+}$. 
This allows to write (\ref{app-Tk-pm-inf-v0}) as 
\be
\label{T_e}
\mathcal{T}_{\kappa,\infty}^{\pm} 
=
\sum_{n\in\mathbb{Z}}\, \sum_{k=0}^{\infty}
\frac{ (\mp \,\ri \kappa )^{n} \,\e^{\mp \ri 2n\eta}(4\eta)^{2\ri yn} }{ (4\eta)^{n^{2}-n+k} } \; 
2^{k} \, \mathcal{G}_{n}(\tilde{\nu})\,\mathcal{D}_{k}(\pm\tilde{\nu}\mp n)
\ee
where
\be
\label{G_n}
\mathcal{G}_{n}
=
\mathcal{G}_{n}(\tilde{\nu})
\equiv
\frac{G(1+\tilde{\nu}-n)\, G(1-\tilde{\nu}+n)}{G(1+\tilde{\nu})\, G(1-\tilde{\nu})}
\ee
which can be written also as 
\be
\mathcal{G}_{n \leqslant -1}(\tilde{\nu}) = \prod_{j=1}^{-n}\frac{\Gamma(j+\tilde{\nu})}{\Gamma(1-j-\tilde{\nu})} 
\;\qquad\;\;
\mathcal{G}_{n =0}(\tilde{\nu}) = 1
\;\qquad\;\;
\mathcal{G}_{n \geqslant 1}(\tilde{\nu}) = \prod_{j=1}^{n}\frac{\Gamma(j-\tilde{\nu})}{\Gamma(1-j+\tilde{\nu})} \,.
\ee
The expression (\ref{T_e}) suggests to introduce 
$j \equiv n^{2}-n+k$ to replace the index $k$.
Thus, for any $n\in\mathbb{Z}$, we have $j\geqslant n^{2}-n$,
which can be equivalently reformulated by introducing 
$\tilde{n}(j)\equiv\big\lfloor \sqrt{j+1/4}-1/2\big\rfloor $ and 
considering the values of $n$ such that
$-\tilde{n}(j)\leqslant n\leqslant\tilde{n}(j)+1$ for any  $j\in\mathbb{N}_{0}$.
For instance, we have $\tilde{n}(0)=\tilde{n}(1)=0$, 
$\tilde{n}(2)=\tilde{n}(3)=\tilde{n}(4)=\tilde{n}(5)=1$, etc.\,.
These manipulations allow to write \eqref{T_e} as follows
\be
\label{cal-R-pm-def}
\mathcal{T}_{\kappa,\infty}^{\pm} 
= 
\sum_{j=0}^{\infty}\frac{\mathcal{R}_{\kappa,j}^{\pm}}{(4\eta)^{j}}
\;\;\qquad\;\;
\mathcal{R}_{\kappa,j}^{\pm}
\equiv  \!\!
\sum_{n=-\tilde{n}(j)}^{\tilde{n}(j)+1}(\mp \,\ri \kappa )^{n}
\e^{\mp \ri 2n\eta}(4\eta)^{2\ri yn}\,
2^{j+n-n^{2}} \, \mathcal{G}_{n}(\tilde{\nu})
\,\mathcal{D}_{j+n-n^{2}}(\pm\tilde{\nu}\mp n)\,.
\ee
By introducing $\tilde{\mathcal{R}}_{\kappa,0}^{\pm}$ as follows
\be
\mathcal{R}_{\kappa,0}^{\pm}=1+\tilde{\mathcal{R}}_{\kappa,0}^{\pm}
\ee
where
\be
\label{tilde-cal-R-0-def}
\tilde{\mathcal{R}}_{\kappa,0}^{\pm}
\equiv
\mp\, \ri \kappa\;
\e^{\mp \ri 2\eta} \,(4\eta)^{2 \ri y}\,
\Omega(y)
\;\;\qquad\;\;
\Omega(y) \equiv 
\mathcal{G}_{1} (\tilde{\nu})
=
\frac{\Gamma\big(\frac{1}{2}- \ri y\big)}{\Gamma \big(\frac{1}{2}+ \ri y\big)}
\ee
the expression of $\mathcal{R}_{\kappa,j}^{\pm}$ in (\ref{cal-R-pm-def}) can be written as
\be
\label{R_j_new}
\mathcal{R}_{\kappa,j}^{\pm}
\,=\!\!
\sum_{n=-\tilde{n}(j)}^{\tilde{n}(j)+1}
\!\! \big(\tilde{\mathcal{R}}_{\kappa,0}^{\pm}\big)^{n}
\,\mathcal{\tilde{G}}_{n}(\tilde{\nu})
\,2^{j+n-n^{2}}
\mathcal{D}_{j+n-n^{2}}(\pm\tilde{\nu}\mp n)
\;\;\qquad\;\;
\tilde{\mathcal{G}}_{n}(\tilde{\nu}) 
 \equiv 
\frac{\mathcal{G}_{n}(\tilde{\nu})}{\mathcal{G}_{1}(\tilde{\nu})^n}\,.
\ee
By exploiting the identity $\Gamma(x+1) = x\, \Gamma(x)$, one finds that
\be
\label{tilde-G}
\tilde{\mathcal{G}}_{n}(\tilde{\nu})
= 
\begin{cases}
\prod_{k=1}^{n-1}\left(y+\hat{y}_{k}\right)^{2(n-k)}  \hspace{1.2cm} & n\geqslant2
\\
1  & n\in \{0,1\}
\\
\prod_{k=1}^{-n}\left(y-\hat{y}_{k}\right)^{2(-n+1-k)} & n\leqslant-1
\end{cases}
\ee
where
\be
\label{hat-y-def}
\hat{y}_{k} 
\,\equiv\,
 \ri\, \bigg(k-\frac{1}{2}\bigg).
\ee

At this point, let us  consider
\be
\label{log_Te}
\log\! \big( \mathcal{T}_{\kappa,\infty}^{\pm}\big)
=
\log\!\big(1+\tilde{\mathcal{R}}_{\kappa,0}^{\pm}\big)
+
\log\!\Bigg(1+\sum_{N=1}^{\infty}\frac{\mathcal{B}_{\kappa,N}^{\pm}}{(4\eta)^{N}}\Bigg)
=
\sum_{N=0}^{\infty}\frac{\mathcal{Y}_{\kappa,N}^{\pm}}{(4\eta)^{N}}
\ee
where we have introduced
\be
\label{cal-B-def}
\mathcal{B}_{\kappa,N}^{\pm}
\equiv
\frac{\mathcal{R}_{\kappa,N}^{\pm} }{1+\tilde{\mathcal{R}}_{\kappa,0}^{\pm} }
\ee
and
\be
\label{Yk_def}
\mathcal{Y}_{\kappa,0}^{\pm}  
\equiv  
\log\!\big(1+\tilde{\mathcal{R}}_{\kappa,0}^{\pm}\big)
\;\qquad\;
\mathcal{Y}_{\kappa,N \geqslant 1}^{\pm}  
\equiv  
\sum_{\Upsilon_{N}}(-1)^{\sum_{j=1}^k r_j +1}\,
\frac{\big(\sum_{j=1}^k r_j -1 \big)!}{ \prod_{j=1}^k r_{j}! } \,
\prod_{j=1}^k
(\mathcal{B}_{\kappa,p_{j}}^{\pm})^{r_{j}}
\ee
being $\Upsilon_{N}$ defined as  
the set made by the integer decompositions of $N\in\mathbb{N}$, 
namely
\be
\Upsilon_{N}\equiv
\bigg\{ 
\Big( (p_{1},r_{1}),\dots,(p_{k},r_{k})\Big)
\in
\big(\mathbb{N}^{2}\big)^{k}
\;\; \textrm{s.t.}\;\;
p_{1}>\cdots>p_{k}
\;\; \textrm{and}\;\;
\sum_{j=1}^k  p_{j} r_{j} =N\,
\bigg\} \,.
\ee
For instance, for $N \in \{1,2,3,4\}$  we have
\bea
 & & 
\Upsilon_{1}=\Big\{ \big((1,1)\big) \Big\} 
\hspace{3.5cm}
\Upsilon_{2}=\Big\{ \big((2,1)\big) , \big((1,2)\big) \Big\} 
\\
& &
\Upsilon_{3}=\Big\{ \big((3,1)\big) , \big((2,1), (1,1) \big) , \big((1,3)\big) \Big\} 
\\
& &
\Upsilon_{4}=\Big\{ \big((4,1)\big) , \big((3,1), (1,1) \big) , \big((2,2)\big) , \big((2,1) , (1,2)\big), \big((1,4)\big) \Big\} 
\eea
which respectively provide the following expression for $\mathcal{Y}_{\kappa,N}^{\pm}$ (from (\ref{Yk_def}))
\bea
& &
\mathcal{Y}_{\kappa,1}^{\pm}=\mathcal{B}_{\kappa,1}^{\pm}
\hspace{4.1cm}
\mathcal{Y}_{\kappa,2}^{\pm}=\mathcal{B}_{\kappa,2}^{\pm}-\frac{1}{2} \big(\mathcal{B}_{\kappa,1}^{\pm}\big)^{2}
\\
\rule{0pt}{.7cm}
& &
\mathcal{Y}_{\kappa,3}^{\pm}
=\mathcal{B}_{\kappa,3}^{\pm}-\mathcal{B}_{\kappa,2}^{\pm}\,\mathcal{B}_{\kappa,1}^{\pm}
+\frac{1}{3}\big(\mathcal{B}_{\kappa,1}^{\pm}\big)^{3}
\\
\rule{0pt}{.7cm}
& &
\mathcal{Y}_{\kappa,4}^{\pm}=\mathcal{B}_{\kappa,4}^{\pm}-\mathcal{B}_{\kappa,3}^{\pm}\mathcal{B}_{\kappa,1}^{\pm}-\frac{1}{2}(\mathcal{B}_{\kappa,2}^{\pm})^{2}+\mathcal{B}_{\kappa,2}^{\pm}(\mathcal{B}_{\kappa,1}^{\pm})^{2}-\frac{1}{4}(\mathcal{B}_{\kappa,1}^{\pm})^{4}\,.
\eea
In our analysis only $N=1$ and $N=2$ have been employed.

Finally, by employing the expansions \eqref{log_Te} into \eqref{S_sub_app_2}, 
the subleading terms of the entanglement entropies 
in (\ref{ee-large-eta-dec})
can be written in the form \eqref{S_larg_exp} with
\be
\label{ent_N}
\widetilde{S}_{A,\kappa,\infty,N}^{(\alpha)}
=
\frac{1}{2\pi \ri}
\int_{-\infty}^{+\infty}\!
\hat{s}'_{\alpha}(y)\Big(\mathcal{Y}_{\kappa,N}^{-}+\mathcal{Y}_{\kappa,N}^{+}\Big)\, \rd y\,
\ee
where the function $\hat{s}'_{\alpha}(y)$ is given in (\ref{s-alpha-der-app}).

\subsubsection{Useful integrals}

In the forthcoming analyses, we systematically encounter the integrals
\be
\label{int_I}
\mathcal{I}_{\alpha,j,\kappa}^{\pm}[\mathcal{P}]
\equiv 
\frac{\e^{\ri \frac{\pi}{4}(1\mp1)}}{2\pi \ri}
\int_{-\infty}^{+\infty} \! 
\hat{s}'_{\alpha}(y)
\Big[
\big(\tilde{\mathcal{R}}_{\kappa,0}^{-}\big)^{j}
\pm
\big(\tilde{\mathcal{R}}_{\kappa,0}^{+}\big)^{j}
\,\Big]
\,\mathcal{P}(y)\, \rd y
\;\; \qquad \;\;
j\in\mathbb{Z}
\ee
where $\mathcal{P}(y)$ is a polynomial 
and the factor $\e^{\ri\frac{\pi}{4}(1\mp1)}$ 
has been introduced for later convenience
(in order to facilitate the construction of the trigonometric functions).

Since the integral \eqref{int_I} for $j=0$ can be performed analytically,
in the following we consider only the cases where $j\neq0$. 
The integrand in \eqref{int_I} involves an integer power of
$\tilde{\mathcal{R}}_{\kappa,0}^{\pm}$ defined in (\ref{tilde-cal-R-0-def});
hence the integral \eqref{int_I} can be evaluated 
by applying the residue theorem. 
It is convenient to choose a closed integration path 
that includes a half circumference at infinity lying
either to the upper half plane or to the lower half plane,
for $j>0$ or $j<0$ respectively. 

As for the singularities of the integrand occurring in \eqref{int_I},
the function $\Omega(y)$ has simple zeros in the upper half plane for $y=\hat{y}_{k}$ 
and simple poles in the lower half plane for $y=-\hat{y}_{k}$, where $k\in\mathbb{N}$.
The opposite holds for $1/\Omega(y)$. 
Furthermore, for $k\in\mathbb{N}$ we have
\bea
\label{Omega_exp}
\Omega(y) 
& = & 
\ri (-1)^{k+1} \big[(k-1)! \big]^{2}
\big(y-\hat{y}_{k}\big)
+
O\big((y-\hat{y}_{k})^{2} \big)
\\
\label{Omega2_exp}
1/\Omega(y)
& = & 
\ri (-1)^{k}
\big[(k-1)!\big]^{2}
\big(y+\hat{y}_{k}\big)
+
O\big((y+\hat{y}_{k})^{2} \big)\,.
\eea
The function $\hat{s}'(y)$ has double poles for $y=\pm\,\hat{y}_{k}$ with $k\in\mathbb{N}$,
and
\be
\label{s_1_sing2}
\hat{s}'(y) =
\frac{\hat{y}_{k}}{(y-\hat{y}_{k} )^{2}}
+O\big( (y-\hat{y}_{k})^{-1} \big)
\;\;\;\qquad\;\;\;
\hat{s}'(y) 
=
- \frac{\hat{y}_{k}}{(y+\hat{y}_{k} )^{2}}
+O\big( (y+\hat{y}_{k})^{-1} \big)\,.
\ee
Instead, considering the function $\hat{s}_{\alpha}'(y)$ for finite $\alpha\neq 1$ 
and decomposing it as follows
\be
\label{s_alpha_split}
\hat{s}_{\alpha}'(y)
=
\frac{\pi\alpha}{\alpha-1}\, \tanh(\pi y)
+
\frac{\pi\alpha}{1-\alpha}\, \tanh(\alpha\pi y)
\ee
one observes that 
the first term has simple poles for $y=\pm\,\hat{y}_{k}$, 
with residues equal to $\alpha/(\alpha-1)$,
while the second term has simple poles for $y=\pm\, \hat{y}_{k} /\alpha $, 
with $k\in\mathbb{N}$ and residues equal to $1/(1-\alpha)$.

The above observations lead us to evaluate the integral \eqref{int_I}
by considering $\alpha = 1$ and finite $\alpha \neq 1$ separately.

When $\alpha=1$, one finds 
 $\mathcal{I}_{1,j,\kappa}^{\pm} [\mathcal{P}] =0$ for $\left|j\right|\geqslant2$
 because the zeros of $\tilde{\mathcal{R}}_{\kappa,0}^{\pm}$ 
 are simple while the poles of $\hat{s}'(y)$ are double.
For  $j \in \{1,-1 \}$,  by using (\ref{Omega_exp}), (\ref{Omega2_exp}) and \ref{s_1_sing2}), 
we get
\bea
\hat{s}'(y) \, \tilde{\mathcal{R}}_{\kappa,0}^{\pm}\,\mathcal{P}(y) 
& = & 
\mp \, \ri \kappa\, (-1)^{k}
(2k-1)\big[(k-1)!\big]^{2}\;
\frac{\e^{\mp \ri 2\eta}\, \mathcal{P}(\hat{y}_{k})}{2\, (4\eta)^{2k-1}}\;
\frac{1}{y-\hat{y}_{k}} 
+O\big( (y-\hat{y}_{k})^{0} \big)
\hspace{1.5cm}
\\
\rule{0pt}{.8cm}
\hat{s}'(y)  \, \frac{ \mathcal{P}(y) }{\tilde{\mathcal{R}}_{\kappa,0}^{\pm}}
& = & 
\pm\,\ri \kappa(-1)^{k}
(2k-1) \big[(k-1)!\big]^{2}\;
\frac{\e^{\pm \ri 2\eta}\,\mathcal{P}(-\hat{y}_{k})}{2\, (4\eta)^{2k-1}}\;\frac{1}{y+\hat{y}_{k}} 
+O\big( (y+\hat{y}_{k})^{0} \big)
\eea
where $k\in\mathbb{N}$. 
Thus, for the integral (\ref{int_I}) in these cases we obtain
(here $j \in \{ -1\,, 1 \}$ and $\xi = \textrm{sign}(j)$)
\be
\label{I_11}
\mathcal{I}_{1, j,\kappa}^{\pm} [\mathcal{P}]
\,=\,
\kappa\sum_{k=1}^{\infty}(-1)^{k+1}
(2k-1) \big[(k-1)!\big]^{2}\,
\frac{\mathcal{P}(\xi\,\hat{y}_{k})}{(4\eta)^{2k-1}} 
\cdot
\bigg\{
\begin{array}{ll}
\sin(2\eta) \hspace{.6cm} &  \textrm{for $\mathcal{I}_{1, j,\kappa}^{+}$}
\\
\cos(2\eta) & \textrm{for $\mathcal{I}_{1,j,\kappa}^{-}\,$.}
\end{array}
\ee

For finite $\alpha\neq 1$,
the contribution of the first term in the r.h.s. of \eqref{s_alpha_split} vanish
because its simple poles cancel with the simple zeros of $\tilde{\mathcal{R}}_{\kappa,0}^{\pm}$.
As for the contribution of the second term in the r.h.s. of \eqref{s_alpha_split} to (\ref{int_I}),
we find 
\bea
\label{I_alpha_pos_neg}
\mathcal{I}_{\alpha,j,\kappa}^{\pm} [\mathcal{P}]
& = & 
\frac{\xi\, \e^{\ri \frac{\pi}{4}(1\mp1)}}{1-\alpha}\sum_{k=1}^{\infty}
\mathcal{P}(\xi y_{k} / \alpha)
\left[
\big(\tilde{\mathcal{R}}_{\kappa,0}^{-}\big)^{j}\big|_{y= \xi y_{k} / \alpha}
\pm
\big(\tilde{\mathcal{R}}_{\kappa,0}^{+}\big)^{j}\big|_{y=\xi y_{k} / \alpha}
\,\right]
\\
\rule{0pt}{.8cm}
 & = & 
 \xi\,\frac{2\,(-\,\kappa)^{j}}{1-\alpha}
 \sum_{k=1}^{\infty} 
\mathcal{P}(\xi\,y_{k} /\alpha)\,
 \bigg( \frac{\Omega(\xi\,\hat{y}_{k} / \alpha)}{(4\eta)^{(2k-1) /\alpha}} \bigg)^j
\cdot
\bigg\{
\begin{array}{ll}
\cos\!\big[(2\eta-\frac{\pi}{2})j\big] \hspace{1cm} &  \textrm{for $\mathcal{I}_{1,j,\kappa}^{+}$}
\\
\rule{0pt}{.45cm}
- \sin\!\big[(2\eta-\frac{\pi}{2})j\big]& \textrm{for $\mathcal{I}_{1,j,\kappa}^{-}$}
\end{array}
\nn
\eea
where $\xi = \textrm{sign}(j)$ again and 
\be
\label{Omega_eva}
\Omega(\hat{y}_{k} / \alpha)
=
\frac{1}{\Omega(-\hat{y}_{k} / \alpha)}
=
\frac{\Gamma\!\left(\frac{1}{2}-\frac{1}{2\alpha}+\frac{k}{\alpha}\right)}{
\Gamma\!\left(\frac{1}{2}+\frac{1}{2\alpha}-\frac{k}{\alpha}\right)}\,.
\ee

\subsubsection{$N=0$ term}
\label{app-large-eta-N-0}

The first term in (\ref{S_larg_exp}) corresponds to $N=0$ and its coefficient 
$\widetilde{S}_{A,\pm,\infty,0}^{(\alpha)}$ can be evaluated from
(\ref{ent_N}) and (\ref{Yk_def}) specialised to $N=0$.
First one expands \eqref{Yk_def} as follows
\be
\mathcal{Y}_{\kappa,0}^{\pm}
=
\log\!\big(1+\tilde{\mathcal{R}}_{\kappa,0}^{\pm}\big)
=
\sum_{j=1}^{\infty}\frac{(-1)^{j+1}}{j}\, \big(\tilde{\mathcal{R}}_{\kappa,0}^{\pm}\big)^{j}
\ee
finding that (\ref{ent_N}) for $N=0$ can be written as
\be
\label{S0}
\widetilde{S}_{A,\kappa,\infty,0}^{(\alpha)}
\,=\,
\frac{1}{2\pi \ri}
\sum_{j=1}^{\infty}\frac{(-1)^{j+1}}{j}
\int_{-\infty}^{+\infty}\!
\hat{s}'_{\alpha}(y)
\left[
\big(\tilde{\mathcal{R}}_{\kappa,0}^{-}\big)^{j}
+
\big(\tilde{\mathcal{R}}_{\kappa,0}^{+}\big)^{j}
\,\right] \rd y
\ee
which is a series whose coefficients 
are the integrals $\mathcal{I}_{\alpha,j}^{+}[\mathcal{P}]$ in \eqref{int_I}
with $j \geqslant 1$ and $\mathcal{P}(y) = 1$ identically.
When $\alpha = 1$, we can employ \eqref{I_11} specialised to this case
(i.e. for $\mathcal{P}(y) = 1$, $\kappa = +1$ and $\xi =+1$), 
finding (\ref{S-0-large-eta-sub}).
For finite $\alpha\neq 1$, 
from (\ref{S0})  and  \eqref{I_alpha_pos_neg} specialised to this case, 
we obtain (\ref{Ren-0-large-eta-sub}).

\subsubsection{$N=1$ term}
\label{app-large-eta-N-1}

As for the $N=1$ term in (\ref{S_larg_exp}),
its coefficient $\widetilde{S}_{A,\pm,\infty,1}^{(\alpha)}$ can be found  
through (\ref{ent_N}) and (\ref{Yk_def}) specialised to this case
(the expression of $\mathcal{R}_{\kappa,1}^{\pm} $ can be obtained from \eqref{R_j_new}), that give
\be
\label{cal-Y-k-1-app}
\mathcal{Y}_{\kappa,1}^{\pm} 
\,=\,
\mathcal{B}_{\kappa,1}^{\pm}
\,=\,
2\,\mathcal{D}_{1}(\pm\tilde{\nu})
+
\frac{2\,\tilde{\mathcal{R}}_{\kappa,0}^{\pm}}{1+\tilde{\mathcal{R}}_{\kappa,0}^{\pm}}\,
\big[
\mathcal{D}_{1}(\pm\tilde{\nu}\mp1)-\mathcal{D}_{1}(\pm\tilde{\nu})
\big]\,.
\ee
From this expression and the expansion 
\be
\label{series-geom-R}
\frac{1}{1+\tilde{\mathcal{R}}_{\kappa,0}^{\pm}}
=
\sum_{j=1}^{\infty}
(-1)^{j+1}\big(\tilde{\mathcal{R}}_{\kappa,0}^{\pm}\big)^{j-1}
\ee
we find that (\ref{ent_N}) in this case becomes
\bea
\label{tilde-S-infty-1-v0}
\widetilde{S}_{A,\kappa,\infty,1}^{(\alpha)} 
& = & 
\frac{1}{2\pi \ri}\, \Bigg\{
2 \int_{-\infty}^{+\infty}\!
\hat{s}'_{\alpha}(y) 
\big[\mathcal{D}_{1}(\tilde{\nu})+\mathcal{D}_{1}(-\tilde{\nu})\big] \rd y
 +
 \sum_{j=1}^{\infty}(-1)^{j+1}
 \int_{-\infty}^{+\infty}\! \hat{s}'_{\alpha}(y)
\\
 &  & 
 \hspace{1.1cm}
\times \, 
2\,
 \bigg[
 \big(\tilde{\mathcal{R}}_{\kappa,0}^{-}\big)^{j}
 \big(  \mathcal{D}_{1}(-\tilde{\nu}+1)-\mathcal{D}_{1}(-\tilde{\nu})  \big)
 +
 \big(\tilde{\mathcal{R}}_{\kappa,0}^{+}\big)^{j}
 \big(\mathcal{D}_{1}(\tilde{\nu}-1)-\mathcal{D}_{1}(\tilde{\nu})\big)
 \bigg]
\, \rd y
 \Bigg\}\,.
 \nonumber 
\eea
By observing that $\mathcal{D}_{1}(\tilde{\nu})+\mathcal{D}_{1}(-\tilde{\nu})=0$
and introducing
\be
\label{P1_def}
2\left[\mathcal{D}_{1}(\pm\tilde{\nu}\mp1)-\mathcal{D}_{1}(\pm\tilde{\nu})\right]
\equiv
\mp \,\ri \,\mathcal{P}_{1}(y)
\;\;\; \qquad \;\;\;
\mathcal{P}_{1}(y)
\equiv
6y^{2}-\frac{1}{2}
\ee
we find that (\ref{tilde-S-infty-1-v0}) can be written as 
\be
\label{tilde-S-infty-1-v1}
\widetilde{S}_{A,\kappa,\infty,1}^{(\alpha)} 
=\,
\frac{1}{2\pi}
 \sum_{j=1}^{\infty}(-1)^{j+1}
 \int_{-\infty}^{+\infty}\!
 \hat{s}'_{\alpha}(y)
\Big[
 \big(\tilde{\mathcal{R}}_{\kappa,0}^{-}\big)^{j}
 -
 \big(\tilde{\mathcal{R}}_{\kappa,0}^{+}\big)^{j}
\Big]
  \,\mathcal{P}_{1}(y) \, \rd y
\ee
whose summand takes the form  \eqref{int_I}.
Thus, when $\alpha=1$, from \eqref{I_11} we obtain (\ref{S-1-large-eta-sub});
while for finite $\alpha \neq 1$ we arrive to (\ref{Ren-1-large-eta-sub})
by employing \eqref{I_alpha_pos_neg} in (\ref{tilde-S-infty-1-v1}).

\subsubsection{$N=2$ term}
\label{app-large-eta-N-2}

The coefficient $\widetilde{S}_{A,\pm,\infty,2}^{(\alpha)}$ 
occurring in the term labelled by $N=2$ in the r.h.s. of (\ref{S_larg_exp})
is given by (\ref{ent_N}) and (\ref{Yk_def}) specialised to $N=2$.

In order to obtain $\mathcal{B}_{\kappa,2}^{\pm} $,
first we construct $\mathcal{R}_{\kappa,2}^{\pm}$ from \eqref{R_j_new},
finding 
\be
\mathcal{R}_{\kappa,2}^{\pm}
=
\frac{ \mathcal{\tilde{G}}_{-1}(\tilde{\nu}) }{ \tilde{\mathcal{R}}_{\kappa,0}^{\pm} }
+
4\,\mathcal{D}_{2}(\pm\tilde{\nu})
+
4\,\tilde{\mathcal{R}}_{\kappa,0}^{\pm}\,\mathcal{D}_{2}(\pm\tilde{\nu}\mp1)
+
\big(\tilde{\mathcal{R}}_{\kappa,0}^{\pm}\big)^{2}\,
\mathcal{\tilde{G}}_{2}(\tilde{\nu}) 
\ee
and then use (\ref{cal-B-def}), which leads to
\bea
\mathcal{B}_{\kappa,2}^{\pm} 
&=&
\frac{ \mathcal{\tilde{G}}_{-1}(\tilde{\nu})  }{ \tilde{\mathcal{R}}_{\kappa,0}^{\pm} }
+
4\,\mathcal{D}_{2}(\pm\tilde{\nu})-\mathcal{\tilde{G}}_{-1}(\tilde{\nu}) 
\\
& &
+ \,
 \frac{\tilde{\mathcal{R}}_{\kappa,0}^{\pm}}{1+\tilde{\mathcal{R}}_{\kappa,0}^{\pm}}\,
\big[ 
 \mathcal{\tilde{G}}_{-1}(\tilde{\nu}) 
 -4\,\mathcal{D}_{2}(\pm\tilde{\nu})+4\,\mathcal{D}_{2}(\pm\tilde{\nu}\mp1)
\big]
 +
 \frac{ 
 (\tilde{\mathcal{R}}_{\kappa,0}^{\pm} )^2 }{ 1+\tilde{\mathcal{R}}_{\kappa,0}^{\pm}} \;
 \mathcal{\tilde{G}}_{2} (\tilde{\nu}) \,.
 \nonumber
\eea
By using this expression and (\ref{cal-Y-k-1-app}) into \eqref{Yk_def} for $N=2$, one obtains
\be
\label{Y_2}
\mathcal{Y}_{\kappa,2}^{\pm} 
=
\mathcal{B}_{\kappa,2}^{\pm}-\frac{1}{2} \, (\mathcal{B}_{\kappa,1}^{\pm})^{2}
\,\equiv\,
\mathcal{Y}_{\kappa,2, a}^{\pm} 
+ \mathcal{Y}_{\kappa,2, b}^{\pm} 
+ \mathcal{Y}_{\kappa,2, c}^{\pm} 
+ \mathcal{Y}_{\kappa,2, d}^{\pm} 
+ \mathcal{Y}_{\kappa,2, e}^{\pm} 
\ee
where
\bea
\label{Y-2-ab-def}
\mathcal{Y}_{\kappa,2, a}^{\pm} 
 & \equiv & 
 \frac{\mathcal{\tilde{G}}_{-1}(\tilde{\nu}) }{ \tilde{\mathcal{R}}_{\kappa,0}^{\pm} }
\hspace{2.5cm}
 \mathcal{Y}_{\kappa,2, b}^{\pm} 
\,\equiv\,
4\,\mathcal{D}_{2}(\pm\tilde{\nu})-2\,\mathcal{D}_{1}(\pm\tilde{\nu})^{2}
 -\mathcal{\tilde{G}}_{-1}(\tilde{\nu}) 
 \\
  \rule{0pt}{.8cm}
  \label{Y-2-c-def}
  \mathcal{Y}_{\kappa,2, c}^{\pm} 
 & \equiv & 
 \frac{\tilde{\mathcal{R}}_{\kappa,0}^{\pm}}{1+\tilde{\mathcal{R}}_{\kappa,0}^{\pm}}\;
\Big\{
 \mathcal{\tilde{G}}_{-1}(\tilde{\nu})  -4\,\mathcal{D}_{2}(\pm\tilde{\nu})
 +4\,\mathcal{D}_{2}(\pm\tilde{\nu}\mp1)
 -4\,\mathcal{D}_{1}(\pm\tilde{\nu})
 \big[\mathcal{D}_{1}(\pm\tilde{\nu}\mp1)  - \mathcal{D}_{1}(\pm\tilde{\nu}) \big]
\Big\}
  \nn
 \\
 & &
 \\
 \label{Y-2-de-def}
   \mathcal{Y}_{\kappa,2, d}^{\pm} 
 & \equiv & 
 \frac{\big(\tilde{\mathcal{R}}_{\kappa,0}^{\pm}\big)^{2}}{1+\tilde{\mathcal{R}}_{\kappa,0}^{\pm}}\;
 \mathcal{\tilde{G}}_{2}(\tilde{\nu}) 
\hspace{1.2cm}
   \mathcal{Y}_{\kappa,2, e}^{\pm} 
\,\equiv\,
 -\, 2\,
\bigg(\frac{\tilde{\mathcal{R}}_{\kappa,0}^{\pm}}{1+\tilde{\mathcal{R}}_{\kappa,0}^{\pm}}\bigg)^{2}
\, \Big[ \mathcal{D}_{1}(\pm\tilde{\nu}\mp1)-\mathcal{D}_{1}(\pm\tilde{\nu}) \Big]^{2}\,.
\eea
In (\ref{Y-2-c-def}) and (\ref{Y-2-de-def}), 
we can employ (\ref{series-geom-R}) and 
\be
\label{series}
\frac{1}{\big(1+\tilde{\mathcal{R}}_{\kappa,0}^{\pm}\big)^{2}}
\,=
\sum_{j=2}^{\infty}
(-1)^{j}(j-1)\big(\tilde{\mathcal{R}}_{\kappa,0}^{\pm}\big)^{j-2}\,.
\ee

Plugging (\ref{Y_2}) into (\ref{ent_N}),
we find that 
\be
\label{ee-infty-2-dec-app}
\widetilde{S}_{A,\kappa,\infty,2}^{(\alpha)}
\,=\,
\mathcal{J}_{\alpha, a} 
+ \mathcal{J}_{\alpha, b} 
+ \mathcal{J}_{\alpha, c} 
+ \mathcal{J}_{\alpha, d} 
+ \mathcal{J}_{\alpha, e} 
\ee
where the terms in the r.h.s. are the integrals
provided by (\ref{Y-2-ab-def}), (\ref{Y-2-c-def}) and (\ref{Y-2-de-def}),
that are defined respectively by 
\bea
\label{Y2_1}
\mathcal{J}_{\alpha, a}
& \equiv &
\frac{1}{2\pi \ri}
\int_{-\infty}^{+\infty} \! \hat{s}'_{\alpha}(y)
\Big[\,
\big(\tilde{\mathcal{R}}_{\kappa,0}^{-}\big)^{-1}
+
\big(\tilde{\mathcal{R}}_{\kappa,0}^{+}\big)^{-1}
\,\Big]
\mathcal{P}_{2,a}(y) \, \rd y
\\
\label{Y2_2}
\rule{0pt}{.7cm}
\mathcal{J}_{\alpha, b}
& \equiv &
\frac{1}{2\pi \ri}\int_{-\infty}^{+\infty}
\! \hat{s}'_{\alpha}(y) \, \mathcal{P}_{2,b}(y)\, \rd y
\\
\label{Y2_3}
\rule{0pt}{.7cm}
\mathcal{J}_{\alpha, c}
& \equiv &
\sum_{j=1}^{\infty}(-1)^{j+1}
\frac{1}{2\pi \ri}\int_{-\infty}^{+\infty}\! \hat{s}'_{\alpha}(y)
\Big[\,
\big(\tilde{\mathcal{R}}_{\kappa,0}^{-}\big)^{j}
+
\big(\tilde{\mathcal{R}}_{\kappa,0}^{+}\big)^{j}
\,\Big] \,
\mathcal{P}_{2,c}(y)\, \rd y
\\
\label{Y2_4}
\rule{0pt}{.7cm}
\mathcal{J}_{\alpha, d}
& \equiv &
\frac{1}{2\pi \ri}\,
\sum_{j=2}^{\infty}(-1)^{j}
\int_{-\infty}^{+\infty}\! \hat{s}'_{\alpha}(y)
\Big[\,
\big(\tilde{\mathcal{R}}_{\kappa,0}^{-}\big)^{j}
+
\big(\tilde{\mathcal{R}}_{\kappa,0}^{+}\big)^{j}
\,\Big] \,
\mathcal{P}_{2,d}(y)
\,\rd y
\\
\label{Y2_5}
\rule{0pt}{.7cm}
\mathcal{J}_{\alpha, e}
& \equiv &
\frac{1}{2\pi \ri}\,
\sum_{j=2}^{\infty}(-1)^{j+1}(j-1)
\int_{-\infty}^{+\infty}\! \hat{s}'_{\alpha}(y)
\Big[\,
\big(\tilde{\mathcal{R}}_{\kappa,0}^{-}\big)^{j}
+
\big(\tilde{\mathcal{R}}_{\kappa,0}^{+}\big)^{j}
\,\Big] \,
\mathcal{P}_{2,e}(y) \, \rd y
\eea
in terms of the polynomials given respectively by 
\bea
\label{P2-a-def}
\mathcal{P}_{2,a}(y)
& \equiv&
\mathcal{\tilde{G}}_{-1}(\tilde{\nu})  \,=\, (y-\hat{y}_{1})^{2} \, =\, (y - \ri /2)^{2}
\\
\rule{0pt}{.6cm}
\mathcal{P}_{2,b}(y)
& \equiv&
4\,\mathcal{D}_{2}(-\tilde{\nu})-2\,\mathcal{D}_{1}(-\tilde{\nu})^{2}
+
4\,\mathcal{D}_{2}(\tilde{\nu})-2\,\mathcal{D}_{1}(\tilde{\nu})^{2}
-
2\,\mathcal{\tilde{G}}_{-1}(\tilde{\nu}) 
\hspace{3.5cm}
\nonumber \\
 & = & -10\,y^{4}+20\ri \,y^{3}+15\,y^{2}-5 \ri \,y-5/8
\\
\label{P2-c-def}
\rule{0pt}{.6cm}
\mathcal{P}_{2,c}(y)
& \equiv&
\mathcal{\tilde{G}}_{-1}(\tilde{\nu})
-4\,\mathcal{D}_{2}(\pm\tilde{\nu}) +4\,\mathcal{D}_{2}(\pm\tilde{\nu}\mp1)
-4\,\mathcal{D}_{1}(\pm\tilde{\nu})
\big[\mathcal{D}_{1}(\pm\tilde{\nu}\mp1)-\mathcal{D}_{1}(\pm\tilde{\nu})\big]
\nonumber \\
 & = & -18\,y^{4}-20\ri \,y^{3}+4\,y^{2}+6\ri \,y - 3/8
\\
\label{P2-d-def}
\rule{0pt}{.6cm}
\mathcal{P}_{2,d}(y)
& \equiv&
\mathcal{\tilde{G}}_{2}(\tilde{\nu}) =\big(y+\hat{y}_{1}\big)^{2} = \big(y+\ri /2\big)^{2}
\\
\label{P2-e-def}
\rule{0pt}{.6cm}
\mathcal{P}_{2,e}(y)
& \equiv&
2\big[
\mathcal{D}_{1}(\pm\tilde{\nu}\mp1)-\mathcal{D}_{1}(\pm\tilde{\nu})
\big]^{2}
=
-18\,y^{4}+3\,y^{2}-1/8\,.
\eea
The integrals occurring in (\ref{Y2_1})-(\ref{Y2_5}) have the form (\ref{int_I}).
Hence, when $\alpha=1$ w can use (\ref{I_11}) for these integrals, finding
\bea
\label{J1-a-b}
\mathcal{J}_{1, a}
&=&
\kappa\,\sin(2\eta)
\sum_{k=1}^{\infty}(-1)^{k}\,
\frac{(2k-1)\big[ (k-1)!\big]^{2}\,\mathcal{P}_{2,a}(-\hat{y}_{k})}{ (4\eta)^{2k-1} }
\hspace{1.5cm}
\mathcal{J}_{1, b} \,=\, - \, \frac{1}{6}
\hspace{2cm}
\\
\label{J1-c-d-e}
\mathcal{J}_{1, c}
&=&
-\,\kappa\,\sin(2\eta)
\sum_{k=1}^{\infty}(-1)^{k}\,
\frac{(2k-1)\big[ (k-1)!\big]^{2}\,\mathcal{P}_{2,c}(\hat{y}_{k})}{ (4\eta)^{2k-1} }
\hspace{1.5cm}
\mathcal{J}_{1, d} \,=\, \mathcal{J}_{1, e} \,=\, 0
\eea
where 
$\,\mathcal{P}_{2,a}(-\hat{y}_{k})=(\hat{y}_{k}+\hat{y}_{1})^{2}=-k^{2}$
and $\mathcal{P}_{2,c}(\hat{y}_{k})=-18k^{4}+16k^{3}-k^{2}-8k+3$.
Then, the expression (\ref{S-2-large-eta-sub}) is obtained
by plugging (\ref{J1-a-b}) and (\ref{J1-c-d-e}) into (\ref{ee-infty-2-dec-app})
specialised to $\alpha =1$.

For positive and finite $\alpha \neq 1$, 
by applying (\ref{I_alpha_pos_neg}) 
to the integrals occurring in (\ref{Y2_1})-(\ref{Y2_5}),
we find 
\bea
\label{J-alpha-a-b}
\mathcal{J}_{\alpha, a}
&=&
\frac{2\, \kappa}{1-\alpha}\; \sin\left(2\eta\right)
\sum_{k=1}^{\infty}
\mathcal{P}_{2,a}(-y_{k} / \alpha)\,
 \frac{\Omega(\hat{y}_{k} / \alpha)}{(4\eta)^{(2k-1) /\alpha}} 
 \hspace{1cm}
\mathcal{J}_{\alpha, b}
\,=\,
\frac{(\alpha+1)(3\alpha^{2}-7)}{48\alpha^{3}}
\hspace{1cm}
\\
\label{J-alpha-c}
\mathcal{J}_{\alpha, c}
&=&
-\,\frac{2}{1-\alpha}\,
\sum_{j=1}^{\infty} \cos\!\big[(2\eta-\tfrac{\pi}{2})j\big] \, \sum_{k=1}^{\infty}
\mathcal{P}_{2,c}(y_{k} / \alpha)\,
\bigg( \frac{\kappa\,\Omega(\hat{y}_{k} / \alpha)}{(4\eta)^{(2k-1) /\alpha}} \bigg)^j
\\
\label{J-alpha-d}
\mathcal{J}_{\alpha, d}
&=&
\frac{2}{1-\alpha} \,
\sum_{j=2}^{\infty} \cos\!\big[(2\eta-\tfrac{\pi}{2})j\big]  \, \sum_{k=1}^{\infty}
\mathcal{P}_{2,d}(y_{k} / \alpha)\,
\bigg( \frac{\kappa\,\Omega(\hat{y}_{k} / \alpha)}{(4\eta)^{(2k-1) /\alpha}} \bigg)^j
\\
\label{J-alpha-e}
\mathcal{J}_{\alpha, e}
&=&
\frac{2}{1-\alpha} \,
\sum_{j=2}^{\infty}(1-j) \, \cos\!\big[(2\eta-\tfrac{\pi}{2})j\big] \sum_{k=1}^{\infty}
\mathcal{P}_{2,d}(y_{k} / \alpha)\,
\bigg( \frac{\kappa\,\Omega(\hat{y}_{k} / \alpha)}{(4\eta)^{(2k-1) /\alpha}} \bigg)^j
\eea
in terms of the polynomials in (\ref{P2-a-def})-(\ref{P2-e-def}) 
evaluated at $y_k /\alpha$.
Finally, 
the expression (\ref{Ren-2-large-eta-sub}) is obtained
by combining (\ref{J-alpha-a-b})-(\ref{J-alpha-e}) into (\ref{ee-infty-2-dec-app}).

\subsubsection{Consistency checks}
\label{app-large-eta-checks}

It is important to provide some consistency checks 
for the analytic expressions of the subleading terms obtained in this Appendix
and reported in Sec.\,\ref{sec_large_eta}.
In the following we consider the replica limit (\ref{replica limit-sec}),
the relation (\ref{sum-entropies-line-pm}) 
and the double scaling limit of some lattice results.

As for the replica limit (\ref{replica limit-sec}),
in the expansion (\ref{S_larg_exp})
it means that $\widetilde{S}_{A,\pm,\infty,N}^{(\alpha)} \to \widetilde{S}_{A,\pm,\infty,N}$ 
as $\alpha \to 1$ for any $N \in \mathbb{N}_0$.
When $N=0$, from (\ref{S-0-large-eta-sub}), (\ref{Ren-0-large-eta-sub}) and 
\be
\lim_{\alpha \to 1}
\frac{\Omega(\hat{y}_{k} / \alpha)}{\alpha-1}
\,=\,
(-1)^{k+1} (k-1/2)\big[ (k-1)!\big]^{2}
\label{limit_al}
\ee
we conclude that only the term corresponding to $j=1$ gives a non vanishing result
in the sum over $j$ occurring in (\ref{Ren-0-large-eta-sub});
hence $\widetilde{S}_{A,\pm,\infty,0}^{(\alpha)} \to \widetilde{S}_{A,\pm,\infty,0}$ 
as $\alpha \to 1$.
Similarly,  from (\ref{S-1-large-eta-sub})-(\ref{Ren-2-large-eta-sub}), 
we have checked that 
$\widetilde{S}_{A,\pm,\infty,N}^{(\alpha)} \to \widetilde{S}_{A,\pm,\infty,N}$ as $\alpha \to 1$
also for $N \in \{1,2\}$.

Another consistency check is 
The validity of the relation (\ref{sum-entropies-line-pm}) 
order by order in the large $\eta$ expansion
in another consistency check.
From (\ref{S-infty-leading-sec}), it is straightforward to realise that this relation holds for the leading terms;
hence (\ref{sum-entropies-line-pm}) can be verified by checking that
\be
\label{check-app-full-line}
\widetilde{S}_{A,+,\infty,N}^{(\alpha)}+\widetilde{S}_{A,-,\infty,N}^{(\alpha)} 
\,=\,
4^N\,\widetilde{S}_{A,\infty,N}^{(\alpha)}
\;\;\;\qquad\;\;
N \in \mathbb{N}_0
\ee
for the coefficients of the expansions (\ref{S_larg_exp}),
where $\widetilde{S}_{A,\infty,N}^{(\alpha)}$ for the interval $[-R,R] \subset \mathbb{R}$
on the line have been determined in \cite{Mintchev:2022xqh} for $N \in \{0,1,2\}$.

When $N=0$, first we split (\ref{Ren-0-large-eta-sub}) as follows
\bea
\label{app-tilde-S-0-split}
\widetilde{S}_{A,\kappa,\infty,0}^{(\alpha)} 
& = & 
\frac{2}{\alpha-1}
\sum_{j =1}^{\infty} \frac{\cos\!\big[(2\eta-\tfrac{\pi}{2})2j\big]}{2j} \, 
\sum_{k=1}^{\infty}
\bigg( \frac{\Omega(\hat{y}_{k} / \alpha)}{(4\eta)^{(2k-1) /\alpha}} \bigg)^{2j}
\\
 &  & 
 +\, \frac{2\, \kappa}{\alpha-1}
 \sum_{j =1}^{\infty}  \frac{ \cos\!\big[(2\eta-\tfrac{\pi}{2})(2j-1)\big] }{2j-1}\, 
 \sum_{k=1}^{\infty}
 \bigg( \frac{\Omega(\hat{y}_{k} / \alpha)}{(4\eta)^{(2k-1) /\alpha}} \bigg)^{2j-1}
 \nonumber
\eea
because it straightforwardly leads to observe 
that the terms coming from the second line of (\ref{app-tilde-S-0-split})
cancel in (\ref{check-app-full-line}),
while the remaining ones 
combine into $\widetilde{S}_{A,\infty,0}^{(\alpha)}$, as expected. 
The validity of (\ref{check-app-full-line}) for $N =1$ has been checked
by employing (\ref{Ren-1-large-eta-sub}).
Finally, we have checked (\ref{check-app-full-line}) for $N=2$ 
by first observing that
the terms corresponding to odd  values of $j$ cancel
in  the sums over $j$ (see (\ref{Ren-2-large-eta-sub}))
occurring in the l.h.s. of (\ref{check-app-full-line}),
and then that the remaining  terms give
\bea
\widetilde{S}_{A,+,\infty,2}^{(\alpha)}+\widetilde{S}_{A,-,\infty,2}^{(\alpha)} 
 & = &
  \frac{(\alpha+1)(3\alpha^{2}-7)}{24\alpha^{3}} 
  \\
  \rule{0pt}{.8cm}
 &  & 
 +\,
 \frac{4}{1-\alpha}\sum_{j=1}^{\infty}(-1)^{j} \cos(4\eta j) 
 \sum_{k=1}^{\infty}
\widetilde{\mathcal{P}}_{2}(2j;\hat{y}_{k} / \alpha)\,
 \bigg( \frac{\Omega(\hat{y}_{k} / \alpha)}{(4\eta)^{(2k-1) /\alpha}} \bigg)^{2j}
\nn
\eea
which agrees
with the result for $\widetilde{S}_{A,\infty,2}^{(\alpha)}$ 
found in \cite{Mintchev:2022xqh}.

It is important to verify that our results agree 
with the proper limit of the corresponding ones obtained on the lattice. 
In particular, taking the double scaling limit 
(defined in the text above \eqref{fd_FC_dsl})
of the expression in Eq.\,(57) of \cite{Fagotti:2010cc}, 
we find that $d_{\alpha, \textrm{\tiny there}}  \to \widetilde{S}^{(\alpha)}_{A,-,\infty,0}$ 
(see (\ref{Ren-0-large-eta-sub})), as expected.

In the literature we have not found  lattice results
whose continuum limit provide the subleading terms corresponding to 
 \eqref{Ren-1-large-eta-sub} and \eqref{Ren-2-large-eta-sub}.
 These lattice results can be obtained  by studying 
 the subleading corrections to \eqref{D_from_FC_proposal},
 as done e.g. in \cite{Calabrese-Essler-10} for the block in the infinite XX chain.

\bibliographystyle{nb}

\bibliography{refsSchrodBdy}

\end{document}